\RequirePackage{fix-cm}
\documentclass[aps,prd,10pt,twocolumn,amsmath,superscriptaddress,amssymb,showpacs,floatfix,nofootinbib,longbibliography]{revtex4-2}

\pdfoutput=1
\usepackage{hyperref}
\usepackage{graphicx}
\usepackage{amsfonts,amsmath,amssymb,bm,bbm,mathrsfs}
\usepackage{color}
\usepackage{url}
\usepackage{float}
\usepackage{slashed}
\usepackage{enumitem}
\usepackage{leftindex}
\usepackage[compat=1.1.0]{tikz-feynman}
\usepackage{feynmp-auto}
\usepackage[capitalise]{cleveref}
\usepackage{braket}

\hypersetup{
    pdfnewwindow=true,      
    colorlinks=true,       
    linkcolor=blue,          
    citecolor=blue,        
    filecolor=blue,      
    urlcolor=blue        
}

\newcommand{\orcid}[1]{\begingroup
  \!\hypersetup{hidelinks}\href{https://orcid.org/#1}{\includegraphics[width=10pt]{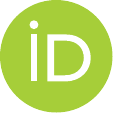}\!\!} \endgroup}

\newcommand{\appropto}{\mathrel{\vcenter{
  \offinterlineskip\halign{\hfil$##$\cr
    \propto\cr\noalign{\kern2pt}\sim\cr\noalign{\kern-2pt}}}}}

\DeclareMathSizes{10}{10}{5}{5}

\DeclareMathOperator{\arccot}{arccot}

\begin{document}

\title{Minimal Dark Matter: Generalized Framework and Direct-Detection Sensitivity}

\author{Spencer~Griffith \orcid{0009-0002-1988-0768}\,}
\email{griffith.1037@osu.edu}
\affiliation{Center for Cosmology and AstroParticle Physics (CCAPP), \href{https://ror.org/00rs6vg23}{Ohio State University}, Columbus, OH 43210}
\affiliation{Department of Physics, \href{https://ror.org/00rs6vg23}{Ohio State University}, Columbus, OH 43210}

\author{Juri Smirnov \orcid{0000-0002-3082-0929}\,}
\email{juri.smirnov@liverpool.ac.uk}
\affiliation{Department of Mathematical Sciences, 
\href{https://ror.org/04xs57h96}{University of Liverpool}, Liverpool, L69 7ZL, United Kingdom}

\author{Laura Lopez-Honorez \orcid{0000-0002-4149-3935}\,}
\email{Laura.Lopez.Honorez@ulb.be}
\affiliation{Service de Physique Th\'eorique, \href{https://ror.org/01r9htc13}{ Universit\'e Libre de Bruxelles}, B-1050 Brussels, Belgium}
\affiliation{Theoretische Natuurkunde \& The International Solvay Institutes, \href{https://ror.org/006e5kg04}{Vrije Universiteit Brussel}, B-1050 Brussels, Belgium}

\author{John F. Beacom \orcid{0000-0002-0005-2631}\,}
\email{beacom.7@osu.edu}
\affiliation{Center for Cosmology and AstroParticle Physics (CCAPP), \href{https://ror.org/00rs6vg23}{Ohio State University}, Columbus, OH 43210}
\affiliation{Department of Physics, \href{https://ror.org/00rs6vg23}{Ohio State University}, Columbus, OH 43210}
\affiliation{Department of Astronomy, \href{https://ror.org/00rs6vg23}{Ohio State University}, Columbus, OH 43210}

\date{\today}


\begin{abstract}
Minimal electroweak dark matter models are compelling due to their simplicity, though calculations of their freezeout abundance are complicated by nonperturbative effects due to Sommerfeld enhancement and bound-state formation.  It has been shown that all individual multiplet scenarios beyond the doublet lead to direct-detection signals above the neutrino floor and thus within the reach of next-generation experiments.  If no signals are found, would minimal dark matter be excluded? Yes for the simplest models, but it has been unknown for the important extension of two multiplets coupled by Higgs interactions (Higgs-coupled minimal dark matter). We present a generalized framework for calculating nonperturbative effects for such models that also covers the case of individual multiplets. In this framework, we calculate nonperturbative effects on freezeout as well as the prospects for direct detection, correcting shortcomings and omissions in the literature. Importantly, for the mixed Majorana (odd) and Dirac (even) multiplet combination $3M2D$ (and marginally the $5M4D$), we find that the predicted direct-detection signals can extend below the neutrino floor. \textit{Fully testing minimal dark matter will thus require more than direct-detection experiments.}
\end{abstract}

\maketitle


\section{Introduction}
\label{sec:introduction}

The particle nature of dark matter (DM) is one of the longest-running questions in physics~\cite{Bertone_2018}. The resolution of this question will have profound implications. For particle physics, understanding DM will reveal physics beyond the standard model (BSM); for examples, see Refs.~\cite{Carr:2017jsz, Ge:2019voa, Cirelli:2018iax, Berges:2019dgr, Blennow:2019fhy, Arcadi:2019lka}. For cosmology, understanding DM is crucial for understanding the overall evolution of the universe and the clustering of matter, see e.g. Refs.~\cite{Allen:2011zs, Salucci:2018hqu, Planck:2018vyg, Simon:2019nxf, Cirelli:2024ssz}.

Minimal dark matter (MDM) ~\cite{Cirelli:2005uq, Cirelli:2007xd, Cirelli:2018iax} is an especially attractive class of weakly interacting massive particle (WIMP) models~\cite{Steigman:1984ac, Bertone:2004pz, Steigman:2012nb, Arcadi:2017kky, Roszkowski:2017nbc}.  These models take the WIMP miracle seriously and propose that DM is a component of an electroweak $SU(2)_L$ multiplet that can be added to the standard model (SM) with little fuss.  At tree level, calculation of the the DM annihilation cross section to SM particles and the DM freezeout abundance is simple~\cite{Cirelli:2005uq}. However, because the DM masses are typically quite large, the weak bosons appear relatively massless, which makes it important to account for nonperturbative effects on the cross section due to Sommerfeld enhancement and bound-state formation.  This is analogous to the enhancement of $e^+ + e^-$ annihilation due to the attractive Coulomb potential and positronium formation~\cite{Smirnov:2022tcg}.  The Sommerfeld effect for DM has been calculated in, e.g., Refs.~\cite{Hisano:2003ec, Hisano:2004ds, Hisano:2006nn, Arkani-Hamed:2008hhe, Cassel:2009wt}, while bound-state effects have been calculated in, e.g., Refs.~\cite{March-Russell:2008klu, vonHarling:2014kha, An:2016gad, Mitridate:2017izz}.  A distinctive aspect of MDM models is that they appear to be fully testable by next-generation direct-detection experiments~\cite{Bottaro:2021snn, Bloch:2024suj, Baudis:2024jnk, PANDA-X:2024dlo}. Furthermore, current bounds from direct detection eliminate even multiplets due to their tree level coupling to the $Z$~\cite{Goodman:1984dc, CDMS:2005rss}, though this can be evaded by adding interactions beyond those of strictly MDM (see, e.g., Ref.~\cite{Cirelli:2024ssz}).

Higgs-coupled minimal dark matter (HC-MDM) is a modest and important extension to this model. Here, DM is a combination of two different multiplets that couple through the Higgs.  This is the simplest way of introducing Higgs-mediated interactions for fermionic DM, thus making use of all the players in the electroweak sector.  This model is still minimal in the sense that it contains only $SU(2)_L$ couplings. Such models have been studied extensively in the specific cases of the singlet-doublet~\cite{Mahbubani:2005pt, DEramo:2007anh, Enberg:2007rp, Cohen:2011ec, Cheung:2013dua, Calibbi:2015nha, Freitas:2015hsa, Banerjee:2016hsk}, doublet-triplet~\cite{Dedes:2014hga, Freitas:2015hsa, Beneke:2016jpw}, and the triplet-quadruplet~\cite{Tait:2016qbg}. In Ref.~\cite{LopezHonorez:2017zrd}, it was shown generally that if one combines two multiplets with a dimension difference of one and at most a small mass difference, the resulting mixed state behaves like an odd multiplet.  Under certain constraints, these particles can have large Higgs couplings while still evading direct-detection constraints. However, large Higgs couplings have a profound effect on the annihilation cross section through Sommerfeld effects and bound-state formation. This has not been well explored. For this model, Ref.~\cite{LopezHonorez:2017zrd} calculated the Sommerfeld effect but only roughly estimated bound-state formation effects. Then, in Refs.~\cite{Oncala:2021tkz, Oncala:2021swy}, these nonperturbative effects were calculated in detail, but only for the singlet-doublet case.

In this paper, our goal is to systematically explore mixed-multiplet HC-MDM models for higher multiplets, to calculate the effects on the freezeout predictions, and to assess how well these models can be probed by direct-detection experiments.  This is significantly different from the exclusively singlet-doublet case considered in  Refs.~\cite{Oncala:2021tkz, Oncala:2021swy} because the singlet has no $SU(2)_L$ coupling, which means that the potentials and cross sections lack some features found in larger multiplet combinations.  We begin by establishing the first complete framework for HC-MDM calculations, accounting for Sommerfeld and bound-state effects on DM freezeout and thus the allowed DM masses.  Our framework, which corrects various shortcomings and omissions in the literature, also applies (in appropriate limits) to MDM with single multiplets.  We reproduce previous results for the prediction of MDM masses~\cite{Asadi:2016ybp, Mitridate:2017izz, Smirnov:2019ngs, Bottaro:2021snn} up to the 13-plet, which is the limiting representation due to unitarity constraints~\cite{Smirnov:2019ngs, Bottaro:2021snn}. With our framework established, we address the key questions of the paper: \textit{what masses are required for HC-MDM to comprise the entirety of observed DM and what are the direct-detection prospects for HC-MDM models?}

The remainder of this paper is organized as follows. In Sec.~\ref{sec:model}, we review the MDM and HC-MDM models as well as establish notation that will be used throughout. In Sec.~\ref{sec:freezeout}, we present our formalism for determining the freezeout abundance and accounting for nonperturbative effects. In Sec.~\ref{sec:potentials and cross sections}, we calculate the long-range potentials of the DM particles that generate the nonperturbative corrections as well as the annihilation cross sections. In Sec.~\ref{sec:scattering states}, we find the Sommerfeld-enhanced annihilation cross section of scattering states. In Sec.~\ref{sec:bound states}, we determine the cross sections for bound-state formation. In Sec.~\ref{sec:results}, we apply these calculations to specific multiplet combinations to determine the DM masses and discuss the implications for direct detection. Finally, in Sec.~\ref{sec:conclusion} we conclude and discuss ways forward.


\section{MDM and HC-MDM models}
\label{sec:model}

In this section, we review the MDM and HC-MDM models and discuss some current experimental constraints. We also introduce the simplifications that we use in the remainder of our discussion. Finally, we establish the notation used in the following sections.

The MDM model proposes adding a single $SU(2)_L$ multiplet to the SM Lagrangian, which may be either a scalar or a fermion. We focus on the fermion case. (In the scalar case, the Higgs portal does not need the introduction of a second multiplet~\cite{Cirelli:2005uq, Hambye:2009pw}.) The SM Lagrangian is modified by adding the term:
\begin{equation}
    \mathscr{L} = \mathscr{L}_{\rm{SM}} + C \overline{X}(i \slashed{\mathscr{D}} + m_X)X,
    \label{eq:MDM lagrangian}
\end{equation}
where $\mathscr{D}$ is the covariant derivative, $m_X$ is the mass of the DM particle $X$, and  $C=1/2$ for self-conjugate (Majorana) $X$ or $C=1$ for non self-conjugate (Dirac) $X$.

The distinction between the Majorana and Dirac cases depends on the assignment of the weak hypercharge ($Y_X$), because a particle with $Y_X \ne 0$ clearly cannot be its own antiparticle. DM with non-zero hypercharge is already ruled out by direct-detection experiments (or minimality must be abandoned to appropriately cancel the offending interactions~\cite{Goodman:1984dc, CDMS:2005rss}). Furthermore, from the Gell-Mann--Nishijima relation:
\begin{equation}
    Q_X = I^3_X + Y_X,
    \label{eq:Gell-Mann}
\end{equation}
where $Q_X$ is the electric charge and $I^3_X$ is the 3rd component of isospin, we see that $Y_X$ is constrained by the requirement that one component of the multiplet is electrically neutral and, consequently, that only odd multiplets can meet the requirement of $Y_X=0$. Additionally, multiplets of dimension five or greater are stable against decay to the SM, while smaller multiplets can be stable if we impose a symmetry under which the DM is odd and the SM is even~\cite{Cirelli:2005uq}.

Reference~\cite{Cirelli:2005uq} showed that odd multiplets of dimension greater than five produce a Landau pole below the Planck mass. However, this does not rule out these multiplets. Instead, it points to the necessity for a UV completion scenario that eliminates the offending operators~\cite{Bottaro:2021snn}. Furthermore, Ref.~\cite{Smirnov:2019ngs} proposes a solution based on resummation techniques. We can, however, determine an upper limit on the multiplet size by demanding $s$-wave unitarity of the cross section. Under this constraint, multiplets up to the 13-plet are acceptable DM candidates~\cite{Smirnov:2019ngs, Bottaro:2021snn}. 

The odd multiplets up to the 13-plet lie slightly below current direct-detection bounds. Conveniently, as shown in Ref.~\cite{Bottaro:2021snn}, they also lie above the ``neutrino floor," the combination of cross section and mass below which neutrino backgrounds make direct detection extremely difficult~\cite{Billard:2013qya}. The proposed next generation of direct-detection experiments (such as XLZD~\cite{Baudis:2024jnk} and PandaX-xT~\cite{PANDA-X:2024dlo}) will probe down to neutrino floor. Indirect-detection bounds also put pressure on MDM models, as recent analyses of Fermi data (see e.g. Refs.~\cite{Fan:2013faa,Cohen:2013ama,Rodd:2024qsi,Safdi:2025sfs, Aghaie:2025iyn}) strongly disfavor the pure triplet scenario, even considering a large Milky way core. For larger multiplets, the exclusion bounds are less stringent as they depend on the exact DM mass and the core radius~\cite{Aghaie:2025iyn}. Furthermore, Ref.~\cite{Baumgart:2025dov} finds that the upcoming Cherenkov Telescope Array Observatory (CTAO) can test up to the 11-plet and place strong pressure on the 13-plet.

In the HC-MDM model, we add Majorana \textit{and} Dirac fermions to $\mathscr{L}_{\rm{SM}}$ ($M$ and $D$, respectively) with multiplet sizes differing by one. By requiring each multiplet to generically have a neutral member, we must have $Y_M = 0$ and $Y_D = 1/2$. The Lagrangian then acquires the terms:
\begin{equation}
    \begin{aligned}
    \mathscr{L} = &\mathscr{L}_{\rm{SM}} + \overline{D}(i \slashed{\mathscr{D}} + m_{D})D + \frac{1}{2} \overline{M}(i \slashed{\mathscr{D}} + m_{M})M \\
    &- y_1 D M H^* - y_2 \overline{D} M H,
    \label{eq:lagrangian}
\end{aligned}
\end{equation}
which exhausts the possible $M,~D$ interactions. As mentioned, if the representation size of either is less than five, we must impose an additional discrete symmetry to prevent decay.
\begin{figure}
\includegraphics[width=0.98\columnwidth]{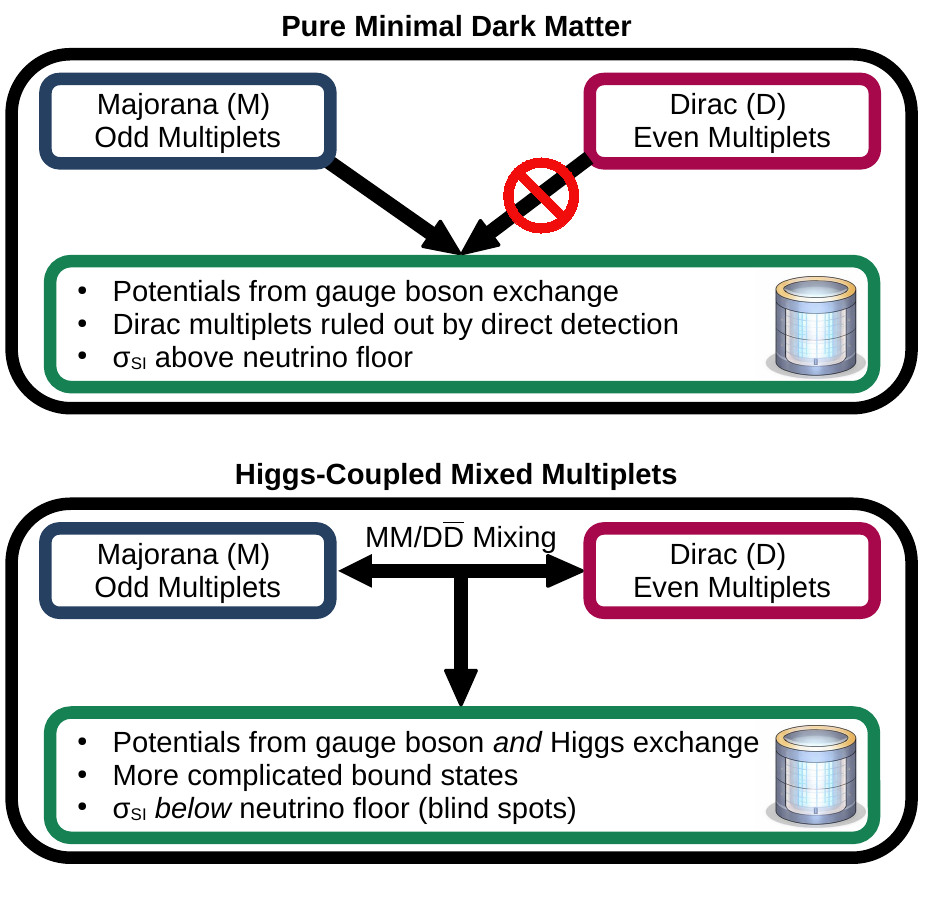}
\caption{Schematic diagram illustrating the salient differences between the MDM and HC-MDM models.}
\label{fig:schematic}
\end{figure}

Figure~\ref{fig:schematic} illustrates the transition from MDM to HC-MDM and highlights some important differences. Our assertion that the HC-MDM parameter space can extend below the neutrino floor is demonstrated in Sec.~\ref{sec:results}.

As shown in Ref.~\cite{LopezHonorez:2017zrd}, after electoweak symmetry breaking $M$ and $D$ can mix to a combined state where the lightest neutral particle behaves as a Majorana fermion and is therefore a viable DM candidate. For this to occur, we must have $|y_1| \approx |y_2|$. In the limit of equality $y_1=\pm y_2$, the Lagrangian of Eq.~(\ref{eq:lagrangian}) possesses a global symmetry, referred to as custodial symmetry (see, e.g., Refs.~\cite{Tait:2016qbg, LopezHonorez:2017zrd}). In this case, it is fairly easy to extract the mass eigenstates and their mixing as well as the couplings to the $Z$ and $H$ bosons. In particular, it can be shown that   when $y_1 = -y_2$ and $m_D = m_M$, the tree-level coupling of the lightest neutral component (which comprises the entirety of DM in the present universe) to the Higgs is zero (see Ref.~\cite{LopezHonorez:2017zrd}, Sec. 2.2.1). This avoids Higgs-induced DM scattering on nucleons at tree level. If this were non-zero, it would add to the scattering cross-section from gauge boson exchanges that controls the amplitude of the direct detection cross section of MDM~\cite{Hisano:2015rsa} (see, e.g., early studies of ``blind spots" within the singlet-doublet model~\cite{Cohen:2011ec, Cheung:2013dua, Calibbi:2015nha, LopezHonorez:2017zrd}). Loop-induced corrections involving $H$ exchanges may provide extra contributions to the scattering cross section. However, as $y_1=-y_2$ corresponds to a custodial symmetry, we still expect the gauge induced contributions to dominate the amplitude of the scattering cross section on nucleons.

To simplify our calculations, we restrict ourselves to the custodial point where $|y_1| = |y_2| = y$, $y_1=-y_2$, and $m_M = m_D = m$. In this case, because the direct detection cross-section is only due to the gauge boson loop-induced contributions as with MDM, we can use direct detection cross sections computed for MDM. We do not consider $H$ induced loop effects and simply note that this is a potential direction of future work. 

When dealing with negligible Higgs coupling, the requirement that $y_1 = -y_2$ is not strictly required and allowing both couplings to have the same sign has no effect on our results or on the direct detection cross section. However, we will maintain this choice to present our results in a unified way, noting that our conclusions are valid also in the $y_1=y_2$ case for small $y$.

These simplifications make sense when we consider mixed multiplets in generality. In Sec.~\ref{sec:results}, we choose reference values of $y_{1,2}$ to demonstrate the magnitude of the nonperturbative effects on freezeout abundance. However, if one has a particularly well-motivated model that includes specific values of $y_{1,2}$ and $m_{M,D}$, then it is straightforward to generalize our results, albeit with increased complexity and computational expense.

We also work in unbroken $SU(2)_L$ symmetry. This is because of our focus on the annihilation rates in the early universe to determine the relic abundance. We justify this in Sec.~\ref{sec:results}.

For convenience, we summarize our notation:
\begin{itemize}
    \item $X$ is a generic DM multiplet ($M$ or $D$).
    \item $R_X$ is the representation size of the $X$ multiplet. 
    \item $g_X$ is the degrees of freedom of $X$ ($2R_X$ and $4R_X$ for Majorana and Dirac particles, respectively). 
    \item $I_X$ is the isospin of $X$.
    \item $I^3_X(n)$ is the 3 component of isospin of the nth element of $X$ in ascending order. For example, for a triplet $X$, $I^3_X(1,2,3) = -1,0,1$.
    \item $t^a_X$ is the a-th $SU(2)$ generator of size of the representation of $X$.
    \item $C_2(R_X)$ is the quadratic Casimir of $R_X$.
    \item $V$ is a non-relativistic potential (the $r$ dependence is generall omitted). When necessary, a superscript indicates the representation size and subscripts indicate particles involved (e.g., $V_{X_1 X_2}^{R}$).
    \item For a process that changes the particle content (e.g., $X_1 + X_2 \rightarrow X_3 + X_4$), we use $X_1 X_2;X_3 X_4$. 
    \item The coupling strength $\alpha$ is defined by $V = -\alpha/r$, with $\alpha_{1,2}$ the $U(1)_Y$ and $SU(2)_L$ gauge couplings and  $\alpha_H = y^2/4 \pi$.
    \item Clebsch-Gordon (CG) coefficients are written $\langle J, M |j_1,m_1;j_2,m_2 \rangle$, where $J,~j_i$ denote angular momentum quantum numbers and $M,~m_i$ the projections onto the z-axis.
    \item $\hat{\epsilon}_{X}$ is the generalization of the Levi-Civita symbol to the dimension of $X$ used to contract indices (see Appendix~\ref{sec:contracting indices}), e.g.:
        \begin{equation*}
            \hat{\epsilon}_{3} = 
            \begin{pmatrix}
            0 && 0 && 1 \\
            0 && -1 && 0 \\
            1 && 0 && 0
            \end{pmatrix}.
        \end{equation*}
    \item Latin subscripts denote a multiplet component (e.g., $D_i$).
    \item Lower case $s$ generally denotes spin. However, $s$ will also be used for entropy density in Sec.~\ref{sec:freezeout}.
    \item Upper case $S$ is the Sommerfeld factor (see Sec.~\ref{sec:scattering states}).
    \item Cross sections are specified by $s$ and $I$: $\leftindex^s{(\sigma v_{\rm{rel}})}^{I}$.
    \item $n_f$ is the number of SM fermions.
    \item $IS$ and $FS$ denote initial and final states. $BS$ denotes a bound state.
    \item $z = m/T$ is a dimensionless time variable, where $T$ is the temperature and $z_f$ denotes $z$ at the freezeout temperature. Note that $z$ is \textit{not} used for redshift.
\end{itemize} 
%


\section{Freezeout abundance formalism}
\label{sec:freezeout}

In this section, we summarize results necessary for calculating the relic abundance of DM from the annihilation and bound state formation cross sections. We also justify the simplifications used to make the calculation of the relic abundance tenable. For the reader interested in the final result, note that~\cref{eq:YDM infinity,eq:zf,eq:omega_DM} are the necessary equations to reproduce our freeze-out calculations (for Eq.~(\ref{eq:YDM infinity}), we discuss the integration limit in Sec.~\ref{sec:results}).

The relic abundance of electroweak DM is determined by its
chemical decoupling from the SM plasma. We describe the evolution of the DM abundance using the comoving abundance $Y_{DM} = n_{DM} / s$, where $n_{DM}$ is the number density and $s = (2\pi^2/45) g_{*s} T^3$ is the entropy density. In terms of $Y_{DM}$, the Boltzmann equation is
\begin{equation}
\frac{dY_{DM}}{dz}
= - \frac{s}{z H}
  \, \langle \sigma v_{\rm rel} \rangle_{\rm eff}
  \left( Y_{DM}^2 - Y_{DM,{\rm eq}}^2 \right),
\label{eq:YDM_evo}
\end{equation}
where $H$ is the Hubble rate during radiation domination, $H = 1.66 \, g_*^{1/2} T^2 / M_{\rm Pl}$. Here, $\langle \sigma v_{\rm rel} \rangle_{\rm eff}$ is the effective thermally averaged inelastic cross section for annihilation to the SM (including nonperturbative effects) (see Refs.~\cite{Mitridate:2017izz,Mitridate:2017oky,Dondi:2019olm,Smirnov:2019ngs,Garny:2021qsr}):
\begin{equation}
\langle \sigma v_{\rm rel} \rangle_{\rm eff}
= \langle \sigma v_{\rm rel} \rangle_{\rm ann}
+ \sum_{BS_i} \langle \sigma v_{\rm rel} \rangle_{\rm form}^{BS_i} \, BR(BS_i; SM).
\label{eq:sigma_eff_Y}
\end{equation}
The first term corresponds to the Sommerfeld-enhanced annihilation of free $XX$ pairs,
while the second term accounts for the formation of unstable $XX$ bound states ($BS_i$)
that subsequently decay to the SM with branching ratio $BR(BS_i; SM)$. The branching ratio is given by
\begin{equation}
\begin{split}
    BR(BS_i; SM) &= \frac{\Gamma_{\rm ann}}{\Gamma_{\rm break} + \Gamma_{\rm ann}} \\
    &=\left( 1 + \frac{\langle \sigma v_{\rm rel} \rangle_{\rm ann}^{BS_i} m^3 e^{-z E_{BS_i}/m}}{2 R_{BS}(2s+1)(4\pi z)^{3/2}) \Gamma_i} \right)^{-1},
\end{split}
\label{eq:branching ratio}
\end{equation}
where $\Gamma_i$ is the annihilation rate of $BS_i$, $R_{BS}$ is the representation size of the bound state, and $E_{BS_i}$ is the binding energy~\cite{Smirnov:2019ngs}. Note the absence of the degrees of freedom squared factor in the second term of Eq.~(\ref{eq:branching ratio}) when compared to Ref.~\cite{Smirnov:2019ngs}. This is due to a difference in conventions. We choose not to average the annihilation cross sections over the DM degrees of freedom here and instead do so when we compute the contribution to the effective annihilation cross section (see Sec.~\ref{sec:scattering states}). However, in Ref.~\cite{Smirnov:2019ngs} the cross sections \textit{are} averaged and this factor serves to undo this averaging to apply to bound states. The branching ratio interpolates smoothly between the regimes of complete breaking of bound states (at $T \gg E_{BS_i}$) and efficient decay of bound states (at $T \ll E_{BS_i}$).

\vspace{0.4em}
\noindent
\textbf{Formation and breaking of bound states.}
Schematically, the relevant processes are 
\begin{equation}
X + X \;\xrightleftharpoons[\,\Gamma_{\rm break}\,]{\langle \sigma v_{\rm rel}\rangle^{BS}_{\rm form}}
(XX)_{BS}
\,\xrightarrow{\;\Gamma_{\rm ann}\;}
{\rm SM},
\label{eq:EW_BSF}
\end{equation}
where $(XX)_{BS}$ denotes a colorless electroweak bound state with quantum numbers fixed by the gauge representation. At early times, the thermal bath of SM particles breaks the newly formed
bound states on a timescale $\Gamma_{\rm break} \gg H$,
leading to detailed balance between formation and breaking:
$n_{BS} \simeq n_{BS,{\rm eq}}$, and thus negligible net depletion.
As the temperature drops below the binding energy, $T \lesssim E_{BS}$,
the breaking rate becomes Boltzmann suppressed,
$\Gamma_{\rm break} \propto e^{-E_{BS}/T}$,
and bound states annihilate promptly with $\Gamma_{\rm ann} \gg H$.
In this limit the bound-state population does not accumulate
and its effect can be incorporated through the effective rate
in Eq.~(\ref{eq:sigma_eff_Y}).

\vspace{0.4em}
\noindent
\textbf{Thermal average of the bound state formation rate.}
For a potential induced by the exchange of an electroweak gauge boson
of mass $m_V$, when $E_{BS} \gtrsim m_V$, the bound-state formation cross section is approximately (see Sec.~\ref{sec:bound states})
\begin{equation}
(\sigma v_{\rm rel})^{BS}_{\rm form}
\simeq \frac{2^9 \pi}{3} \,
\frac{\alpha_{\rm eff}}{m_X^2 v_{\rm rel}} \,
\frac{e^{-4\xi \arccot{\xi}}}{1 - e^{-2\pi\xi}},
\qquad
\xi = \frac{\alpha_{\rm eff}}{v_{\rm rel}},
\label{eq:EW_BSF_rate}
\end{equation}
where $\alpha_{\rm{eff}}$ is the coupling strength of the relevant potential.  
The thermally averaged rate $\langle \sigma v_{\rm rel} \rangle^{\rm BS}_{\rm form}$
is then evaluated by integrating over the Maxwell–Boltzmann distribution of relative velocities.
In the Coulombic regime, $\langle \sigma v_{\rm rel} \rangle^{BS}_{\rm form}$
scales approximately as
\begin{equation}
\langle \sigma v_{\rm rel} \rangle^{BS}_{\rm form}
\simeq C_{BS} \,
\frac{\alpha_{\rm eff}^3}{m_X^2 \sqrt{z}},
\label{eq:BSF_thermal_scaling}
\end{equation}
with $C_{BS}$ a numerical factor of order unity that depends on the gauge representation
and the available emission channels ($\gamma$, $W$ or $Z$ bosons, or Higgs).

\vspace{0.4em}
\noindent
\textbf{Effect on the relic yield.}
When Eq.~(\ref{eq:YDM_evo}) is solved with the effective cross section
of Eq.~(\ref{eq:sigma_eff_Y}), 
the enhanced depletion through bound-state formation shifts the freezeout
to larger values of $z$ (lower temperatures)
and reduces the final yield $Y_{DM}(\infty)$ relative to the standard case:
\begin{equation}
Y_{DM}(\infty) \simeq
\frac{Y_{DM}^{\rm ann}(\infty)}{1 + \Delta_{BS}},
\qquad
\Delta_{BS} \approx
\frac{\sum_{BS_i}\langle \sigma v_{\rm rel} \rangle^{BS_i}_{\rm form}}
{\langle \sigma v_{\rm rel} \rangle_{\rm ann}}
\bigg|_{z = z_f} .
\label{eq:Yinf_BS}
\end{equation}
For heavy electroweak multiplets, where $\alpha_{\rm eff} m_X / m_V \gg 1$,
the bound-state contribution is significant and must be included
in determining the relic abundance and the corresponding unitarity limit on the thermal DM mass.

More quantitatively, Eq.~(\ref{eq:YDM_evo}) has the approximate asymptotic solution (see Ref.~\cite{Smirnov:2019ngs}),
\begin{equation}
    Y_{DM}(\infty) = \frac{1}{\lambda} \left(\int_{z_f}^\infty \frac{\langle\sigma v_{\rm rel}\rangle_{\rm eff}(z)}{z^2} dz + \frac{\langle \sigma v_{\rm rel}\rangle_{\rm eff}(z_f)}{z_f^2} \right)^{-1},
    \label{eq:YDM infinity}
\end{equation}
where $\lambda=\sqrt{g_{SM} \pi/45}M_{\rm Pl} m$. $z_f$ is given by
\begin{equation}
    z_f = \ln{\left( \frac{2 (g_M + g_D) \langle \sigma v_{\rm rel}\rangle_{\rm eff}(z_f) \lambda}{(2 \pi z_f)^{3/2}}\right)}.
    \label{eq:zf}
\end{equation}
The DM density is then
\begin{equation}
\begin{split}
    \Omega_{\rm DM} &\equiv \frac{\rho_{\rm DM}}{\rho_{\rm crit}} = \frac{s Y_{\rm DM}(\infty) m}{3 H_0^2 / 8 \pi G}\\
    &= \frac{0.110}{h^2} \cdot \frac{Y_{\rm DM}(\infty) m}{0.4\,{\rm eV}},
    \label{eq:omega_DM}
\end{split}
\end{equation}
where $H_0$ and $G$ are the Hubble and gravitational constants.

In summary, the inclusion of electroweak bound-state formation and thermal breaking
modifies the effective inelastic rate governing the chemical decoupling of heavy DM.
Equation~(\ref{eq:YDM_evo}) provides a compact formulation that self-consistently accounts
for the transition from the regime ($T \gg E_{BS}$) to the annihilation-dominated regime
($T \ll E_{BS}$),
thereby capturing the dominant impact of bound-state dynamics on the
thermal freezeout of electroweak DM.


\section{Potentials and annihilation cross sections}
\label{sec:potentials and cross sections}

In this section, we present the long-range potentials between DM particles that generate Sommerfeld corrections and bound states. We go on to discuss our formalism for handling the mixed $MM/D \overline{D}$ states that arise due to Higgs exchange (analogous to the mixing of the $K^0$ and $\overline{K}^0$ mesons arising from $W^\pm$ exchange~\cite{ParticleDataGroup:2024cfk}). Finally, we show the annihilation cross sections to the SM for the various combinations of DM particles, with additional details in Appendix~\ref{sec:cross section calculations}.

\subsection{Potentials}
\label{subsec:potentials}
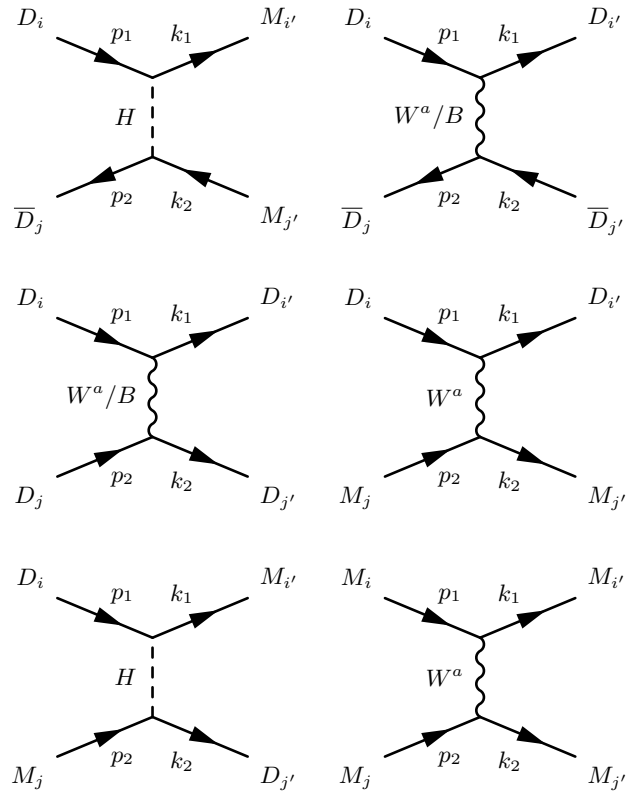
\begin{figure}
\begin{fmffile}{diagram1}
 \begin{fmfgraph*}(90,60)
   \fmfleft{i1,i2}
   \fmfright{o1,o2}
   \fmf{fermion, label=$p_2$,l.side=left}{v1,i1}
   \fmf{fermion,label=$k_2$,l.side=left}{o1,v1}
   \fmf{fermion,label=$p_1$,l.side=left}{i2,v2}
   \fmf{fermion,label=$k_1$,l.side=left}{v2,o2}
   \fmf{dashes, label=$H$}{v1,v2}
   \fmflabel{$\overline{D}_j$}{i1}
   \fmflabel{$M_{j'}$}{o1}
   \fmflabel{$D_i$}{i2}
   \fmflabel{$M_{i'}$}{o2}
 \end{fmfgraph*}
 \,\,\,\,\,\,\,\,\,\,\,\,\,\,\,\,\,\,
 \begin{fmfgraph*}(90,60)
   \fmfleft{i1,i2}
   \fmfright{o1,o2}
   \fmf{fermion, label=$p_2$,l.side=left}{v1,i1}
   \fmf{fermion,label=$k_2$,l.side=left}{o1,v1}
   \fmf{fermion,label=$p_1$,l.side=left}{i2,v2}
   \fmf{fermion,label=$k_1$,l.side=left}{v2,o2}
   \fmf{photon, label=$W^a/B$}{v1,v2}
   \fmflabel{$\overline{D}_j$}{i1}
   \fmflabel{$\overline{D}_{j'}$}{o1}
   \fmflabel{$D_i$}{i2}
   \fmflabel{$D_{i'}$}{o2}
 \end{fmfgraph*}
\end{fmffile}

\vspace{45pt}

\begin{fmffile}{diagram2}
 \begin{fmfgraph*}(90,60)
   \fmfleft{i1,i2}
   \fmfright{o1,o2}
   \fmf{fermion, label=$p_2$,l.side=right}{i1,v1}
   \fmf{fermion,label=$k_2$,l.side=right}{v1,o1}
   \fmf{fermion,label=$p_1$,l.side=left}{i2,v2}
   \fmf{fermion,label=$k_1$,l.side=left}{v2,o2}
   \fmf{photon, label=$W^a/B$}{v1,v2}
   \fmflabel{$D_j$}{i1}
   \fmflabel{$D_{j'}$}{o1}
   \fmflabel{$D_i$}{i2}
   \fmflabel{$D_{i'}$}{o2}
 \end{fmfgraph*}
 \,\,\,\,\,\,\,\,\,\,\,\,\,\,\,\,\,\,
 \begin{fmfgraph*}(90,60)
   \fmfleft{i1,i2}
   \fmfright{o1,o2}
   \fmf{fermion, label=$p_2$,l.side=right}{i1,v1}
   \fmf{fermion,label=$k_2$,l.side=right}{v1,o1}
   \fmf{fermion,label=$p_1$,l.side=left}{i2,v2}
   \fmf{fermion,label=$k_1$,l.side=left}{v2,o2}
   \fmf{photon, label=$W^a$}{v1,v2}
   \fmflabel{$M_j$}{i1}
   \fmflabel{$M_{j'}$}{o1}
   \fmflabel{$D_i$}{i2}
   \fmflabel{$D_{i'}$}{o2}
 \end{fmfgraph*}
\end{fmffile}

\vspace{45pt}

\begin{fmffile}{diagram3}
 \begin{fmfgraph*}(90,60)
   \fmfleft{i1,i2}
   \fmfright{o1,o2}
   \fmf{fermion, label=$p_2$,l.side=right}{i1,v1}
   \fmf{fermion,label=$k_2$,l.side=right}{v1,o1}
   \fmf{fermion,label=$p_1$,l.side=left}{i2,v2}
   \fmf{fermion,label=$k_1$,l.side=left}{v2,o2}
   \fmf{dashes, label=$H$}{v1,v2}
   \fmflabel{$M_j$}{i1}
   \fmflabel{$D_{j'}$}{o1}
   \fmflabel{$D_i$}{i2}
   \fmflabel{$M_{i'}$}{o2}
 \end{fmfgraph*}
 \,\,\,\,\,\,\,\,\,\,\,\,\,\,\,\,\,\,
 \begin{fmfgraph*}(90,60)
   \fmfleft{i1,i2}
   \fmfright{o1,o2}
   \fmf{fermion, label=$p_2$,l.side=right}{i1,v1}
   \fmf{fermion,label=$k_2$,l.side=right}{v1,o1}
   \fmf{fermion,label=$p_1$,l.side=left}{i2,v2}
   \fmf{fermion,label=$k_1$,l.side=left}{v2,o2}
   \fmf{photon, label=$W^a$}{v1,v2}
   \fmflabel{$M_j$}{i1}
   \fmflabel{$M_{j'}$}{o1}
   \fmflabel{$M_i$}{i2}
   \fmflabel{$M_{i'}$}{o2}
 \end{fmfgraph*}
\end{fmffile}

\vspace{20pt}

\caption{Diagrams for the interactions generating long-range potentials among $M$ and $D$. The spin indices are omitted because particle spins are individually conserved in the non-relativistic limit. The $M \overline{D}$ diagrams are omitted but follow straightforwardly from the $MD$ cases.}
\label{fig:potential diagrams}
\end{figure}

Figure~\ref{fig:potential diagrams} shows the diagrams arising from the Lagrangian in Eq.~(\ref{eq:lagrangian}) that generate long-range potentials. Because we are considering heavy DM annihilation after $T \ll m$, we are interested in the non-relativistic limit of these processes.

We specify the kinematics with $\vec{P}$ as the total momentum of the system and $\vec{p},~\vec{k}$ as the momenta $\vec{p}_1,~\vec{k}_1$ in the center of momentum frame. Then
\begin{equation}
    \vec{p}_1=\frac{\vec{P}}{2} + \vec{p},
    \,\,\, \vec{p}_2=\frac{\vec{P}}{2} - \vec{p},
    \,\,\, \vec{k}_1=\frac{\vec{P}}{2} + \vec{k},
    \,\,\, \vec{k}_2=\frac{\vec{P}}{2} - \vec{k}.
    \label{eq:potential kinematics}
\end{equation}
When considering long-range interactions, in the non-relativistic and $SU(2)_L$ symmetric ($m_H=m_W=m_B=0$) limits, the amplitudes are
\begin{equation}
\begin{split}
    &\mathcal{M}_{D \overline{D} ; MM} = \frac{4 m^2}{|\vec{p} - \vec{k}|^2} y_1 y_2 \\
    &\mathcal{M}_{D \overline{D}} = \frac{4 m^2}{|\vec{p} - \vec{k}|^2} (g_1^2 Y_D^2 - g_2^2 t^a_D t^a_{\overline{D}}) \\
    &\mathcal{M}_{DD} = -\frac{4 m^2}{|\vec{p} - \vec{k}|^2} (g_1^2 Y_D^2 + g_2^2 t^a_D t^a_D) \\
    &\mathcal{M}_{MD} = -  \frac{4 m^2}{|\vec{p} - \vec{k}|^2} (g_2^2 t^a_D t^a_M - y_1 y_2) \\
    &\mathcal{M}_{M \overline{D}} = -  \frac{4 m^2}{|\vec{p} - \vec{k}|^2} (g_2^2 t^a_{\overline{D}} t^a_M - y^2) \\
    &\mathcal{M}_{MM} = -  \frac{4 m^2}{|\vec{p} - \vec{k}|^2} g_2^2 t^a_M t^a_M,
    \label{eq:potential amplitudes}
\end{split}
\end{equation}
where we have omitted the $u$-channel diagrams where they appear (see below).

Relating the scattering amplitude to the Born approximation, we find that the non-relativistic potential is
\begin{equation}
        V = - \frac{1}{4 m^2} \int \frac{d^3 \vec{q}}{(2 \pi)^3} \mathcal{M}(\vec{q}) e^{i \vec{q} \cdot \vec{r}}.
        \label{eq:potential from amplitudes}
\end{equation}

Note that for identical particles, we should in principle include $u$-channel diagrams. However, to obtain the correct non-relativistic, long-range potentials, we consider only the $t$-channel diagrams. This is because the $u$-channel (exchange) interaction does not generate a contribution to the non-relativistic potential, but is instead encoded in the symmetry properties of the wave-function~\cite{sakurai1967advanced,Berestetskii:1982qgu}. As a first check on the validity of this approach, we note that this choice ensures that the non-relativistic, long-range Coulomb potential is the same for $e^- e^-$ ($t$- and $u$-channel diagrams) and $e^- \mu^-$ (t-channel diagram only), in accordance with the requirement that we recover Coulomb's law in the classical limit. In short, the reasoning behind this is that the $t$-channel diagram generates the correct long-range $1/r$ potential in the Coulomb limit, while the $u$-channel diagram generates only short-range contact interactions as we show below.

We present two complementary perspectives on the role of exchange diagrams. First, a localized wave-packet argument from non-relativistic quantum mechanics shows that the exchange matrix elements are suppressed at large separations. Then, beginning with the relativistic quantum field theory amplitudes for gauge boson exchange and matching onto a local long-range potential, we show that only the direct $t$-channel produces the soft singularity responsible for the Coulomb/Yukawa tail and we recover the same suppression of the $u$-channel contribution as in the non-relativistic case.

For the wave-packet argument, we begin with two identical, spatially localized fermions at large spatial separation in states $\ket{a}, \, \ket{b}$. By this, we mean two particles whose spatial wave-functions do not significantly overlap and are known to be separated in two regions of space. Alternatively, we can think of two particles sufficiently separated that one can conceivably construct an experiment that is sensitive to only one particle. We can then write the appropriately antisymmetrized state of the system as
\begin{equation}
    \ket{a,b}_{-} = \frac{1}{\sqrt{2}} \left( \ket{a}\ket{b} - \ket{b} \ket{a}\right).
    \label{eq:antisym}
\end{equation}

We decompose the Hamiltonian into 
\begin{equation}
    \hat{H} = \hat{H}_0 + \hat{V},
\end{equation}
where $\hat{H}_0$ contains the kinetic terms and $\hat{V}$ the potential due to the interaction. Because $\hat{H}_0$ is symmetric under particle exchange, the interesting behavior arises from the $\hat{V}$ term. The expectation value of this operator is
\begin{equation}
\begin{split}
    &E_{\hat{V}} = E_{\hat{V},{\rm dir}} + E_{\hat{V},{\rm ex}} \\
    &E_{\hat{V},{\rm dir}} = \frac{1}{2} \left(\bra{b}\bra{a}\hat{V}\ket{a}\ket{b} + \bra{a}\bra{b}\hat{V}\ket{b}\ket{a}\right) \\
    &E_{\hat{V},{\rm ex}} = - \frac{1}{2} \left(\bra{b}\bra{a}\hat{V}\ket{b}\ket{a} + \bra{a}\bra{b}\hat{V}\ket{a}\ket{b}\right).
    \label{eq:V expectation}
\end{split}
\end{equation}

$E_{\hat{V},{\rm dir}}$ corresponds to the $t$-channel process. When we transition to position space (using $\phi_{a/b}(\vec x) = \braket{\vec x | a/b}$) and employ the Born approximation, $E_{\hat{V},{\rm dir}}$ contains terms like
\begin{equation*}
    \int \int d^3\vec x_1 d^3\vec x_2 V(\vec x_1, \vec x_2) |\phi_a (\vec x_1)|^2 |\phi_b (\vec x_2)|^2.
\end{equation*}
Meanwhile, $E_{\hat{V},{\rm ex}}$ corresponds to the $u$-channel process. When we transition to position space and employ the Born approximation, $E_{\hat{V},{\rm ex}}$ contains terms like
\begin{equation*}
    \int \int d^3\vec x_1 d^3\vec x_2 V(\vec x_1, \vec x_2) \phi_a^* (\vec x_1) \phi_b (\vec x_1) \phi_a^* (\vec x_2) \phi_b (\vec x_2).
\end{equation*}
These terms contain two factors of the spatial overlap between $\ket{a}$ and $\ket{b}$. Because the particles are widely separated, $E_{\hat{V},{\rm ex}} \rightarrow 0$ and the $u$-channel process is irrelevant to the long-range potential, as expected.
 
We will now show the correspondence between the non-relativistic quantum mechanics formalism and the approach taken when we derive the potentials in a relativistic quantum field theory. Here, we start with the amplitudes for gauge boson exchange between identical fermions and take their Fourier transform. Working in the center-of-momentum frame, we define the momentum transfers for the $t$ and $u$-channel processes
\begin{equation}
\begin{split}
   \vec q_t = \vec p - \vec k \\
    \vec q_u = \vec p + \vec k.
\end{split}
\end{equation}
For exchange of a mediator of mass $m_V$, the non-relativistic $t$ and $u$-channel
kernels have the form
\begin{equation}
\begin{split}
    K_t(\vec q_t) &\propto \frac{1}{\vec q_t^2+m_V^2} \\
    K_u(\vec q_u)&\propto \frac{A}{\vec q_u^2+m_V^2},
\end{split}
\end{equation}
where $A$ is the appropriate anti-symmetrization factor for the $u$-channel diagram.

We then use Eq.~(\ref{eq:potential from amplitudes}) to obtain the local potential
\begin{equation}
   V\propto
-\int \frac{d^3\vec q_t}{(2\pi)^3} \left( \frac{1}{\vec q_t^{\,2}+m_V^2} + \frac{A}{\vec q_u^{\,2}+m_V^2} \right) e^{i\vec q_t \cdot \vec r}.
\end{equation}
The first term yields the familiar Yukawa potential
\begin{equation}
V_1 = -\int \frac{d^3\vec q_t}{(2\pi)^3} \frac{e^{i\vec q_t \cdot \vec r}}{\vec q_t^{\,2}+m_V^2}
=
-\frac{1}{4\pi r}e^{-m_V r},
\end{equation}
which reduces to the Coulomb form, $V_1 \propto -1/r$, in the SU(2)$_L$-symmetric ($m_V =0$) limit.

Unlike the first term, the second term is not a function of $\vec q_t$ alone. For the purpose of identifying
the long-range behavior, one may expand in the soft-transfer regime
$|\vec q_t|\ll |\vec p|$
\begin{equation}
\begin{split}
    \frac{1}{\vec q_u^2+m_V^2} &=  \frac{1}{(2\vec p-\vec q_t)^2+m_V^2}\\
    &= \frac{1}{4\vec p^{\,2}+m_V^2} + \frac{4\vec p\cdot \vec q_t -\vec q_t^{\,2}} {(4\vec p^{\,2}+m_V^2)^2} + \cdots.
\end{split}
\end{equation}
In the $m_V = 0$ limit, this expansion is analytic in $\vec q_t$ around $\vec q_t=0$. Therefore,
the $u$-channel kernel contains no soft singularity of the form
$1/\vec q_t^{\,2}$ and hence does not generate a long-range $1/r$ tail.
After transformation to coordinate space, such analytic terms correspond
to short-range exchange contributions, represented by contact operators
and their derivatives in an effective long-distance description.

To see this, we can bring the second term into a form similar to the first using $\vec{q}_t = \vec{q}_u - 2 \vec{k}$
\begin{equation}
\begin{split}
    V_2 &= - A e^{-2i \vec{k} \cdot \vec{r}} \int \frac{d^3\vec q_u}{(2\pi)^3} \frac{e^{i\vec q_u \cdot \vec r}}{\vec q_u^{\,2}+m_V^2} \\
    &= Ae^{-2i \vec{k} \cdot \vec{r}} V_{1}.
\end{split}
\end{equation}
We can then interpret the factor $e^{-2i \vec{k} \cdot \vec{r}}$ in terms of the free-particle wave-functions. Using $\vec{r} = \vec{x}_1 - \vec{x}_2$, where $\vec{x}_{1,2}$ are the positions of the two particles
\begin{equation}
\begin{split}
    V_2 &= A \left( e^{-i \vec{k} \cdot \vec{x}_1} e^{i \vec{k} \cdot \vec{x}_2} \right)^2  V_{1} \\
    &= A \left| \bra{\vec{k}} \overline{\psi}(\vec{x}_1) \ket{0} \bra{0} \psi(\vec{x}_2) \ket{\vec{k}} \right|^2 V_{1},
\end{split}
\end{equation}
where $\psi(\vec{x}),~\overline{\psi}(\vec{x})$ are the fermion field operators and we have omitted the spinor indices. Because the two terms inside the absolute value are the position-space representations of the two particles (see Ref.~\cite{Peskin:1995ev}), we have recovered the same dependence on the square of the overlap between the spatial wave-functions for the $u$-channel process that we found using non-relativistic quantum mechanics. Consequently, our conclusion about the irrelevance of the exchange interaction in the long-range potential is upheld.

The neglect of $u$-channel diagrams in the derivation of the long-range
potential is therefore not the statement that exchange effects are
absent. Rather, it reflects that these terms do not generate an
additional long-range Coulomb/Yukawa potential.

We note that operationally, the effect of ignoring the $u$-channel diagrams is to consider that the particles are distinguishable.  Arguments along these lines are given in Refs.~\cite{schiff1968quantum, cohen2019quantum}, which consider the particles distinguishable based on the wave-functions not overlapping significantly at long range, and Ref.~\cite{bethe1957}, which works in the context of the helium atom and finds that as the spatial separation between the two identical electrons grows, the exchange interaction becomes increasingly infrequent. The effects of particle identity are instead imposed through the symmetry properties of the two-body states and the corresponding selection rules in the annihilation and bound-state formation rates.

In Eq.~(\ref{eq:potential amplitudes}), the generator combinations can be simplified through the identity
\begin{equation}
\begin{split}
        t^a_{X_1} t^a_{X_2} &\rightarrow -\frac{1}{2} (C_2(R_{X_1}) + C_2(R_{X_2}) - C_2(R)) \\
        &= C_R^{X_1 X_2},
\end{split}
\label{eq:generator combination}
\end{equation}
where we have used $R$ for the representation size of the combined state.

Therefore, the non-relativistic potentials are
\begin{equation}
\begin{split}
    &V_{D \overline{D} ; MM} = - \frac{y_1 y_2}{4 \pi r} \\
    &V_{D \overline{D}} = - \frac{1}{r} \left( \frac{\alpha_1}{4} - \alpha_2 C_R^{D \overline{D}} \right) \\
    &V_{DD} = \frac{1}{r} \left( \frac{\alpha_1}{4} + \alpha_2 C_R^{DD} \right) \\
    &V_{MD} = \frac{1}{r} \left(\alpha_2 C_R^{MD} - \frac{y_1 y_2}{4 \pi} \right) \\
    &V_{M \overline{D}} = \frac{1}{r} \left(\alpha_2 C_R^{M \overline{D}} - \frac{y^2}{4 \pi} \right) \\
    &V_{MM} = \frac{\alpha_2 C_R^{MM}}{r}.
    \label{eq:potentials}
\end{split}
\end{equation}
In particular, $V_{D \overline{D} ; MM}$ and the contribution to $V_{MD}$ from scalar exchange is effectively $\propto y_1 y_2$ and is attractive if $y_1, y_2$ have the same sign as expected (see e.g., Ref.~\cite{Peskin:1995ev}). When $y_1, \, y_2$ have opposite signs, as is the case in the remainder of this work, these potentials become repulsive. We define $\alpha_H = y^2 / 4 \pi$ (because in our analysis $|y_1| = |y_2| = y$) and rewrite
\begin{equation}
\begin{split}
    &V_{D \overline{D} ; MM} = \frac{\alpha_H}{r} \\
    &V_{MD} = \frac{1}{r} \left(\alpha_2 C_R^{MD} + \alpha_H \right).
    \label{eq:potentials2}
\end{split}
\end{equation}
However, $V_{M \overline{D}}$ is insensitive to this change and does not acquire a sign change when written in terms of $\alpha_H$.

Furthermore, the $V_{DD}$ and $V_{D \overline{D};MM}$ long-range potentials due to gauge boson exchange are always nonzero, independent of spin and angular momentum quantum numbers. Reference~\cite{Oncala:2021tkz} obtained different results by considering the $u$-channel diagrams in their calculation of the potentials, which should only be valid in the short-range limit. This is irrelevant to the formation of bound states which still obey Pauli exclusion from the selection rules in the bound-state formation cross sections (Sec.~\ref{sec:bound states}). However, the Sommerfeld corrections are affected.


\subsection{Mixed states}
\label{subsec:mixed states}

Much of the methodology in this subsection is borrowed from Ref.~\cite{Oncala:2021tkz}, which considered the specific case of a Majorana singlet and Dirac doublet.

The first diagram in Fig.~\ref{fig:potential diagrams} allows for mixing between the $MM$ and $D \overline{D}$ states. Because the representation must match on both sides, and
\begin{equation}
\begin{split}
    &M \otimes M = 1 \oplus ... \oplus (2 R_M -1) \\
    &D \otimes \overline{D} = 1 \oplus ... \oplus (2 R_D -1),
\end{split}
\label{eq:tensor product}
\end{equation}
mixing can occur in the representations
\begin{equation}
    R \leq \min \{ 2 R_M -1,~ 2 R_D - 1 \}.
\label{eq:mixing reps}
\end{equation}

We form the vector
\begin{equation}
    \Phi(R)=\begin{pmatrix}
        \phi^{MM}(R) \\
        \phi^{D \overline{D}}(R)
    \end{pmatrix},
\label{eq:mix vector}
\end{equation}
where $\phi^{X_1 X_2}$ are the two particle wavefunctions and we have omitted the $\vec{r}$ dependence. This obeys the coupled Schrodinger equation
\begin{equation}
    \left( - \frac{\nabla^2}{m} + \hat{V}^R \right) \Phi(R) = \mathscr{E}(R) \Phi(R),
\label{eq:coupled schrodinger}
\end{equation}
where $\hat{V}^R$ is the potential matrix
\begin{equation}
    \hat{V}^R = - \frac{1}{r} \begin{pmatrix}
        \alpha_{MM}^R && \alpha_{D \overline{D} ; MM}^R \\
        \alpha_{D \overline{D} ; MM}^R && \alpha_{D \overline{D}}^R
    \end{pmatrix}.
\label{eq:mixing matrix}
\end{equation}

When $\alpha_H \ll \alpha_2, \alpha_1$,
\begin{equation}
    \hat{V}^R \rightarrow - \frac{1}{r} \begin{pmatrix}
        \alpha_{MM}^R && 0 \\
        0 && \alpha_{D \overline{D}}^R
    \end{pmatrix},
\end{equation}
so there is no $MM \rightleftharpoons D \overline{D}$ mixing, as expected.

When $\alpha_H$ is non-negligible ($\alpha_H \sim \alpha_2$), this matrix has eigenvalues
\begin{equation}
\begin{split}
    &\alpha_{\rm{mix1,2}}^R = \\
    &\frac{1}{2} \left(\alpha_{D \overline{D}}^R + \alpha_{MM}^R \mp \sqrt{(\alpha_{D \overline{D}}^R - \alpha_{MM}^R)^2 + 4 (\alpha_{D \overline{D} ; MM}^R)^2} \right)
    \label{eq:mixing eigenvalues}
\end{split}
\end{equation}
and eigenvectors
\begin{equation}
\begin{split}
    &\vec{E}_{\rm{mix1,2}}^R = C_{\rm{mix1,2}} \begin{pmatrix}
    \frac{\alpha_{MM}^R - \alpha_{D \overline{D}}^R \mp \sqrt{(\alpha_{D \overline{D}}^R - \alpha_{MM}^R)^2 + 4 (\alpha_{D \overline{D} ; MM}^R)^2}}{2 \alpha_{D \overline{D} ; MM}^R} \\
        1
    \end{pmatrix} \\
    &C_{\rm{mix1,2}} = \frac{1}{\sqrt{1 + \left( \frac{\alpha_{MM}^R - \alpha_{D \overline{D}}^R \mp \sqrt{(\alpha_{D \overline{D}}^R - \alpha_{MM}^R)^2 + 4 (\alpha_{D \overline{D} ; MM}}^R)^2}{2 \alpha_{D \overline{D} ; MM}^R} \right)^2}}
    \label{eq:mixing eigenvectors}
\end{split}
\end{equation}
where upper (lower) signs correspond to the mix1 (mix2) eigenstates.

We can use the matrix
\begin{equation}
    \hat{P}^R = \left( \vec{E}_{\rm{mix1}}^R, \vec{E}_{\rm{mix2}}^R \right) 
    \label{eq:P matrix}
\end{equation}
to diagonalize $\hat{V}^R$
\begin{equation}
\begin{split}
    \tilde{V}^R &= \left( \hat{P}^R \right)^{-1} \hat{V}^R \hat{P}^R \\
    &= - \frac{1}{r} \begin{pmatrix}
        \alpha_{\rm{mix1}}^R && 0 \\
        0 && \alpha_{\rm{mix2}}
    \end{pmatrix}.
\end{split}
\label{eq:V twidle}
\end{equation}
Then the vector
\begin{equation}
\begin{split}
    \tilde{\Phi}(R) &= \begin{pmatrix}
        \tilde{\phi}_1(R) \\
        \tilde{\phi}_2(R)
    \end{pmatrix} = \begin{pmatrix}
        w_{\rm{mix1MM}} \phi^{MM} + w_{\rm{mix1D \overline{D}}} \phi^{D \overline{D}} \\
        w_{\rm{mix2MM}} \phi^{MM} + w_{\rm{mix2D \overline{D}}} \phi^{D \overline{D}}
    \end{pmatrix}\\
    &= \left( \hat{P}^R \right)^{-1} \Phi(R) \hat{P}^R
\end{split}
\label{eq:phi twidle}
\end{equation}
obeys the uncoupled Schrodinger equation
\begin{equation}
    \left( - \frac{\nabla^2}{m} + \tilde{V}^R \right) \tilde{\Phi}(R) = \mathscr{E}(R) \tilde{\Phi}(R).
\label{eq:uncoupled schrodinger}
\end{equation}
The two states $\tilde{\phi}_{1,2}$ evolve in the potentials of magnitude $-\alpha_{\rm{mix1,2}}/r$, respectively.

When we compute the Sommerfeld corrections and bound-state formation rates, we need to work in terms of definite potentials. Therefore, we use the $\tilde{\phi}$ states. We then project back onto the $\phi^{MM,D \overline{D}}$ states to calculate annihilation rates.


\subsection{Annihilation cross sections}
\label{subsec:annihilation cross sections}

Here we compile the tree-level annihilation cross sections for $M$ and $D$ in the $SU(2)_L$ symmetric limit, which we apply later to the annihilation of scattering and bound states. We do not average over the initial degrees of freedom here because these are different between scattering and bound states.

For a single multiplet $X$, the tree-level annihilation is given in Ref.~\cite{Cirelli:2024ssz}:
\begin{equation}
\begin{split}
    (\sigma v_{\rm{rel}})_{XX} &= \frac{1}{16m_X^2 g_X} (\pi \alpha_2^2 (2 R_{X}^4 + 17 R_{X}^2 - 19) \\
    &+ 4 Y_{X}^2 \alpha_1^2 (41 + 8 Y_{X}^2) + 16 \alpha_1 \alpha_2 Y_{X}^2 (R_{X}^2 - 1)).
    \label{eq:tree level annihilation single multiplet}
\end{split}
\end{equation}

The contributions due to the hypercharge gauge boson, $B$, emission and mediation have factors of $Y_X$. For a Dirac fermion with $Y_X = 1/2$, we can compute the relative contribution of $B$ processes to the total cross section. For a doublet,
\begin{equation}
        \frac{(\sigma v_{\rm{rel}})_{_{XX}}^B}{(\sigma v_{\rm{rel}})_{_{XX}}^{\rm{tot}}} \approx 0.08.
        \label{eq:relative B contribution}
\end{equation}
The relative contribution is small for the doublet and decreases with increasing multiplet size, so we neglect this contribution. This is straightforward to add back in if one wants increased accuracy for a specific model.

Additionally, we only consider processes with zero angular momentum ($\ell=0$). Annihilation for $\ell>0$ is suppressed by higher powers of $\alpha$ relative to $\ell = 0$, so their contribution is sub-dominant~\cite{Mitridate:2017izz}. This is a conservative choice which underpredicts the overall annihilation cross section and therefore the mass.

For $M,\,D$ in the in the non-relativistic limit, we have:
\begin{equation}
\begin{split}
    &\leftindex^0( \sigma v_{\rm{rel}})_{D \overline{D};WW}^{0,2} =
    \frac{\pi \alpha_2^2}{m^2} \left( G^{ab}_{D \overline{D};WW} + G^{ba}_{D \overline{D};WW} \right)^2 \\
    &\leftindex^1( \sigma v_{\rm{rel}})_{D \overline{D};HH^*}^{0,1} =
    \frac{4 \pi \alpha_H^2}{m^2} \\
    &\leftindex^1( \sigma v_{\rm{rel}})_{D \overline{D};f \overline{f}}^{1} =
    \frac{2 \pi \alpha_2^2}{m^2} n_f G^{2}_{D \overline{D};f \overline{f}} \\
    &\leftindex^0( \sigma v_{\rm{rel}})_{MM;WW}^{0,2} =
    \frac{\pi \alpha_2^2}{2 m^2} \left( G^{ab}_{MM;WW} + G^{ba}_{MM;WW} \right)^2 \\
    &\leftindex^1( \sigma v_{\rm{rel}})_{MM;HH^*}^{0,1} =
    \frac{8 \pi \alpha_H^2}{m^2} \\
    &\leftindex^1( \sigma v_{\rm{rel}})_{MM;f \overline{f}}^{1} =
    \frac{\pi \alpha_2^2}{m^2} n_f G^{2}_{MM;f \overline{f}} \\
    &\leftindex^1( \sigma v_{\rm{rel}})_{DD;HH}^{0,1} =
    \frac{4 \pi \alpha_H^2}{m^2} \\
    &\leftindex^1( \sigma v_{\rm{rel}})_{\overline{D} \overline{D};H^* H^*}^{0,1} =
    \frac{4 \pi \alpha_H^2}{m^2} \\
    &\leftindex^1( \sigma v_{\rm{rel}})_{MD;WH}^{1/2,3/2} =
    \frac{2 \pi \alpha_H \alpha_2}{m^2} \left( \leftindex^\pm_1 G_{MD;WH} + \leftindex^\pm_2 G_{MD;WH} \right)^2 \\
    &\leftindex^1( \sigma v_{\rm{rel}})_{M \overline{D};WH^*}^{1/2,3/2} =
    \frac{2 \pi \alpha_H \alpha_2}{m^2} \left( \leftindex^\pm_1 G_{M \overline{D};WH^*} + \leftindex^\pm_2 G_{M \overline{D};WH^*} \right)^2
    \label{eq:annihilation cross sections}
\end{split}
\end{equation}
where the $G$ terms are gauge factors that depend on Clebsch-Gordon coefficients. For example,
\begin{equation}
    G^{ab}_{D \overline{D};WW} = \langle I_{IS},M_{IS}|I_D, I_D^3(i);I_D,I_D^3(j) \rangle (\hat{\epsilon}_{D} t^a_D t^b_D)_{ij}.
\end{equation}
The remaining gauge factors and derivations of the amplitudes are given in Appendix~\ref{sec:cross section calculations}. The cross sections for $MM/D \overline{D};HH^*$ specifically refer to the $t$-channel process. To account for $MM/D \overline{D}\rightarrow W \rightarrow HH^*$, we use the $MM/D \overline{D}\rightarrow W \rightarrow \bar f f$ cross sections, replacing $n_f=12 \rightarrow n_p=25/2$, where $ n_p$ is the number of SM fermions plus an additional $1/2$ that arises from this process.

\section{Annihilation of scattering states}
\label{sec:scattering states}

In this section, we discuss the direct annihilation process of two DM particles to SM particles (direct in that the annihilation proceeds without the formation of an intermediate bound state). We then detail how this will be applied to the effective cross section introduced in Sec.~\ref{sec:freezeout}.

For a two-particle process ($X_1+X_2 \rightarrow X_3+X_4$), the cross section is modified by the Sommerfeld factor:
\begin{equation}
    \sigma v \rightarrow S \sigma v.
    \label{eq:sommerfeld correction}
\end{equation}
This factor accounts for the long-range interactions driven by the potentials listed in Eq.~(\ref{eq:potentials}). In the $SU(2)_L$ symmetric limit with massless Higgs and vector bosons,
\begin{equation}
    S = \frac{2 \pi \alpha_{\rm{eff}}/v_{\rm{rel}}}{1 - e^{-2 \pi \alpha_{\rm{eff}} / v_{\rm{rel}} }}.
    \label{eq:sommerfeld factor}
\end{equation}

We now apply the Sommerfeld  corrections to the annihilation of scattering states. Because the Sommerfeld factor depends on $\alpha_{\rm eff}$, which itself depends on the particle combination and representation, we need to use the isospin decomposition of cross sections in Eq.~(\ref{eq:annihilation cross sections}). These cross sections also need to be averaged over the initial degrees of freedom. For scattering states, this gives a $(g_M + g_D)^2$ factor in the denominator.  Therefore,
\begin{equation}
    (\sigma v_{\rm{rel}})_{\rm{ann}} = \frac{2}{(g_M + g_D)^2} \sum_{j \geq i} \sum_i \sum_I \sum_{s=0,1}  \leftindex^s{\left( S \sigma v_{\rm{rel}} \right)}_{ij}^I,
    \label{eq:scattering state annihilation}
\end{equation}
where $i,\,j= 1,\,2,\,3$ correspond to $M,D,\overline{D}$ and $I=0,\,1/2,\,1,\,3/2,\,2$ to account for all possible SM final states. The factor of $2$ arises in different ways for $i \neq j$ and $i = j$. When $i \neq j$, this accounts for the two possible ways to form a combination of distinguishable particles (see Ref.~\cite{Griest:1990kh}). When $i=j$, this factor accounts for the two identical particles lost in the annihilation and effectively cancels a symmetry factor of $1/2$ in the cross section (see Ref.~\cite{Cirelli:2024ssz} Sec. 4.1.3). In the latter case, this factor would be more logically included in the Boltzmann equation. However, because we obtain a common factor in both cases, we include it in the effective cross section for convenience. We use $(\sigma v_{\rm{rel}})_{ij}$ to mean the cross section for annihilation of $i,j$ to any SM final state.

When $\alpha_H$ is large, the $MM$ and $D \overline{D}$ states do not have defined potentials due to mixing. Consequently, we exclude $MM$ and $D \overline{D}$ states from Eq.~(\ref{eq:scattering state annihilation}) and instead include the terms $\leftindex^s{\left( S \sigma v_{\rm{rel}} \right)}^I_{\rm{mix1,2}}$. From Eq.~(\ref{eq:phi twidle}), we see that
\begin{equation}
\begin{split}
    &\leftindex^s{\left( S \sigma v_{\rm{rel}} \right)}^I_{\rm{mix1,2}} = \\
    &2 \leftindex^s{\left( S_{\rm{mix1,2}} \left[ w_{mix1,2MM}^2 (\sigma v_{\rm{rel}})_{MM} +  w_{mix1,2D \overline{D}}^2 (\sigma v_{\rm{rel}})_{D \overline{D}}\right] \right)}^I.
\end{split}
\label{eq:mix state projection}
\end{equation}
%


\section{Formation and annihilation of bound states}
\label{sec:bound states}

In this section, we calculate the cross sections for formation of DM bound states as well as their annihilation rate to the SM. We conclude with the implementation for the mixed $MM / D \overline{D}$ states.

As with the scattering state annihilation cross sections, we only consider bound states with $\ell=0$. States with higher angular momentum annihilate inefficiently and generally break before annihilating. These states can transition to $\ell=0$ and then annihilate, but we ignore this because it is subdominant. This is a conservative choice whose overall effect is to underestimate the DM mass necessary to reproduce the observed density.


\subsection{Formation}
\label{subsec:bound-state formation}

The amplitudes and overlap integrals used in this section, are derived in detail in Appendix~\ref{sec:bound-state formation amplitudes}. We denote bound-state formation cross sections with a left subscript indicating the boson emitted (e.g., $\leftindex^s_{W}{\left(\sigma v_{\rm{rel}} \right)}_{X_1 X_2}^I$). The bound state also depends on the quantum numbers $n \ell m$, however these are suppressed in our notation.

We consider formation through emission of $W$, $B$ and $H$ bosons in Secs.~\ref{sec:Wemission}--\ref{sec:Hemission}, respectively.


\subsubsection{W emission}
\label{sec:Wemission}

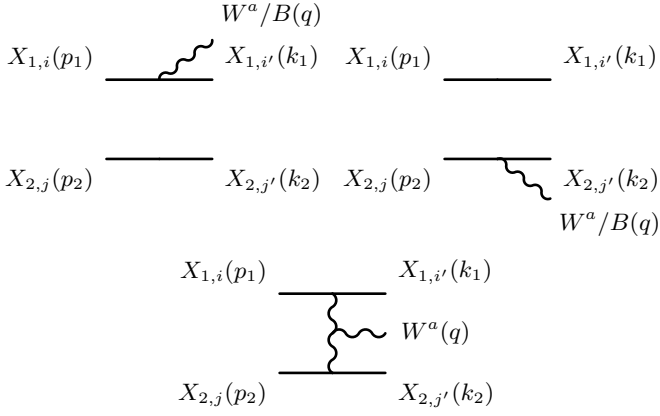
\begin{figure}
\begin{center}
\begin{fmffile}{diagram4}
\begin{fmfgraph*}(40,60)
   \fmfstraight
   \fmfleft{i1,i2,i3,i4,i5}
   \fmfright{o1,o2,o3,o4,o5}
   \fmf{plain}{i2,v2}
   \fmf{plain}{v2,o2}
   \fmf{plain}{i4,v4}
   \fmf{plain}{v4,o4}
   \fmf{photon,tension=0}{v4,o5}
   \fmflabel{$X_{2,j}(p_2)$}{i2}
   \fmflabel{$X_{2,j'}(k_2)$}{o2}
   \fmflabel{$X_{1,i}(p_1)$}{i4}
   \fmflabel{$X_{1,i'}(k_1)$}{o4}
   \fmflabel{$W^a/B(q)$}{o5}
\end{fmfgraph*}
\,\,\,\,\,\,\,\,\,\,\,\,\,\,\,\,\,\,\,\,\,\,\,\,\,\,\,\,\,\,\,\,\,\,\,\,\,\,\,\,\,\,\,\,\,\,\,\,\,\,\,\,\,
\begin{fmfgraph*}(40,60)
   \fmfstraight
   \fmfleft{i1,i2,i3,i4,i5}
   \fmfright{o1,o2,o3,o4,i5}
   \fmf{plain}{i2,v2}
   \fmf{plain}{v2,o2}
   \fmf{plain}{i4,v4}
   \fmf{plain}{v4,o4}
   \fmf{photon,tension=0}{v2,o1}
   \fmflabel{$X_{2,j}(p_2)$}{i2}
   \fmflabel{$X_{2,j'}(k_2)$}{o2}
   \fmflabel{$X_{1,i}(p_1)$}{i4}
   \fmflabel{$X_{1,i'}(k_1)$}{o4}
   \fmflabel{$W^a/B(q)$}{o1}
\end{fmfgraph*}
\end{fmffile}

\vspace{20pt}

\begin{fmffile}{diagram5}
\begin{fmfgraph*}(40,60)
   \fmfstraight
   \fmfleft{i1,i2,i3,i4,i5}
   \fmfright{o1,o2,o3,o4,o5}
   \fmf{plain}{i2,v2}
   \fmf{plain}{v2,o2}
   \fmf{plain}{i4,v4}
   \fmf{plain}{v4,o4}
   \fmf{phantom}{i3,v3}
   \fmf{photon}{v3,o3}
   \fmf{photon,tension=0}{v2,v4}
   \fmflabel{$X_{2,j}(p_2)$}{i2}
   \fmflabel{$X_{2,j'}(k_2)$}{o2}
   \fmflabel{$X_{1,i}(p_1)$}{i4}
   \fmflabel{$X_{1,i'}(k_1)$}{o4}
   \fmflabel{$W^a(q)$}{o3}
\end{fmfgraph*}
\end{fmffile}
\end{center}
\caption{Diagrams for bound-state formation through vector-boson emission. Mediators of the potential are not shown.}
\label{fig:vector emmision diagrams}
\end{figure}

Figure~\ref{fig:vector emmision diagrams} shows the processes $X_1 X_2 \rightarrow BS(X_1 X_2) + W$. Because $W$ has spin one and the individual particle spins are conserved in the process, we must have $|\Delta \ell| = 1$ from the initial to bound state. 
In the center of momentum frame, we have
\begin{equation}
    \sigma v_{\rm{rel}} = \frac{|\vec{q}|}{128 \pi^2 m^3} \int d \Omega \left| \epsilon_\mu^{a*}(q) \mathcal{M}^\mu\right|^2,
    \label{eq:BS W formation cross section 1}
\end{equation}
where $q$ is the $W^a$ 4-momentum, while  $\vec{q}$ denotes its 3-momentum, $\epsilon_\mu^a(q)$ is the polarization vector, and we have implicitly summed over the $W$ boson polarizations. We temporarily omit all indices on the cross sections and restore them at the end.

In the massless $W$ limit considered here, the sum over polarizations is $\sum_{\rm pol} \epsilon_\nu^a \epsilon_\mu^{a*} = -\eta_{\mu \nu}$. As in Ref.~\cite{Mitridate:2017izz}, we use the Ward identity, $q_\mu \mathcal{M}^\mu = 0$, to put the resulting expression entirely in terms of the spatial part of $\mathcal{M}^\mu$, denoted with $\vec{\mathcal{M}}$,
\begin{equation}
\begin{split}
    \sigma v_{\rm{rel}} &= \frac{|\vec{q}|}{128 \pi^2 m^3} \int d \Omega \left( | \vec{\mathcal{M}} |^2 - \frac{| \vec{q} \cdot \vec{\mathcal{M}}|^2}{|\vec{q}|^2} \right).
    \label{eq:BS W formation cross section 2}
\end{split}
\end{equation}
Using $\vec{\mathcal{A}} = \vec{\mathcal{M} }/( 8 \sqrt{m}g_2)$, we perform the angular integral to get
\begin{equation}
\begin{split}
    \sigma v_{\rm{rel}} &= \frac{8 |\vec{q}| \alpha_2}{m^2} \frac{2}{3} |\mathcal{\vec{A}}|^2.
    \label{eq:BS W formation cross section 3}
\end{split}
\end{equation}
Because the gauge boson carries the binding energy,
\begin{equation}
    |\vec{q}| \approx \frac{m v_{\rm{rel}}^2}{2} \left( 1 + \frac{\alpha_{BS}^2}{n^2 v_{\rm{rel}}^2}\right)\,.
\end{equation}

The diagrams in Fig.~\ref{fig:vector emmision diagrams} produce amplitudes with spatial parts of the form:
\begin{equation}
\begin{split}
    &i \vec{\mathcal{A}} = \\
    &\frac{1}{2} \left( (t^a_{X_1})_{i'k} (\hat{\epsilon}_{X_1})_{ki} (\hat{\epsilon}_{X_2})_{j'j} - (t^a_{X_2})_{j'k} (\hat{\epsilon}_{X_2})_{kj} (\hat{\epsilon}_{X_1})_{i'i}\right)\vec{J}^{ij,i'j'} \\
    &- i (t^b_{X_1})_{i'k} (\hat{\epsilon}_{X_1})_{ki} (t^c_{X_2})_{j'l} (\hat{\epsilon}_{X_2})_{lj} f^{abc} \vec{T}^{ij,i'j'},
    \label{eq:BS W formation amplitude}
\end{split}
\end{equation}
where $\vec{J}, \, \vec{T}$ are overlap integrals that depend on $p, \, n \ell m$ (see Appendix~\ref{sec:bound-state formation amplitudes}). We can decompose this into isospin channels and extract the $ij, \, i'j'$ dependence from the overlap integrals. Let
\begin{equation}
\begin{split}
   \leftindex^{W}_{J}{C}^{IS,BS}_{X_1 X_2} = &\frac{1}{2} \left( (t^a_{X_1} \hat{\epsilon}_{X_1})_{i'i} (\hat{\epsilon}_{X_2})_{j'j} - (\hat{\epsilon}_{X_1})_{i'i} (t^a_{X_2} \hat{\epsilon}_{X_2})_{j'j} \right) \\
   &\langle I_{IS}, M_{IS}|I_{X_1}, I^3_{X_1}(i);I_{X_2}, I^3_{X_2}(j) \rangle \\
   &\langle I_{BS}, M_{BS}|I_{X_1},I^3_{X_1}(i');I_{X_2},I^3_{X_2}(j') \rangle
\end{split}
\label{J coefficient W}
\end{equation}
and
\begin{equation}
\begin{split}
   \leftindex^{W}_{T}{C}^{IS,BS}_{X_1 X_2} = &i f^{abc}  \left( t^b_{X_1} \hat{\epsilon}_{X_1} \right)_{i'i} \left( t^c_{X_2} \hat{\epsilon}_{X_2} \right)_{j'j} \\
   &\langle I_{IS}, M_{IS}|I_{X_1}, I^3_{X_1}(i);I_{X_2}, I^3_{X_2}(j) \rangle \\
   & \langle I_{BS}, M_{BS}|I_{X_1},I^3_{X_1}(i');I_{X_2},I^3_{X_2}(j') \rangle.
\end{split}
\label{T coefficient W}
\end{equation}
Then
\begin{equation}
    i \vec{\mathcal{A}} = \leftindex^{W}_{J}{C}_{X_1 X_2}^{IS,BS} \vec{J} - \leftindex^{W}_{T}{C}_{X_1 X_2}^{IS,BS} \vec{T},
\end{equation}
and the overlap integrals have no dependence on the $SU(2)_L$ indices. We have omitted the $M_{IS,BS}$ arguments of $\leftindex^{W}_{J,T}{C}_{X_1 X_2}^{IS,BS}$, because these coefficients are contracted and summed over when $\vec{\mathcal{A}}$ is squared.

If $X_1 = X_2=X$, we also have $u$-channel diagrams. These get a factor of $-1$ from fermion exchange as a well as  a factor
\begin{equation}
    (-1)^{\ell + s + 1 + 2I_X - I},
    \label{eq:BSuchannel}
\end{equation}
where $I$ refers to the isospin of the combined state, from the symmetries of the wavefunction.  Note in particular that we have used the identity for CG coefficients 
\begin{equation}
    \langle J,M|j_1,m_1;j_2,m_2 \rangle = (-1)^{J - j_1 - j_2} \langle J,M|j_2,m_2;j_1,m_2 \rangle.
    \label{eq:CGid}
\end{equation} 
which gives rise to the power $2I_X-I$ in Eq.~(\ref{eq:BSuchannel}). Because $I_{BS} = I_{IS} \pm 1, \ell_{BS} = \ell_{IS} \pm 1,$ and $ s_{IS}=s_{BS}$, it does not matter whether we use the initial-state or bound-state quantum numbers, as long as we use choose one state consistently for all factors.

We also divide by a symmetry factor for the identical particles in the initial and final states. Therefore, we multiply the cross section by 
\begin{equation}
    \rho_{X_1X_2} = \begin{cases}
        1 & X_1 \neq X_2 \\
        \left( 1 - (-1)^{\ell + s + 1 + 2 I_X - I} \right)^2 / 4 & X_1 = X_2\,.
    \end{cases}
\end{equation}

Averaging over initial states and summing over spins, we obtain
\begin{equation}
\begin{split}
    \leftindex^s_{W}{\left(\sigma v_{\rm{rel}} \right)}_{X_1 X_2}^{I_{IS} I_{BS}} =& \frac{2s+1}{(g_M + g_D)^2} \frac{8 \alpha_2}{m^2} \frac{m v_{\rm{rel}}^2}{2} \left( 1 + \frac{\alpha_{BS}^2}{n^2 v_{\rm{rel}}^2} \right) \\
    &\frac{2}{3} \rho_{X_1X_2} \left| \leftindex^{W}_{J}{C}_{X_1 X_2}^{IS,BS} \vec{J} - \leftindex^{W}_{T}{C}_{X_1 X_2}^{IS,BS} \vec{T} \right|^2 .
\end{split}
\label{eq:W emission sigma}
\end{equation}
If we explicitly evaluate the overlap integrals, we obtain the results of Ref.~\cite{Mitridate:2017izz}, as expected.


\subsubsection{B emission}
\label{sec:Bemission}
For $B$-emission, we only have the first two diagrams from Fig.~\ref{fig:vector emmision diagrams}. We can read off the cross section from the $W$-emission case. In Eq.~(\ref{eq:BS W formation amplitude}) we replace
\begin{equation}
\begin{split}
    &(t^a_{X_1})_{i'k} (\hat{\epsilon}_{X_1})_{ki} (\hat{\epsilon}_{X_2})_{j'j} \rightarrow Y_{X_1} (\hat{\epsilon}_{X_1})_{i'i} (\hat{\epsilon}_{X_2})_{j'j} \\ &(t^a_{X_2})_{j'k} (\hat{\epsilon}_{X_2})_{kj} (\hat{\epsilon}_{X_1})_{i'i} \rightarrow Y_{X_2} (\hat{\epsilon}_{X_1})_{i'i} (\hat{\epsilon}_{X_2})_{j'j} \\
    &i (t^b_{X_1})_{i'i} (t^c_{X_2})_{j'j} f^{abc} \vec{T}^{ij,i'j'} \rightarrow 0\,.
\end{split}
\label{B replacements}
\end{equation}
We define
\begin{equation}
\begin{split}
   \leftindex^{B}_{J}{C}_{X_1 X_2}^{IS,BS} = &\frac{1}{2} \left( Y_{X_1} - Y_{X_2} \right) (\hat{\epsilon}_{X_1})_{i'i}(\hat{\epsilon}_{X_2})_{j'j} \\
   &\langle I_{IS}, M_{IS}|I_{X_1}, I^3_{X_1}(i);I_{X_2}, I^3_{X_2}(j) \rangle \\
   & \langle I_{BS}, M_{BS}|I_{X_1},I^3_{X_1}(i');I_{X_2},I^3_{X_2}(j') \rangle\,. \\
   &.
\end{split}
\label{J coefficient B}
\end{equation}
When $X_1 =X_2$ this is zero, so we do not need to consider $u$-channel factors. After averaging over initial states and summing over spins, we get
\begin{equation}
\begin{split}
    \leftindex^s_{B}{\left(\sigma v_{\rm{rel}} \right)}_{X_1 X_2}^I =& \frac{2s+1}{(g_M + g_D)^2} \frac{8 \alpha_1}{m^2} \frac{m v_{\rm{rel}}^2}{2} \left( 1 + \frac{\alpha_{BS}^2}{n^2 v_{\rm{rel}}^2} \right) \\
    &\frac{2}{3} \left| \leftindex^{B}_{J}{C}_{X_1 X_2}^{IS,BS} \vec{J} \right|^2 .
\end{split}
\label{eq:B emission sigma}
\end{equation}
%


\subsubsection{H emission}
\label{sec:Hemission}
\begin{figure}
\begin{fmffile}{diagram6}
\begin{fmfgraph*}(40,60)
   \fmfstraight
   \fmfleft{i1,i2,i3,i4,i5}
   \fmfright{o1,o2,o3,o4,o5}
   \fmf{plain}{i2,v2}
   \fmf{plain}{v2,o2}
   \fmf{plain}{i4,v4}
   \fmf{plain}{v4,o4}
   \fmf{dashes,tension=0}{v4,o5}
   \fmflabel{$M_{j}(p_2)$}{i2}
   \fmflabel{$M_{j'}(k_2)$}{o2}
   \fmflabel{$D_{i}(p_1)$}{i4}
   \fmflabel{$M_{i'}(k_1)$}{o4}
   \fmflabel{$H(q)$}{o5}
\end{fmfgraph*}
\,\,\,\,\,\,\,\,\,\,\,\,\,\,\,\,\,\,\,\,\,\,\,\,\,\,\,\,\,\,\,\,\,\,\,\,\,\,\,\,\,\,\,\,\,\,\,\,\,\,\,\,\,
\begin{fmfgraph*}(40,60)
   \fmfstraight
   \fmfleft{i1,i2,i3,i4,i5}
   \fmfright{o1,o2,o3,o4,i5}
   \fmf{plain}{i2,v2}
   \fmf{plain}{v2,o2}
   \fmf{plain}{i4,v4}
   \fmf{plain}{v4,o4}
   \fmf{dashes,tension=0}{v2,o1}
   \fmflabel{$M_{j}(p_2)$}{i2}
   \fmflabel{$M_{j'}(k_2)$}{o2}
   \fmflabel{$\overline{D}_{i}(p_1)$}{i4}
   \fmflabel{$M_{i'}(k_1)$}{o4}
   \fmflabel{$H^*(q)$}{o1}
\end{fmfgraph*}
\end{fmffile}

\vspace{20pt}

\caption{Diagrams for $MM$ bound-state formation through $H$ emission. We follow the convention of Fig.~\ref{fig:vector emmision diagrams} and omit the mediators of the potential.}
\label{fig:MM H emmision diagrams}
\end{figure}
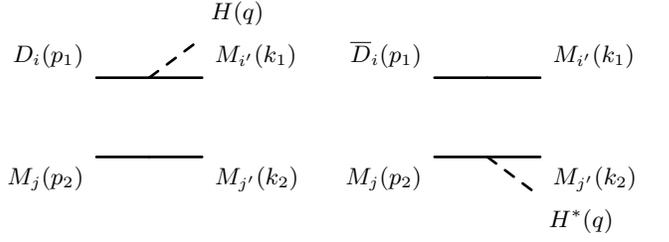

Figures~\ref{fig:MM H emmision diagrams} and~\ref{fig:MDbar H emmision diagrams} show two examples in detail from which we can extract all of the $H$-emission bound-state formation cross sections.

\vspace{0.4em}
\noindent
\textbf{$\mathbf{MM}$ bound states.}
$MM$ bound-state formation with Higgs emission occurs through the two diagrams in Fig.~\ref{fig:MM H emmision diagrams}. These diagrams do not interfere so we can compute the cross sections separately and add them together.

The t-channel contribution to the first diagram for a given initial state produces the cross section (again suppressing indices on the cross section)
\begin{equation}
\begin{split}
    \sigma v_{\rm{rel}} =& \frac{4 \alpha_H}{m^2} \frac{m v_{\rm{rel}}^2}{2} \left( 1 + \frac{\alpha_{BS}^2}{n^2 v_{\rm{rel}}^2} \right) \\
    &\left| (\hat{\epsilon}_{D})_{ki} \delta^{I^3_M(i')}_{I^3_D(k) \pm 1/2}(\hat{\epsilon}_{M})_{j'j} F^{ij,i'j'} \right|^2 \,,
\end{split}
\end{equation}
where $F$ is an overlap integral and we have divided by two for the identical final state particles. We define
\begin{equation}
\begin{split}
   \leftindex^{H}_{F}{C}_{DM;MM}^{IS,BS} = &(\hat{\epsilon}_D)_{i'i} (\hat{\epsilon}_M)_{j'j} \langle I_{IS}, M_{IS}|I_{D}, I^3_{D}(i);I_{M}, I^3_{M}(j) \rangle \\
   & \langle I_{BS}, M_{BS}|I_{M},I^3_{D}(i') \pm 1/2;I_{M},I^3_{M}(j') \rangle
\end{split}
\label{F coefficient H DM;MM case}
\end{equation}
which accounts for the contribution to the transition amplitude from each isospin channel.
We also have a $u$-channel version of this diagram that receives the same factor as in the $W$-emission case. Putting these pieces together,
\begin{equation}
\begin{split}
    \sigma v_{\rm{rel}} = &\frac{4 \alpha_H}{m^2} \frac{m v_{\rm{rel}}^2}{2} \left( 1 + \frac{\alpha_{BS}^2}{n^2 v_{\rm{rel}}^2} \right) \left( 1 - (-1)^{\ell + s + 1 + 2I_M - I_{BS}} \right)^2 \\
    &\left|  \leftindex^{H}_{F}{C}_{DM; MM}^{IS,BS} \right|^2 \left| F \right|^2 .
\end{split}
\label{eq:DM;MM H emission cross section}
\end{equation}

The swap $D\rightarrow \overline{D}$ in the initial state only affects the calculation by changing the initial state potential in the large $\alpha_H$ case. So the second diagram cross section is given by Eq.~(\ref{eq:DM;MM H emission cross section}) as well.

Adding the diagrams, averaging over initial states, and summing spins, we get the total $H$-emission cross section to form a $MM$ bound state,
\begin{equation}
\begin{split}
    \leftindex^s_{H}{\left(\sigma v_{\rm{rel}} \right)}_{MM}^I = &\frac{2s + 1}{(g_M + g_D)^2} \frac{4 \alpha_H}{m^2} \frac{m v_{\rm{rel}}^2}{2} \left( 1 + \frac{\alpha_{BS}^2}{n^2 v_{\rm{rel}}^2} \right) \\
    &\left( 1 - (-1)^{\ell + s + 1 + 2I_M - I_{BS}} \right)^2 \\
    &\left( \left|  \leftindex^{H}_{F}{C}_{DM; MM}^{IS,BS} \right|^2 \left| F \right|^2 + \left| \leftindex^{H}_{F}{C}_{\overline{D}M; MM}^{IS,BS} \right|^2 \left| F \right|^2\right).
\end{split}
\label{eq:MM H emission cross section}
\end{equation}
We have written $F$ separately for each term because it depends on the initial state, which is different between the terms.

\begin{figure}
\begin{center}
\begin{fmffile}{diagram7}
\begin{fmfgraph*}(60,60)
   \fmfstraight
   \fmfleft{i1,i2,i3,i4,i5}
   \fmfright{o1,o2,o3,o4,o5}
   \fmf{plain}{i2,v2}
   \fmf{plain}{v2,o2}
   \fmf{plain}{i4,v4}
   \fmf{plain}{v4,o4}
   \fmf{dashes,tension=0}{v4,o5}
   \fmflabel{$\overline{D}_{j}$}{i2}
   \fmflabel{$\overline{D}_{j'}$}{o2}
   \fmflabel{$D_{i}$}{i4}
   \fmflabel{$M_{i'}$}{o4}
   \fmflabel{$H$}{o5}
\end{fmfgraph*}
\end{fmffile}

\vspace{20pt}

\begin{fmffile}{diagram8}
\begin{fmfgraph*}(60,60)
   \fmfstraight
   \fmfleft{i1,i2,i3,i4,i5}
   \fmfright{o1,o2,o3,o4,o5}
   \fmf{plain}{i2,v2}
   \fmf{plain}{v2,o2}
   \fmf{plain}{i4,v4}
   \fmf{plain}{v4,o4}
   \fmf{dashes,tension=0}{v4,o5}
   \fmflabel{$\overline{D}_{j}$}{i2}
   \fmflabel{$\overline{D}_{j'}$}{o2}
   \fmflabel{$\overline{D}_{i}$}{i4}
   \fmflabel{$M_{i'}$}{o4}
   \fmflabel{$H^*$}{o5}
\end{fmfgraph*}
\,\,\,\,\,\,\,\,\,\,\,\,\,\,\,\,\,\,\,\,\,\,\,\,\,\,\,\,\,\,\,\,\,\,\,\,\,
\begin{fmfgraph*}(60,60)
   \fmfstraight
   \fmfleft{i1,i2,i3,i4,i5}
   \fmfright{o1,o2,o3,o4,i5}
   \fmf{plain}{i2,v2}
   \fmf{plain}{v2,o2}
   \fmf{plain}{i4,v4}
   \fmf{plain}{v4,o4}
   \fmf{dashes,tension=0}{v2,o1}
   \fmflabel{$\overline{D}_{j}$}{i2}
   \fmflabel{$M_{j'}$}{o2}
   \fmflabel{$\overline{D}_{i}$}{i4}
   \fmflabel{$\overline{D}_{i'}$}{o4}
   \fmflabel{$H^*$}{o1}
\end{fmfgraph*}
\end{fmffile}

\vspace{20pt}

\begin{fmffile}{diagram9}
\begin{fmfgraph*}(60,60)
   \fmfstraight
   \fmfleft{i1,i2,i3,i4,i5}
   \fmfright{o1,o2,o3,o4,o5}
   \fmf{plain}{i2,v2}
   \fmf{plain}{v2,o2}
   \fmf{plain}{i4,v4}
   \fmf{plain}{v4,o4}
   \fmf{dashes,tension=0}{v4,o5}
   \fmflabel{$M_{j}$}{i2}
   \fmflabel{$M_{j'}$}{o2}
   \fmflabel{$M_{i}$}{i4}
   \fmflabel{$\overline{D}_{i'}$}{o4}
   \fmflabel{$H$}{o5}
\end{fmfgraph*}
\,\,\,\,\,\,\,\,\,\,\,\,\,\,\,\,\,\,\,\,\,\,\,\,\,\,\,\,\,\,\,\,\,\,\,\,\,
\begin{fmfgraph*}(60,60)
   \fmfstraight
   \fmfleft{i1,i2,i3,i4,i5}
   \fmfright{o1,o2,o3,o4,i5}
   \fmf{plain}{i2,v2}
   \fmf{plain}{v2,o2}
   \fmf{plain}{i4,v4}
   \fmf{plain}{v4,o4}
   \fmf{dashes,tension=0}{v2,o1}
   \fmflabel{$M_{j}$}{i2}
   \fmflabel{$\overline{D}_{j'}$}{o2}
   \fmflabel{$M_{i}$}{i4}
   \fmflabel{$M_{i'}$}{o4}
   \fmflabel{$H$}{o1}
\end{fmfgraph*}
\end{fmffile}
\end{center}
\caption{Diagrams for $M \overline{D}$ bound-state formation through $H$ emission. We follow the convention of Fig.~\ref{fig:vector emmision diagrams} and omit the mediators of the potential.}
\label{fig:MDbar H emmision diagrams}
\end{figure}

\vspace{0.4em}
\noindent
\textbf{$\mathbf{M \overline{D}}$ bound states.}
The diagrams in Fig.~\ref{fig:MDbar H emmision diagrams} show the processes for $M \overline{D}$ bound-state formation with $H$-emission. We note that the diagrams in the second row do interfere and have $u$-channels, as do the diagrams in the third row. 

The diagram in the first row can be read off directly from the $MM$ case. We obtain Eq.~(\ref{eq:MM H emission cross section}) with the $u$-channel factor removed and $\leftindex^{H}_{F}{C}_{DM; MM}^{IS,BS} \rightarrow  \leftindex^{H}_{F}{C}_{D \overline{D}; M \overline{D}}^{IS,BS}$.

For the diagrams in the second row, temporarily suppressing indices on $\sigma v_{\rm{rel}}$, we get
\begin{equation}
\begin{split}
    &\sigma v_{\rm{rel}} = \frac{4 \alpha_H}{m^2} \frac{m v_{\rm{rel}}}{2} \left( 1 + \frac{\alpha_{BS}^2}{v_{\rm{rel}}^2 n^2} \right) \\
    &\left| (\delta^{I^3_M(i')}_{I^3_D(k) \pm 1/2} (\hat{\epsilon}_D)_{ik} (\hat{\epsilon}_D)_{jj'} + \delta^{I^3_M(j')}_{I^3_D(k) \pm 1/2} (\hat{\epsilon}_D)_{jk} (\hat{\epsilon}_D)_{ii'} ) F^{ij,i'j'} \right|^2.
\end{split}
\end{equation}
Here we need to be careful with indices. We accomplish this by making the ordering of subscripts significant. By that we mean that we write the process $\overline{D}_i \overline{D}_j ; M_{i'} \overline{D}_{j'}$ as $\overline{D} \overline{D} ; M \overline{D}$ and $\overline{D}_i \overline{D}_j ; \overline D_{i'} M_{j'}$ as $\overline{D} \overline{D} ; \overline{D} M$, where the placement of each particle on each side of the semi-colon corresponds to its identity.

We again decompose the cross section into isospin channels by defining
\begin{equation}
\begin{split}
   \leftindex^{H}_{F}{C}_{\overline{D} \overline{D};M \overline{D}}^{IS,BS} = &(\hat{\epsilon}_D)_{ii'} (\hat{\epsilon}_D)_{jj'} \langle I_{IS}, M_{IS}|I_{D}, I^3_{D}(i);I_{D}, I^3_{D}(j) \rangle \\
   & \langle I_{BS}, M_{BS}|I_{M},I^3_{D}(i') \pm 1/2;I_{D},I^3_{D}(j') \rangle.
\end{split}
\end{equation}
We see that 
\begin{equation}
    \leftindex^{H}_{F}{C}_{\overline{D} \overline{D} ; \overline{D} M}^{IS,BS} = \leftindex^{H}_{F}{C}_{\overline{D} \overline{D} ; M \overline{D}}^{IS,BS}(-1)^{1 + I_D + I_M - (I_{IS} + I_{BS})}.
\end{equation}
where we have used the identity in Eq.~(\ref{eq:CGid}). 
We can now write the cross section as
\begin{equation}
\begin{split}
    &\leftindex^s_{H}{(\sigma v_{\rm{rel}})}_{\overline{D} \overline{D}; M \overline{D}}^{IS,BS} = \frac{4 \alpha_H}{m^2} \frac{m v_{\rm{rel}}^2}{2} \left( 1 + \frac{\alpha_{BS}^2} {v_{\rm{rel}}^2 n^2} \right) \left| \leftindex^{H}_{F}{C}_{\overline{D} \overline{D};M \overline{D}}^{IS,BS} \right|^2 \\
    &|F|^2 (1 - (-1)^{\ell + s + 1 + 2I_D - I_{IS}})^2 ( 1 - (-1)^{I_D + I_M - (I_{IS} + I_{BS})})^2,
\end{split}
\end{equation}
which includes the $u$-channel factor. We note that $\ell, s$ do not change in the process so we do not need to specify whether they are taken in the initial state or bound state.

For the third row, a similar calculation gives
\begin{equation}
\begin{split}
    &\leftindex^s_{H}{(\sigma v_{\rm{rel}})}_{MM; M \overline{D}}^{IS,BS} = \frac{4 \alpha_H}{m^2} \frac{m v_{\rm{rel}}^2}{2} \left( 1 + \frac{\alpha_{BS}^2} {v_{\rm{rel}}^2 n^2} \right) \left| \leftindex^{H}_{F}{C}_{MM;M \overline{D}}^{IS,BS} \right|^2 \\
    &|F|^2 (1 - (-1)^{\ell + s + 1 + 2I_M - I_{IS}})^2 ( 1 + (-1)^{I_D + I_M - (I_{IS} + I_{BS})})^2 .
\end{split}
\end{equation}

Therefore, the total cross section to form a $M \overline{D}$ bound state through $H$ emission is
\begin{equation}
\begin{split}
    &\leftindex^s_{H}{(\sigma v_{\rm{rel}})}_{M \overline{D}}^{IS,BS} = \frac{2s + 1}{(g_{M} + g_{D})^2} \frac{8 \alpha_H}{m^2} \frac{m v_{\rm rel}^2}{2} \left(1 + \frac{\alpha_{BS}^2}{v_{\rm rel}^2 n^2} \right) \\
    &\bigg( \left| \leftindex^{H}_{F}{C}_{D \overline{D};M \overline{D}}^{IS,BS} \right|^2  \left| F \right|^2 + \frac{\left( 1 - (-1)^{\ell + s + 1 + 2I_D - I_{IS}} \right)^2}{2} \\
    &\left(1 + (-1)^{I_M + I_D - (I_{IS} + I_{BS})} \right)^2 \left| \leftindex^{H}_{F}{C}_{\overline{D} \overline{D};M \overline{D}}^{IS,BS} \right|^2  \left| F \right|^2 \\
    &+ \frac{\left( 1 - (-1)^{\ell + s + 1 + 2I_M - I_{IS}} \right)^2}{2} \left(1 - (-1)^{I_M + I_D - (I_{IS} + I_{BS})} \right)^2 \\
    &\left| \leftindex_{F}^{H}{C}_{MM;M \overline{D}}^{IS,BS} \right|^2  \left| F \right|^2 \bigg)\,.
\end{split}
\label{eq:MDbar H emission cross section}
\end{equation}
We again write $F$ separately for each term because it depends on the initial states.

From these two examples, we can now compile all of the $H$ emission $BS$ formation cross sections. Let
\begin{equation}
    A = \frac{2s + 1}{(g_M + g_D)^2} \frac{4 \alpha_H}{m^2} \frac{m v_{\rm rel}^2}{2} \left(1 + \frac{\alpha_{BS}^2}{v_{\rm rel}^2 n^2} \right).
\end{equation}
Then
\begin{equation}
\begin{split}
    &\leftindex^s_{H}{(\sigma v_{\rm{rel}})}_{M M}^{IS,BS} = A (1 + (-1)^{\ell + s + 2I_M - I_{BS}})^2 \\ &\left( \left|  \leftindex^{H}_{F}{C}_{DM; MM}^{IS,BS} \right|^2 \left| F \right|^2 + \left| \leftindex^{H}_{F}{C}_{\overline{D}M; MM}^{IS,BS} \right|^2 \left| F \right|^2\right) \\
    &\leftindex^s_{H}{(\sigma v_{\rm{rel}})}_{DD}^{IS,BS} = A (1 + (-1)^{\ell + s + 2I_D - I_{BS}})^2 \left| \leftindex_{F}^{H}{C}_{MD;DD}^{IS,BS} \right|^2 |F|^2 \\
    &\leftindex^s_{H}{(\sigma v_{\rm{rel}})}_{\overline{D} \overline{D}}^{IS,BS} = A (1 + (-1)^{\ell + s + 2I_D - I_{BS}})^2 \left| \leftindex_{F}^{H}{C}_{M \overline{D};\overline{D} \overline{D}}^{IS,BS} \right|^2 |F|^2 \\
    &\leftindex^s_{H}{(\sigma v_{\rm{rel}})}_{D \overline{D}}^{IS,BS} = A \left( \left| \leftindex_{F}^{H}{C}_{M D;\overline{D} D}^{IS,BS} \right|^2 |F|^2 + \left| \leftindex_{F}^{H}{C}_{M \overline{D};D \overline{D}}^{IS,BS} \right|^2 |F|^2 \right) \\
    &\leftindex^s_{H}{(\sigma v_{\rm{rel}})}_{M D}^{IS,BS} = {\rm Eq.}~(\ref{eq:MDbar H emission cross section}) \\
    &\leftindex^s_{H}{(\sigma v_{\rm{rel}})}_{M \overline{D}}^{IS,BS} = {\rm Eq.}~(\ref{eq:MDbar H emission cross section}). \\
\end{split}
\label{eq:H emission sigma}
\end{equation}
%


\subsection{Annihilation rates}
\label{subsec:bound state annihilation}

The annihilation rate for a bound state $BS_i$ is given by
\begin{equation}
    \Gamma_i = \left| \psi(0) \right|^2 (\sigma v_{\rm rel})_{\rm ann}^{BS_i},
\label{eq:annihilation rate}
\end{equation}
where $\psi(0)$ is the $BS$ wavefunction evaluated at $r=0$ and $(\sigma v_{\rm rel})_{\rm ann}^{BS_i}$ is the annihilation cross section of $BS_i$ averaged over the bound state degrees of freedom. The radial part of the bound state wavefunction in the $SU(2)_L$ symmetric limit with massless vector and Higgs bosons is given by
\begin{equation}
\begin{split}
    R_{nl} = &\left( \frac{\alpha_{\rm eff} m}{n} \right)^{3/2} \sqrt{\frac{(n-\ell-1)!}{2n(n + \ell)!}} e^{-r \alpha_{\rm eff} m / 2n} \left( \frac{r \alpha_{\rm eff} m}{n} \right)^\ell \\
    &L^{2 \ell + 1}_{n - \ell - 1} \left( \frac{r \alpha_{\rm eff} m}{n} \right),
\end{split}
\label{eq:bound state wavefunction}
\end{equation}
where $L$ are the Laguerre polynomials~\cite{Mitridate:2017izz}. Therefore,
\begin{equation}
    \left| \psi(0) \right|^2 = \frac{m^3 \alpha_{BS}^3}{2^3 \pi n^3}.
\end{equation}
This annihilation rate is then used to determine the branching ratio as described in Sec.~\ref{sec:freezeout}.

The contribution to the effective annihilation cross section is then 
\begin{equation}
    \sum_{BS_i} \langle \sigma v_{\rm rel} \rangle_{\rm form}^{BS_i} BR(BS_i; SM),
\end{equation}
where $(\sigma v_{\rm rel})_{\rm form}^{BS_i}$ is the total formation cross section of $BS_i$. When the bound state consists of identical particles, $BR(BS_i; SM)$ receives an additional factor of 2 to account for the disappearance of two particles of the same type in the annihilation because we have chosen not to include this in the Boltzmann equation (as described in Sec.~\ref{sec:scattering states}).


\subsection{Bound $D \overline{D} / MM$ states}
\label{subsec:bound mixed states}

When we work with mixed $MM \rightleftharpoons D \overline{D}$ states, we project onto the $MM,D \overline{D}$ states to find the formation cross section
\begin{equation}
    (\sigma v_{\rm rel})_{\rm form}^{\rm mix1,2} = w^2_{\rm{mix}1,2 MM} (\sigma v_{\rm rel})_{\rm form}^{MM} + w^2_{\rm{mix}1,2 D \overline{D}} (\sigma v_{\rm rel})_{\rm form}^{D \overline{D}}.
\end{equation}
Here we suppress the spin and isospin dependence. We then thermally average this cross section and project again to find the contribution to the effective cross section
\begin{equation}
\begin{split}
    \langle \sigma v_{\rm rel} \rangle^{\rm mix1,2}_{\rm form} \big( &w^2_{\rm{mix}1,2 MM} BR(MM \rightarrow SM) \\
    &+ w^2_{\rm{mix}1,2 D \overline{D}} BR(D \overline{D} \rightarrow SM) \big).
\end{split}
\end{equation}
%


\section{Results}
\label{sec:results}

In this section, we apply our formalism to determine the DM mass necessary for a viable HC-MDM model.  We consider the $3M2D,\, 5M4D,\, 7M6D,\, 9M8D,\, 11M10D,\,{\rm and}~13M12D$ representations and match the relic abundance to the observed DM mass density. We then use these results to determine the direct-detection prospects of HC-MDM. Finally, we discuss additional constraints on the Higgs coupling imposed by the unitarity bound.


\subsection{Methodology}
\label{subsec:methodology}

We have constrained the couplings $y_{1,2}$ to have identical magnitudes ($y$) and opposite signs. However, these constants can take a continuum of values up to the limit of perturbativity. Therefore, to demonstrate their effect on the DM mass, we choose three values of $y$ covering a range of scenarios and determine the mass of the DM particles in each case. We consider a small coupling ($y_{\rm small} = 10^{-9}$), a mid-range coupling ($y_{\rm mid} = 0.005$), and a large coupling ($y_{\rm large}=1$).

From the dependence of the branching ratio for bound states to the SM on the binding energy (see Eq.~(\ref{eq:branching ratio})), we see that we only need to consider the most deeply bound states for a given Majorana-Dirac multiplet combination. We determine the most deeply bound state for a particular combination, then consider any bound states with binding energy greater than a quarter of this value. Consideration of less deeply bound states is relatively straightforward, however the calculational expense quickly becomes overwhelming, especially for larger multiplets.

We can justify this choice by noting that for the $5M4D,\,7M6D,\,9M8D,\,11M10D, \, {\rm and}\,13M12D$ cases, removing the least deeply bound state that we \textit{do} consider changes the final mass determination by $< 6\%$ in the $y_{\rm large}$ case (where the bound state effects are most significant). For the $3M2D$ case, due to the small number of bound states considered, this is no longer true and the difference is more significant. However, adding the first bound state with $E_{BS}<E_{BS,{\rm max}}/4$ changes the result by $< 1\%$, so our truncation is still justified in this case.

Our choice to only consider states with $E_{BS}>E_{BS,{\rm max}}/4$ excludes states with $n>1$, so all states within our consideration are manifestly $\ell=0$ states. If one includes less deeply bound states, then it becomes necessary to consider states with non-zero angular momentum. As mentioned in Sec.~\ref{sec:bound states}, states with $|\ell|>0$ annihilate innefficiently and generally break in the thermal bath or decay to $\ell=0$ states then annihilate. The latter is subdominant to the direct formation and annihilation of $\ell=0$ states (see Ref.~\cite{Mitridate:2017izz}). Consideration of these states is beyond the scope of this work.

Similarly, bound states of representation size greater than five cannot annihilate directly to the standard model. Instead, these states can decay to lower representations through gauge boson emission, then annihilate to the SM. However, this is again a subdominant effect (considering that these states are necessarily not the most deeply bound states, see~\cref{eq:generator combination,eq:potentials}) and is ignored here.

We also need to consider the maximum value of $z$ that we integrate Eq.~(\ref{eq:YDM infinity}) over to find the relic abundance. Because we are working in the $SU(2)_L$ symmetric limit and a component-wise calculation valid after symmetry breaking is beyond the scope of this work, the obvious answer is to integrate up to $z_{\rm sym}=m/T_{\rm sym}$, where the subscript ``sym" denotes the temperature of symmetry breaking. This choice would greatly underestimate the DM mass because it would ignore a sizeable amount of annihilation after symmetry breaking. However, as noted in Ref.~\cite{Mitridate:2017izz}, the $SU(2)_L$ symmetric approximaton remains approximately valid up $T_{\Delta m} \approx \Delta m \approx \alpha_{\rm em} M_W $, where $\Delta m$ is the mass splitting between members of the $SU(2)_L$ multiplet after symmetry breaking (see Ref.~\cite{Cirelli:2005uq}). So we take as the upper limit of integration $z_{\Delta m} = m / 0.58~{\rm GeV}$. This still ignores some amount of annihilation taking place after the temperature decreases below the mass splitting, and a component-wise calculation becomes necessary for increased accuracy.

The effect of these choices is to mildly underestimate the overall annihilation cross section and, consequently, the DM mass. A more detailed treatement that avoids these simplifications may be justified for cases of particular interest.

\begin{figure}
\includegraphics[width=0.98\columnwidth]{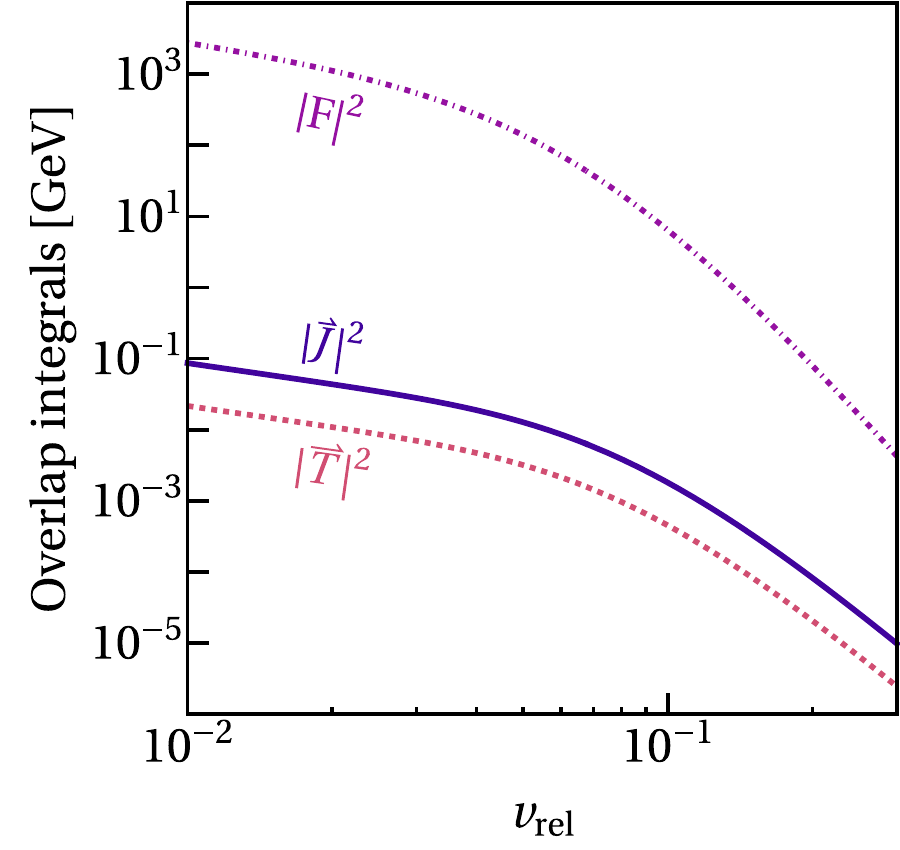}
\caption{Comparison of the overlap integral squared norms. All values taken with $m=50~{\rm TeV}$, $\alpha_{IS}=\alpha_2$, and $\alpha_{BS}=2 \alpha_2$.}
\label{fig:overlap integrals}
\end{figure}

Figure~\ref{fig:overlap integrals} shows the squared norms of the overlap integrals $\vec{J}, \, \vec{T},$ and $F$ as functions of $v_{\rm rel}$ with arbitrary (but reasonable) choices of $m=50~{\rm TeV}$, $\alpha_{IS}=\alpha_2$, and $\alpha_{BS}=2 \alpha_2$ to compare their magnitudes. These significantly affect the $W,B,H$ emission bound state formation amplitudes (see Eqs.~(\ref{eq:W emission sigma}, \ref{eq:B emission sigma}, \ref{eq:H emission sigma})). We see that for values of $\alpha_H$ comparable to $\alpha_2 = 1/29.6$ (taken at the $Z$-boson mass scale for simplicity), the Higgs emission cross sections completely dominate the bound state formation. This is also the motivation behind our choice for $y_{\rm mid}$. This value gives $\alpha_H|F|^2 \sim \alpha_2 |\vec{J}|^2$, so $W$ and $H$ emission cross sections are similar.

\subsection{Small $\alpha_H$}
\label{subsec:small aH}

Table~\ref{table:small aH bound states} shows the relevant bound states meeting the criteria described above for the $y_{\rm small}$ case. As noted in Sec.~\ref{subsec:mixed states}, for small values of $y$, the mixed states $MM \rightleftharpoons D\overline{D}$ are essentially pure $MM$ and $D \overline{D}$, so we specify the states in this limit as ${\rm mix}_1 \rightarrow D \overline{D}$ and ${\rm mix}_2 \rightarrow M M$. We make the additional distinction that states which cannot annihilate efficiently are excluded. For example, the $MD$ doublet has a binding energy above the threshold for the 5M4D and larger multiplets, however, this state cannot efficiently annihilate due to the smallness of $\alpha_H$ (see the cross sections in Eq.~(\ref{eq:annihilation cross sections})), so it is not included.
\begin{table}[b]
\centering
\begin{tabular}{||c | c | c | c | c | c||} 
 \hline
 3M2D & 5M4D & 7M6D & 9M8D & 11M10D & 13M12D\\ [0.5ex] 
 \hline\hline
 $MM^1$ & $MM^1$ & $MM^1$ & $MM^1$ & $MM^1$ & $MM^1$\\ 
 $MM^3$ & $MM^3$ & $MM^3$ & $MM^3$ & $MM^3$ & $MM^3$\\
 & ${D \overline{D}}^1$ & $MM^5$ & $MM^5$ & $MM^5$ & $MM^5$\\
 & $MM^5$ & ${D \overline{D}}^1$ & ${D \overline{D}}^1$ & ${D \overline{D}}^1$ & ${D \overline{D}}^1$\\
 &  & ${D \overline{D}}^3$ & ${D \overline{D}}^3$ & ${D \overline{D}}^3$ & ${D \overline{D}}^3$\\
 &  &  & ${D \overline{D}}^5$ & ${D \overline{D}}^5$ & ${D \overline{D}}^5$\\ [1ex] 
 \hline
\end{tabular}
\bigskip
\caption{Bound states above the $E_{BS}$ consideration threshold for $y_{\rm small}=10^{-9}$. Superscripts denote representation size. States are organized from largest to smallest binding energy. All states are $n=1$, $\ell=0$.}
\label{table:small aH bound states}
\end{table}

\begin{figure}
\includegraphics[width=0.98\columnwidth]{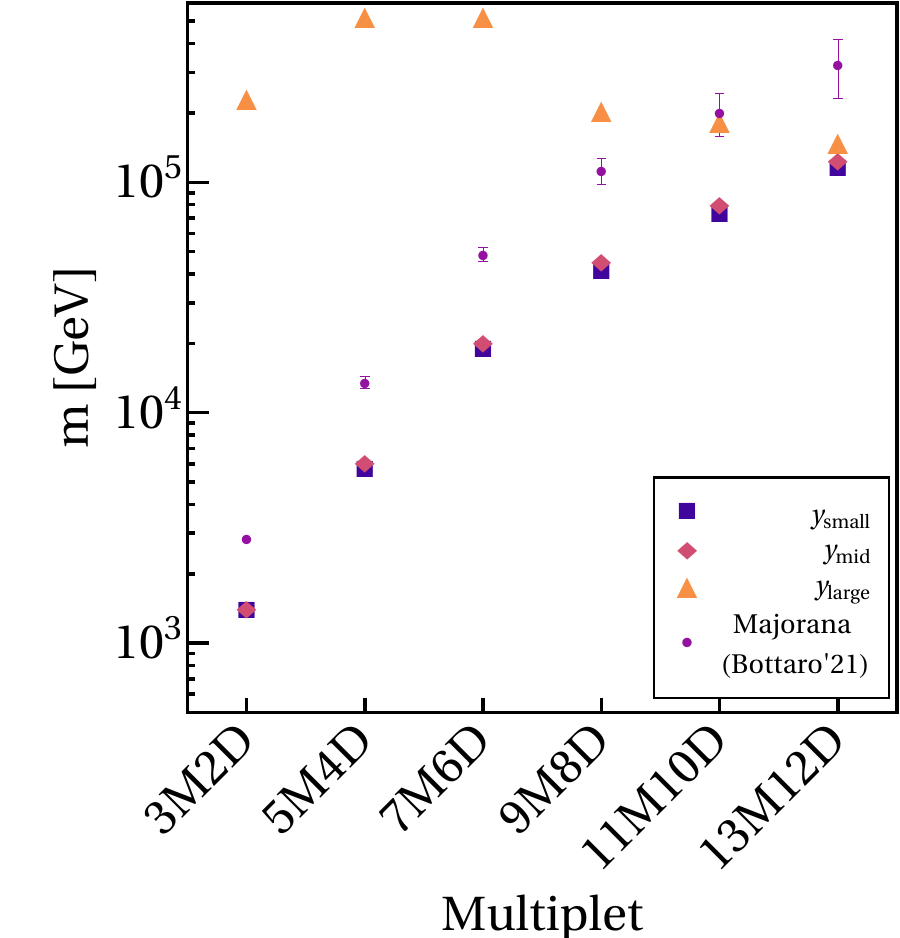}
\caption{DM mass needed to match the observed DM density to the computed relic abundance with $y_{\rm small}$, $y_{\rm mid}$, and $y_{\rm large}$. Also shown are the corresponding pure Majorana states from Ref.~\cite{Bottaro:2021snn}. Our values for the pure Majorana case are discussed in Appendix~\ref{sec:pure majorana multiplets}.}
\label{fig:large mid small aH masses}
\end{figure}

Figure~\ref{fig:large mid small aH masses} shows the DM mass necessary to match the observed DM density to the computed relic abundance for the $3M2D$, $5M4D$, $7M6D$, $9M8D$, $11M10D$, and $13M12D$ combinations as well as the pure Majorana case corresponding to each pair. In all cases, we find that the DM mass for the mixed multiplet case is less than the pure Majorana case. In this situation, the Higgs coupling has a negligible effect on the long-range potentials. Furthermore, annihilation into the Higgs is negligible, so the only particle pairs that can efficiently annihilate are $MM$ and $D \overline{D}$. This means that only a fraction of the potential interacting pairs can annihilate, so this result matches our intuitive expectation. The exact value of the difference between the pure Majorana case and the mixed case is determined by this effect as well as the different potentials experienced by the $MM$ and $D \overline{D}$ pairs.

\begin{figure}
\includegraphics[width=0.98\columnwidth]{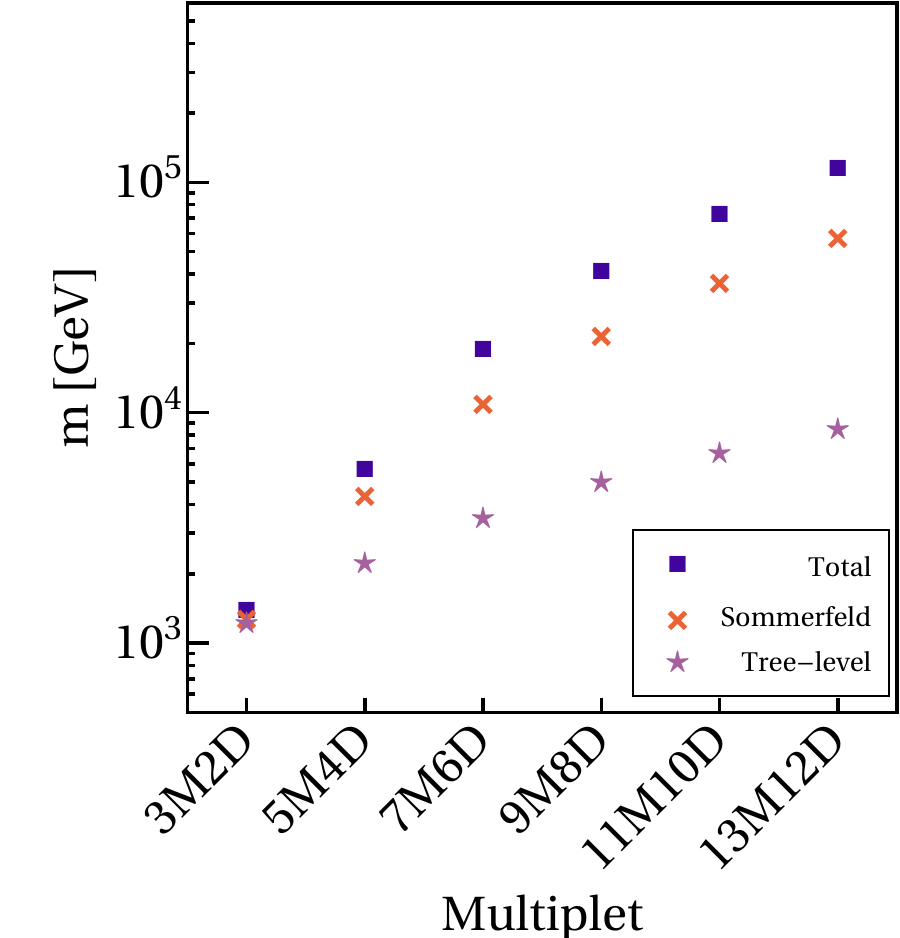}
\caption{Comparison of DM masses obtained from the tree-level, Sommerfeld enhanced, and total (including bound state effects)  annihilation cross sections for $y_{\rm small}$.}
\label{fig:small aH mass comparison}
\end{figure}

Figure~\ref{fig:small aH mass comparison} compares the masses obtained from the tree-level, Sommerfeld enhanced, and total annihilation cross sections. For the $3M2D$ case, we see that the contribution from bound states is subdominant to the Sommerfeld enhancement. This matches the conclusion in Ref.~\cite{Mitridate:2017izz} that for the pure Majorana triplet bound-state formation is less important than Sommerfeld enhancement. For the other multiplets, we see that the two effects are roughly comparable. In greater detail, we see that the relative importance of bound states grows from the $3M2D$, $5M4D$, and $7M6D$ cases and is then essentially constant for larger multiplets. This is influenced by our truncation of which bound states we calculate. A full accounting for all bound states would likely show a monotonically growing bound state contribution with multiplet size.

\subsection{Mid $\alpha_H$}
\label{subsec:mid aH}
\begin{figure}
\includegraphics[width=0.98\columnwidth]{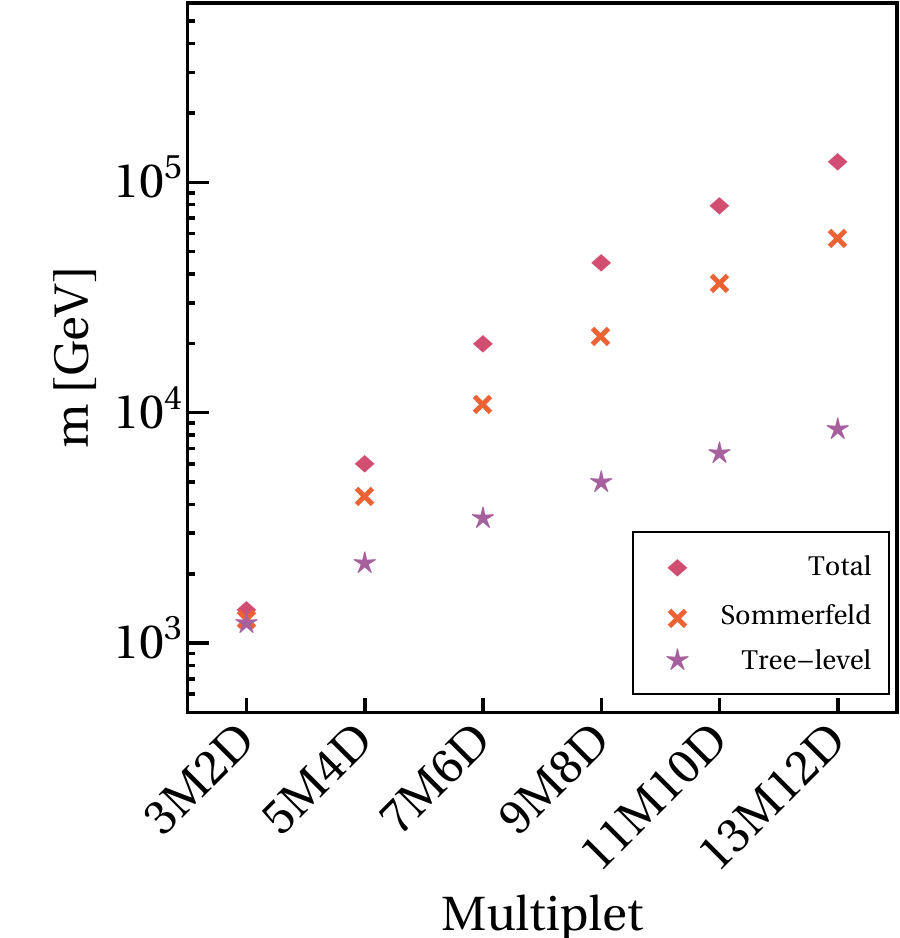}
\caption{Comparison of DM masses obtained from the tree-level, Sommerfeld enhanced, and total (including bound state effects)  annihilation cross sections for $y_{\rm mid}$.}
\label{fig:mid aH mass comparison}
\end{figure}

\begin{table}[b]
\centering
\begin{tabular}{||c | c | c | c | c | c||} 
 \hline
 $3M2D$ & $5M4D$ & $7M6D$ & $9M8D$ & $11M10D$ & $12M12D$\\ [0.5ex] 
 \hline\hline
 ${\rm mix2}^1$ & ${\rm mix2}^1$ & ${\rm mix2}^1$ & ${\rm mix2}^1$ & ${\rm mix2}^1$ & ${\rm mix2}^1$\\ 
 ${M \overline{D}}^2$ & ${\rm mix2}^3$ & ${\rm mix2}^3$ & ${\rm mix2}^3$ & ${\rm mix2}^3$ & ${\rm mix2}^3$\\
 $MD^2$ & ${M \overline{D}}^2$ & ${M \overline{D}}^2$ & ${M \overline{D}}^2$ & ${M \overline{D}}^2$ & ${\rm mix2}^5$\\
 ${\rm mix2}^3$ & $MD^2$ & $MD^2$ & $MD^2$ & $MD^2$ & ${M \overline{D}}^2$\\
 & $M \overline{D}^4$ & $M \overline{D}^4$ & ${\rm mix2}^5$ & ${\rm mix2}^5$ & $MD^2$\\
 & $M D^4$ & $M D^4$ & $M \overline{D}^4$ & $M \overline{D}^4$ & $M \overline{D}^4$\\
 & ${\rm mix1}^1$ & ${\rm mix2}^5$ & $M D^4$ & $M D^4$ & $M D^4$\\
 & $DD^1$ & ${\rm mix1}^1$ & ${\rm mix1}^1$ & ${\rm mix1}^1$ & ${\rm mix1}^1$\\
 & ${\rm mix2}^5$ & $DD^1$ & $DD^1$ & $DD^1$ & $DD^1$\\
 &  & ${\rm mix1}^3$ & ${\rm mix1}^3$ & ${\rm mix1}^3$ & ${\rm mix1}^3$\\
 &  & $DD^3$ & $DD^3$ & $DD^3$ & $DD^3$\\
 &  &  & ${\rm mix1}^5$ & ${\rm mix1}^5$ & ${\rm mix1}^5$\\
 [1ex] 
 \hline
\end{tabular}
\bigskip
\caption{Bound states above the consideration threshold for Higgs coupling $y_{\rm mid}=0.005$. Superscripts denote representation size. States are organized from largest to smallest binding energy. All states are $n=1$, $\ell=0$.}
\label{table:mid aH bound states}
\end{table}

Table~\ref{table:mid aH bound states} lists the bound states relevant for the $y_{\rm mid}$ case. Because $y$ is no longer negligible, we now work in terms of the mixed $MM \rightleftharpoons D \overline{D}$ states. For $y_{\rm mid}$ these states are still essentially pure $MM$ and $D \overline{D}$ states (as can be confirmed from Eqs.~(\ref{eq:mixing eigenvalues}),~(\ref{eq:P matrix}), and~(\ref{eq:phi twidle})), but using this basis allows us to compare our results to the $y_{\rm large}$ case where these states are no longer approximated by the pure case.

Figure~\ref{fig:large mid small aH masses} compares the DM mass for the $y_{\rm mid}$ and the $y_{\rm small}$ cases. We find that for small multiplets the masses are approximately equal, while for larger representations the masses are slightly larger for the $y_{\rm mid}$ condition. In all cases the mass remains lower than the pure Majorana case. This is again a result of the fact that no new annihilation channels are effectively open (because $\alpha_H$ is not multiplied by the overlap integral for the annihilation cross sections and $\alpha_H \ll \alpha_2$). However, for bound-state formation $\alpha_H$ appears in conjunction with $|F|^2$ so the combination is comparable to $\alpha_2 |\vec{J}|^2$ and bound-state formation is enhanced.

Figure~\ref{fig:mid aH mass comparison} compares the masses obtained from the tree-level, Sommerfeld enhanced, and total annihilation cross sections, providing additional insight. We again see that for the $3M2D$ case the contribution from bound states is subdominant to the Sommerfeld enhancement. However, in this case we see that the relative importance of the bound contribution grows monotonically with increasing multiplet size. This effect would be even more pronounced without the bound state truncation used here.

\subsection{Large $\alpha_H$}
\label{subsec:large aH}
\begin{figure}
\includegraphics[width=0.98\columnwidth]{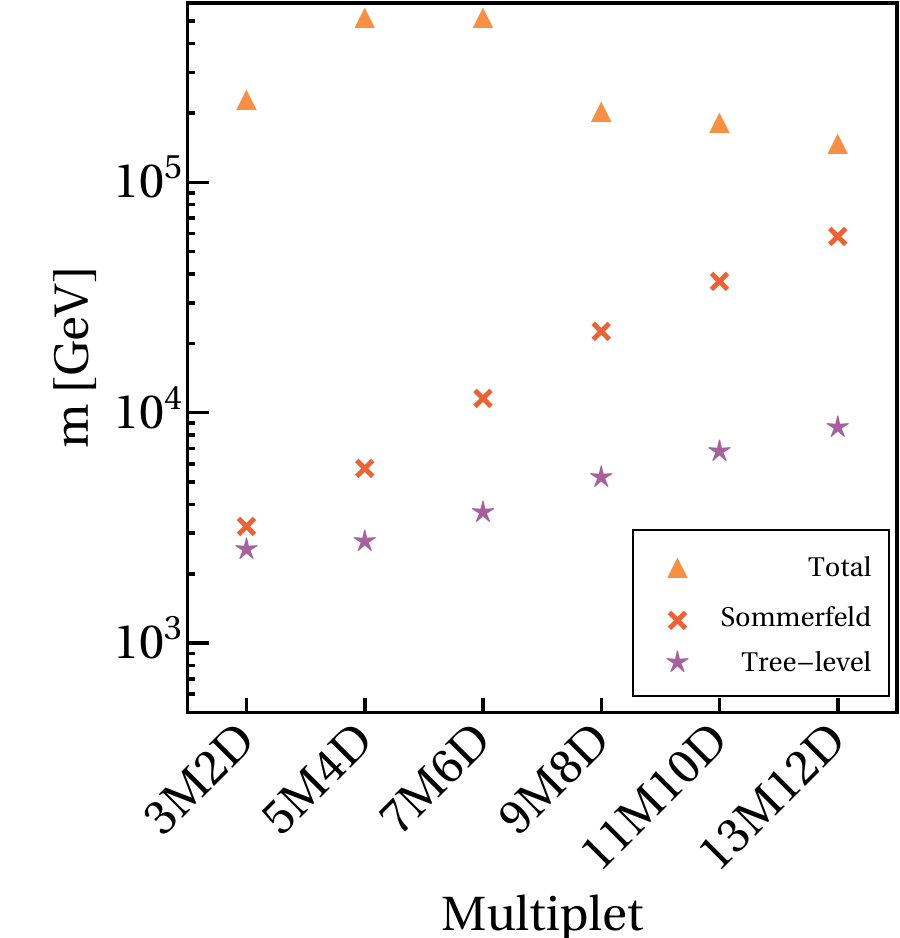}
\caption{Comparison of DM masses obtained from the tree-level, Sommerfeld enhanced, and  (including bound state effects) annihilation cross sections for $y_{\rm large}$.}
\label{fig:large aH mass comparison}
\end{figure}

\begin{table}[b]
\centering
\begin{tabular}{||c | c | c | c | c | c||} 
 \hline
 3M2D & 5M4D & 7M6D & 9M8D & 11M10D & 13M12D \\ [0.5ex] 
 \hline\hline
 ${\rm mix2}^1$ & ${\rm mix2}^1$ & ${\rm mix2}^1$ & ${\rm mix2}^1$ & ${\rm mix2}^1$ & ${\rm mix2}^1$\\ 
 ${M \overline{D}}^2$ & ${M \overline{D}}^2$ & ${M \overline{D}}^2$ & ${M \overline{D}}^2$ & ${\rm mix2}^3$ &  ${\rm mix2}^3$ \\
 ${\rm mix2}^3$ & ${\rm mix2}^3$ & ${\rm mix2}^3$ & ${\rm mix2}^3$ & ${M \overline{D}}^2$ & ${M \overline{D}}^2$\\
 $M \overline{D}^4$ & $M \overline{D}^4$ & $M \overline{D}^4$ & $M \overline{D}^4$ & $M \overline{D}^4$ & $M \overline{D}^4$\\
 & ${\rm mix2}^5$ & ${\rm mix2}^5$ & ${\rm mix2}^5$ & ${\rm mix2}^5$ & ${\rm mix2}^5$ \\
 &  & $DD^1$ & $DD^1$ & $MD^2$ & $MD^2$ \\
 &  & $MD^2$ & $MD^2$ & $DD^1$ & $DD^1$ \\
 &  & $DD^3$ & ${\rm mix1}^1$ & $MD^4$ & $MD^4$ \\
 &  & ${\rm mix1}^1$ & $DD^3$ & ${\rm mix1}^1$ & ${\rm mix1}^1$ \\
 &  & $MD^4$ & $MD^4$ & $DD^3$ & $DD^3$ \\
 &  &  & ${\rm mix1}^3$ & ${\rm mix1}^3$ & ${\rm mix1}^3$ \\
 &  &  & ${\rm mix1}^5$ &  ${\rm mix1}^5$ & ${\rm mix1}^5$ \\
 [1ex] 
 \hline
\end{tabular}
\bigskip
\caption{Bound states above the consideration threshold for $y_{\rm large}=1$. Superscripts denote representation size. States are organized from largest to smallest binding energy. All states are $n=1$, $\ell=0$.}
\label{table:large aH bound states}
\end{table}

Table~\ref{table:large aH bound states} shows the $y_{\rm large}$ bound states. The separation between the $MD$ and $M \overline{D}$ potentials for a given representation is larger than in the $y_{\rm mid}$ case. This is due to the relative sign difference between $y_1$ and $y_2$, which affects these two potentials differently, and the magnitude of $y_{\rm large}$ (see Sec.~\ref{subsec:potentials}).

\begin{figure}
\includegraphics[width=0.98\columnwidth]{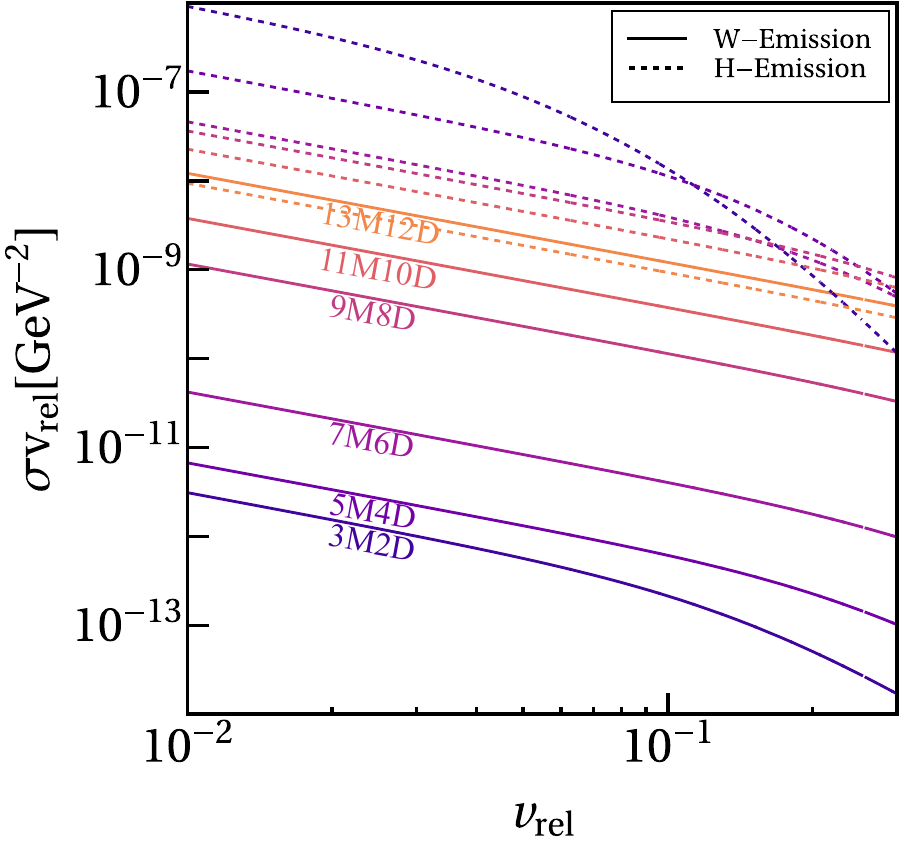}
\caption{Bound-state formation cross sections through $W$ (solid lines) and $H$ (dashed lines) emission for $y_{\rm large}$.}
\label{fig:large aH cross section comparison}
\end{figure}

Figure~\ref{fig:large mid small aH masses} shows that the Higgs coupling now has a large effect on the DM mass for smaller representation sizes compared to the pure Majorana case. For larger representations this effect diminishes and by the $11M10D$ case the mass is actually slightly smaller than for the pure Majorana. 

Figure~\ref{fig:large aH mass comparison} shows the origin of this behavior. First, bound state formation dominates the total cross section for all of the representations considered here. However, we see that the relative importance of bound states actually decreases for larger multiplet combinations. 

Figure~\ref{fig:large aH cross section comparison} shows the behavior of the bound state formation cross sections  as a function of the relative velocity in detail. We see that the $W$-emission bound state formation cross sections grow monotonically with increasing multiplet size over the relevant velocity range as expected from the pure Majorana case ($B$-emission not shown because it is subdominant to $W$-emission and follows the same pattern). However, the $H$-emission cross section actually decreases with the multiplet size, at least for $v_{rel}<0.1$. We also see that for small multiplets the $H$-emission completely dominates the bound state formation cross section. However, due to the behavior noted above, this becomes much less pronounced for large multiplet combinations and for the $13M12D$ case we actually find that the $H$-emission cross section is subdominant to $W$-emission. Therefore, we find that the cross sections, and consequently the masses, begin to approach that of the small $\alpha_H$ case as the multiplet sizes increase. This is exactly what we see in~\cref{fig:large mid small aH masses}.


\subsection{The unitarity bound}
\label{subsec:unitarity}

Above, we used an upper limit on $y$ based on perturbativity. However, we can also impose the constraint that the $s$-wave cross-sections considered here remain under the unitarity bound. When we only consider the $s$-wave, we obtain the limit
\begin{equation}
    (\sigma v)_{\rm ini} \leq \frac{4 \pi}{v m^2},
    \label{eq:unitarity1}
\end{equation}
where $(\sigma v)_{\rm ini}$ refers to the cross section for an initial state ($MM,~DD,$ etc...) to annihilate or form a bound state (which might not match the initial state). This term is given schematically by
\begin{equation}
    (\sigma v)_{X_1 X_2} = \frac{1}{g_{X_1} g_{X_2}}\left(\sum({\rm spin~0}) + \frac{1}{9} \sum({\rm spin~1)}\right),
    \label{eq:unitarity2}
\end{equation}
where the two summations include the cross sections for $X_1 X_2 \rightarrow {\rm SM~or~BS}$ in the relevant spin state. Note that because we have only considered bound states with $\ell=0$ and $|\Delta \ell|=1$ for $W/B$-emission, this only includes bound states formed through $H$-emission.

We note some subtleties unique to our choices in this paper when we do this computation. The cross-section in Eq.~(\ref{eq:scattering state annihilation}) should be divided by two when $i=j$. Furthermore, our expressions for the bound-state formation cross sections include an averaging over all possible DM initial states; see Eqs.~(\ref{eq:W emission sigma}), (\ref{eq:B emission sigma}) and  (\ref{eq:H emission sigma}). Consequently, these must be multiplied by $(g_M + g_D)^2$ so that this average can be replaced by the average over the relevant initial state, as shown in Eq.~(\ref{eq:unitarity2}).

\subsection{Overall results}
\label{subsec:overall results}
\begin{figure*}[t]
\includegraphics[width=0.98\textwidth]{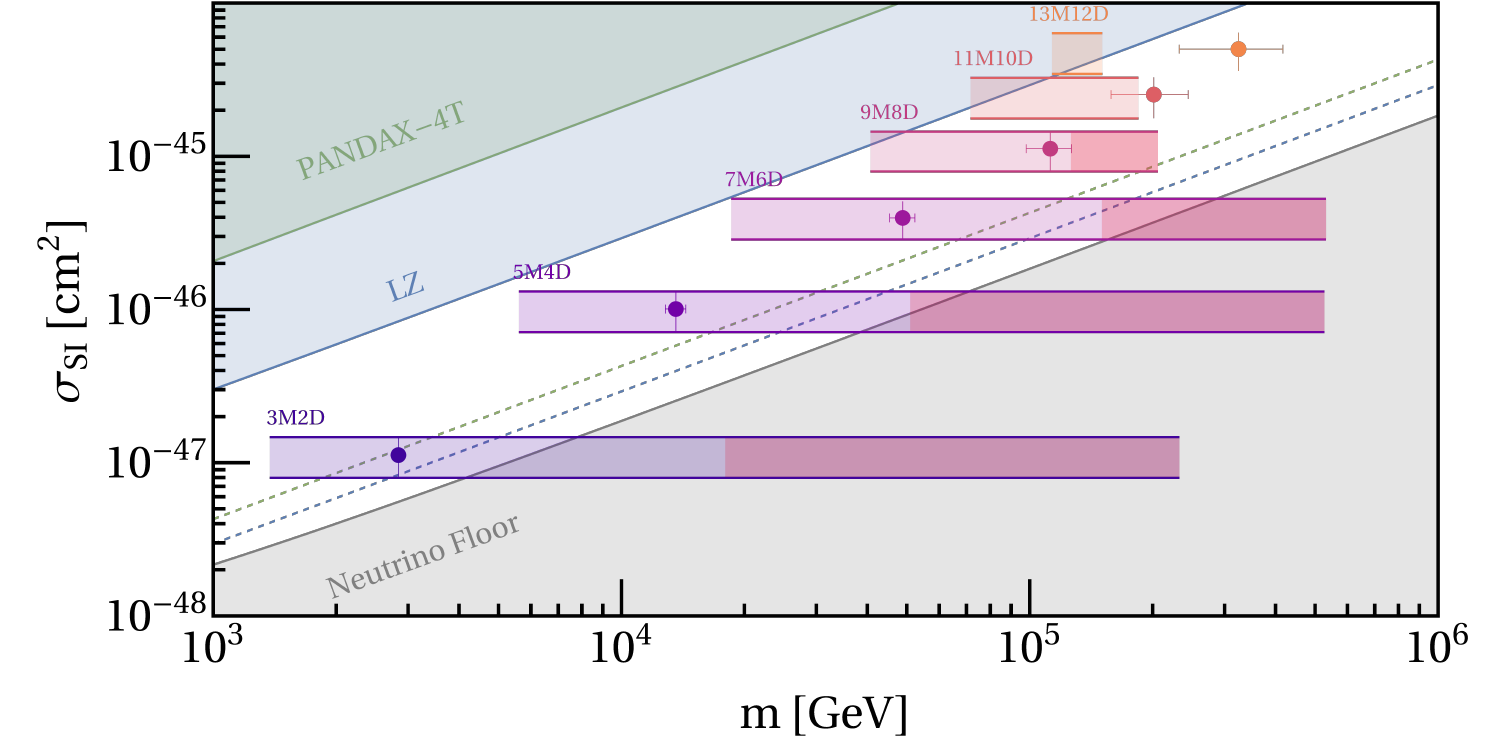}
\caption{Allowed parameter space of masses and spin-independent cross sections on nucleons for different multiplet combinations, shown with horizontal bands. The direct detection cross sections and masses are valid in the $y_1 = -y_2, \, m_D = m_M$ limit necessary to cancel tree-level Higgs induced scattering on nucleons (see Sec.~\ref{sec:model}). The band thickness is determined by the lattice QCD uncertainties for the elastic cross section on nuclei~\cite{Bottaro:2021snn}. Dots indicate the pure Majorana case for comparison. The red-shaded regions indicate parameter space excluded by imposing unitarity of the $s$-wave cross sections.  The gray shaded region indicates the neutrino floor. The blue and green shaded regions are the LUX-ZEPLIN (LZ) and PandaX-4T exclusion regions, respectively~\cite{LZ:2024zvo, PandaX:2024qfu}. The dashed blue and green lines indicate the projected XLZD and PandaX-xT 200 ton-year 90\% C.L. exclusion sensitivities, respectively~\cite{Baudis:2024jnk, PANDA-X:2024dlo}.}
\label{fig:main plot}
\end{figure*}

Our choices for $y$ span the parameter space from an essentially negligible Higgs coupling to the perturbative limit. Because we are considering $y_1 = -y_2$ and $m_D=m_M$ (i.e., the custodial point), the amplitude of the spin-independent cross-section is independent of the Higgs coupling and equal to the case of pure MDM as discussed in Sec.~\ref{sec:model}~\cite{LopezHonorez:2017zrd}. 

Figure~\ref{fig:main plot} shows the possible masses for the $3M2D$, $5M4D$, $7M6D$, $9M8D$, $11M10D$, and $13M12D$ cases and their corresponding spin-independent (SI) cross sections on nuclei. The spin-independent cross sections are taken from Ref.~\cite{Bottaro:2021snn}, which determined the cross-sections and their uncertainties in the pure MDM case using lattice QCD calculations. The red shaded regions of the $3M2D,~5M4D,~7M6D,~9M8D$ parameter spaces indicate violation of $s$-wave unitarity. For the largest multiplets, unitarity is not violated up to our previously imposed limit of $y=1$.

Figure~\ref{fig:main plot} also shows the current exclusion limits from PandaX-4T~\cite{PandaX:2024qfu} and LUX-ZEPLIN (LZ)~\cite{LZ:2024zvo} as well as the proposed sensitivities of XLZD~\cite{Baudis:2024jnk} and PandaX-xT~\cite{PANDA-X:2024dlo}. We see that for the $3M2D$ case, a significant portion of the parameter space lies below the neutrino floor (for the $5M4D$ case, this is true only for a marginal amount of the parameter space). Therefore, we conclude that for lower multiplet mixtures, HC-MDM cannot be excluded currently by standard direct-detection efforts. Conversely, parts of the parameter space for these multiplets are actually easier to probe in future direct-dection experiments than their pure Majorana counterparts. Furthermore, we see that for the $9M8D$, $11M10D$, and especially the $13M12D$ cases, the parameter space is already constrained by LZ results. These multiplets lie entirely in the sensitivity range of the next generation of direct-detection experiments. Finally, we note again that we have repeatedly made conservative choices that underpredict the mass of the DM particles, so the parameter spaces will extend \textit{further} below the neutrino floor in a more detailed calculation.


\section{Conclusions}
\label{sec:conclusion}

Due to its simplicity and predictive power, the minimal DM model is an especially compelling resolution to the particle nature of DM. For individual multiplets, this model is within the reach of the next generation of direct-detection experiments to be either confirmed or ruled out. But whether this remains true  for a modest and important extension of this model, introducing two multiplets coupled by Higgs interactions, has been unknown.

In this paper, we have presented a detailed framework for calculating the relic abundance for HC-MDM. This includes both the tree-level annihilation cross section as well as the enhancement from Sommerfeld effects and bound-state formation. Furthermore, this formalism can be used with slight modification (mainly setting a series of parameters to zero) in individual-multiplet MDM models. We then use this formalism to determine the relic abundance of various viable multiplet combinations. Assuming that a given combination makes up the entirety of DM, we also determine the mass required for each multiplet combination.

Due to the phenomenological similarity between the behavior of HC-MDM and individual-multiplet MDM in the late universe when scattering on nuclei, we can easily determine the viability of detecting this particle in the next generation of direct-detection experiments. We find that the parameter space for some of the lower dimensional multiplet combinations extends beyond the reach of upcoming proposed experiments. Furthermore, the parameter space extends well below the neutrino floor, ruling out complete coverage in standard direct-detection experiments. We emphasize that this conclusion applies to \textit{standard} direct-detection experiments. In principle, detectors with directional capabilities may be able to probe below the neutrino floor~\cite{Billard:2011zj}. Larger multiplet combinations, however, are well within the grasp of the next generation of experiments, with some already constrained by recent LZ results.

Ultimately, the most reliable test of any WIMP model lies in observing its annihilation products in the late universe. Furthermore, this method would evade the limitations on direct-detection experiments which make detection of low dimensional multiplets impossible. To accurately model the spectrum and composition of these annihilation products in the HC-MDM model requires a different formalism from that presented here, valid after $SU(2)_L$ symmetry breaking. This is deferred for a later paper.

There are other potential avenues for future exploration. First, in this paper we have made a series of simplifying assumptions about the magnitude of the Higgs coupling constants of the DM multiplets as well as their individual masses. We have also neglected some of the smaller contributions to the annihilation cross section, mainly those of $\ell>0$ states. Forgoing these simplifications increases the computational complexity, but the formalism developed here can, in principle, be extended to cover this scenario. Additionally, here we have considered large coupling values and masses that cancel the tree-level $H$-mediated scattering cross section contributions to direct detection of HC-MDM particles. As noted in Sec.~\ref{sec:model}, loop effects may modify the direct-detection cross-section. However the approximate custodial symmetry corresponding to $y_1 = -y_2$ should leave our conclusions qualitatively unchanged.  Ultimately, while fully probing the parameter space of HC-MDM will be challenging, it remains an essential goal.


\section*{Acknowledgments}

We are grateful for helpful discussions with Fareed Alasiri, Eric Braaten, Roberto Bruschini, Marco Cirelli, Sean Fleming, Richard Furnstahl, Jianglai Liu, Kalliopi Petraki, Chris Hirata, Obada Nairat, Stuart Raby, Michele Redi, Diego Redigolo, Tracy Slatyer, Alessandro Strumia, Todd Thompson, and Bryan Zaldivar. 

SG and JFB were supported by National Science Foundation Grant No.\ PHY-2310018. JS was supported by the UK Research and Innovation Future Leader Fellowship MR/Y018656/1.  LLH was supported by the Fonds de la Recherche Scientifique F.R.S.-FNRS through a senior research associate position, is a member of BLU-ULB (\href{https://blu.ulb.be/}{Brussels Laboratory of the Universe}), and acknowledges the support of the FNRS research grant number J.0134.24, the ARC program of the Federation Wallonie-Bruxelles, and the IISN convention No.\ 4.4503.15. 


\appendix

\vspace{1cm}
\centerline{\Large {\bf Appendices}} 
\vspace{0.5cm}

Here we collect a series of results necessary to rederive the work in the main body of the paper. Some of these details are already given in other papers referenced throughout this work, however we compile them here to aid the reader by providing a single source for all of the necessary results. We also briefly discuss our results for pure Majorana MDM and compare them to the existing literature.

In the following, we cover our results for the pure Majorana case, contraction of $SU(2)_L$ indices, the annihilation amplitudes for DM to the standard model, and the DM bound state formation amplitudes.


\section{Pure Majorana multiplets} 
\label{sec:pure majorana multiplets}

\begin{figure}[b]
\includegraphics[width=0.98\columnwidth]{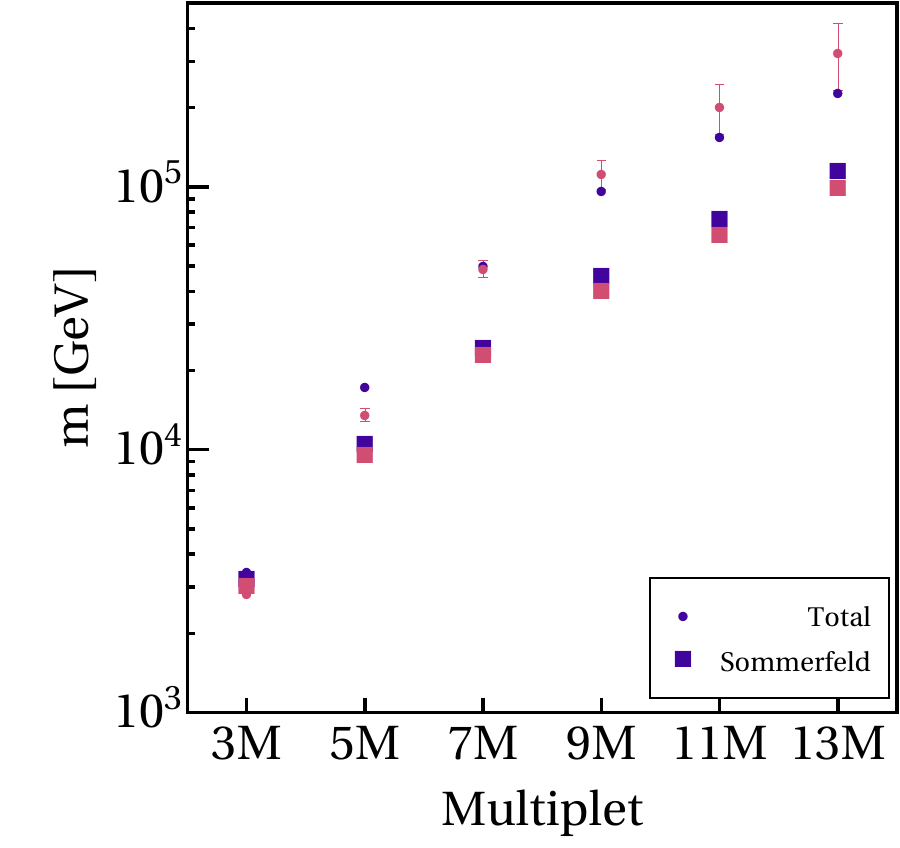}
\caption{Comparison of our results (blue) to Ref.~\cite{Bottaro:2021snn} (pink) for pure Majorana MDM. Shown are the total masses obtained including bound states as well as the result considering Sommerfeld enhancement only.}
\label{fig:bottaro comparison}
\end{figure}

We have chosen to use the results of Ref.~\cite{Bottaro:2021snn} for the pure Majorana case in this paper. As mentioned in Sec.~\ref{sec:results}, we only consider a subset of possible bound states in our calculation of the DM mass. This is due to the computational cost associated with the mixed-multiplet calculation. Therefore, the results of Ref.~\cite{Bottaro:2021snn} offer increased accuracy in the simpler pure Majorana case. For completeness, we compare our results for pure Majorana MDM with those of Ref.~\cite{Bottaro:2021snn}.

Figure~\ref{fig:bottaro comparison} compares our results (blue) for the DM mass giving rise to the right relic abundance to those of Ref.~\cite{Bottaro:2021snn} (pink) with (dots) and without (squares) bound states. We see that when we only consider Sommerfeld enhancement, we have excellent agreement with Ref.~\cite{Bottaro:2021snn}. When we consider bound states, we find that we again have excellent agreement for all cases, the worst being the $5M4D$ case, which is still within $\sim 20 \%$.


\section{Contracting indices} 
\label{sec:contracting indices}

Here we demonstrate the origin of the $\hat{\epsilon}_{X}$ operator used to contract two $SU(2)_L$ multiplets given in Sec.~\ref{sec:model} and used throughout the paper. We use a quadruplet $X$ with $Y_X = 1/2$ for this example.
We can represent $X, \overline{X}$ in two different ways. The first is
\begin{equation}
    X = \begin{pmatrix}
        X^{++} \\
        X^+ \\
        X^0 \\
        X^-
    \end{pmatrix}, \quad
    \overline{X} = \begin{pmatrix}
        \overline{X}^+ \\
        \overline{X}^0 \\
        \overline{X}^- \\
        \overline{X}^{--}
    \end{pmatrix},
\end{equation}
where the superscripts indicate the electric charge from Eq.~(\ref{eq:Gell-Mann}). 

We can also represent the multiplet by a totally symmetric tensor with 3 indices
\begin{equation}
    X = \begin{pmatrix}
        X_{111} \\
        \sqrt{3} X_{112} \\
        \sqrt{3} X_{122} \\
        X_{222}
    \end{pmatrix}, \quad
    \overline{X} = \begin{pmatrix}
        \overline{X}_{111} \\
        \sqrt{3} \overline{X}_{112} \\
        \sqrt{3} \overline{X}_{122} \\
        \overline{X}_{222}
    \end{pmatrix}.
\end{equation}
It is then straightforward to contract $\overline{X} X$ using the $SU(2)_L$ invariant Levi-Civita symbol $\epsilon_{ij}$
\begin{equation}
\begin{split}
    \overline{X} X &= \overline{X}_{ijk} X_{i'j'k'} \epsilon^{ii'} \epsilon^{jj'} \epsilon^{kk'} \\
    &= \overline{X}_{111} X_{222} - (\overline{X}_{112} X_{221} + \overline{X}_{121} X_{212} + \overline{X}_{211} X_{122}) \\
    &+ (\overline{X}_{122} X_{211} + \overline{X}_{212} X_{121} +\overline{X}_{221} X_{112}) - \overline{X}_{222} X_{111} \\
    &= \overline{X}^+ X^- - \overline{X}^0 X^0 + \overline{X}^- X^+ - \overline{X}^{--} X^{++}.
\end{split}
\end{equation}
Defining the matrix
\begin{equation}
    \hat{\epsilon}_{4} = \begin{pmatrix}
        0 & 0 & 0 & 1 \\
        0 & 0 & -1 & 0 \\
        0 & 1 & 0 & 0 \\
        -1 & 0 & 0 & 0
    \end{pmatrix}
\end{equation}
we see that 
\begin{equation}
    \overline{X} X = \overline{X} \cdot \hat{\epsilon}_4 X.
\end{equation}

We can generalize this to any dimension $R$ (written as $\hat{\epsilon}_{R}$). Essentially, $\hat{\epsilon}_{R}$ flips the order of the multiplet to form gauge invariant combinations. We frequently make use of the fact that $\hat{\epsilon}_{R} \hat{\epsilon}_{R}^T = 1$.

For example, we wish to decompose the process $\overline{D}_i D_j \rightarrow W W$ into isospin contributions. Using CG coefficients,
\begin{equation}
\begin{split}
    \mathcal{M} &\propto \langle I_{IS}, M_{IS}|I_D, I^3_D(i);I_D, I^3_D(j) \rangle (\hat{\epsilon_{D}})_{ik} (t^a_D)_{kl} (t^b_D)_{lj} \\
    &= \langle IS \rangle_{ij} (\hat{\epsilon}_{D} t^a_D t^b_D)_{ij},
\end{split}
\end{equation}
where in the second line we have used a shorthand notation for the CG coefficients.


\section{Annihilation amplitudes}
\label{sec:cross section calculations}

In this appendix we derive the amplitudes used in the cross sections in Sec.~\ref{sec:potentials and cross sections}.  Throughout, we use the non-relativistic spinor approximations
\begin{equation}
    u(p) \rightarrow \sqrt{m} \begin{pmatrix}
        \xi \\
        \xi
    \end{pmatrix}, \quad v(p) \rightarrow \sqrt{m} \begin{pmatrix}
        \xi \\
        - \xi
    \end{pmatrix}.
\end{equation}
%

\subsection{$D \overline{D} \rightarrow WW$}

\begin{figure}
\begin{fmffile}{diagram10}
 \begin{fmfgraph*}(90,60)
   \fmfleft{i1,i2}
   \fmfright{o1,o2}
   \fmf{fermion, label=$p_2$,l.side=right}{i1,v1}
   \fmf{photon,label=$k_2$,l.side=left}{o1,v1}
   \fmf{fermion,label=$p_1$,l.side=right}{v2,i2}
   \fmf{photon,label=$k_1$,l.side=left}{v2,o2}
   \fmf{plain, label=$M$,l.side=left}{v1,v2}
   \fmflabel{$D_j$}{i1}
   \fmflabel{$W^b_\nu$}{o1}
   \fmflabel{$\overline{D}_i$}{i2}
   \fmflabel{$W^a_\mu$}{o2}
 \end{fmfgraph*}
 \,\,\,\,\,\,\,\,\,\,\,\,\,\,\,\,\,\,
\begin{fmfgraph*}(90,60)
    \fmfleft{i2,i1}
    \fmfright{o2,o1}
    \fmf{phantom}{i1,v1,o1}
    \fmf{plain,label.side=right,label=$M$}{v1,v2} 
    \fmf{phantom}{i2,v2,o2}
    \fmffreeze
   \fmf{fermion, label=$p_1$,l.side=right}{v1,i1}
   \fmf{photon}{o2,v1}
   \fmf{fermion,label=$p_2$,l.side=right}{i2,v2}
   \fmf{photon}{v2,o1}
   \fmflabel{$D_j$}{i2}
   \fmflabel{$W^b_\nu$}{o2}
   \fmflabel{$\overline{D}_i$}{i1}
   \fmflabel{$W^a_\mu$}{o1}
    \end{fmfgraph*}
\end{fmffile}

\vspace{20pt}

\caption{Diagrams for $D \overline{D}$ annihilation to $WW$. }
\label{fig:DDbar to WW diagrams}
\end{figure}
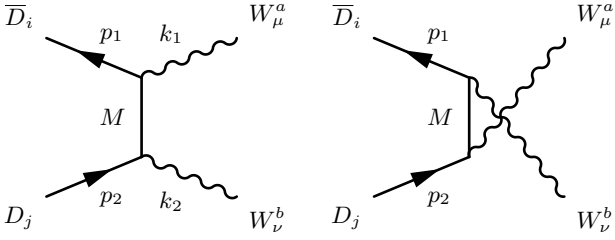

In the non-relativistic limit with massless vector bosons, the kinematics are (see Fig.~\ref{fig:DDbar to WW diagrams}):
\begin{equation}
\begin{split}
    p_1 &\approx (m,\vec{0}), \qquad k_1 \approx (m,0,0,m) \\
    p_2 &\approx (m,\vec{0}), \qquad k_2 \approx (m,0,0,-m).
\end{split}
\end{equation}
This leads to the simplifications
\begin{equation}
    \slashed{p}_1 - \slashed{k}_1 \rightarrow m \begin{pmatrix}
        0 & \sigma^3 \\
        -\sigma^3 & 0
    \end{pmatrix}, \qquad (p_1 - k_1)^2 \rightarrow -m^2.
\end{equation}

For the $t$-channel $D \overline{D}$ annihilation to $WW$ process, the amplitude is
\begin{equation}
\begin{split}
    &i \mathcal{M}_t = \\
    &(i g_2)^2 {\epsilon^a_\mu}^*(k_1) {\epsilon^b_\nu}^*(k_2) \overline{v}(p_1) t^a_D \gamma^\mu \frac{i(\slashed{p_1} - \slashed{k_1} + m)}{(p_1 - k_1)^2 - m^2} t^b_D \gamma^\nu u(p_2).
\end{split}
\end{equation}
Explicitly writing the $SU(2)_L$ indices, this has the structure
\begin{equation}
    \overline{v}_i (\hat{\epsilon}_D)_{ij} (t^a_D)_{jk} (t^b_D)_{kl} u_l.
\end{equation}
We can decompose this into isospins by letting
\begin{equation}
    G^{ab}_{D \overline{D};WW} = \langle I_{IS},M_{IS}|I_D, I_D^3(i);I_D,I_D^3(j) \rangle (\hat{\epsilon}_{D} t^a_D t^b_D)_{ij},
\end{equation}
where we have used the contraction operator from Appendix~\ref{sec:contracting indices}.

Applying the non-relativistic simplifications and using $\epsilon^a_0 = \epsilon^b_0 = 0$,
\begin{equation}
    i \mathcal{M}_t = -i (ig_2)^2 {\rm Tr}(\sigma^i \sigma^3 \sigma^j \xi \xi^{' \dagger}) \epsilon_i^{a*}(k_1) \epsilon_j^{b*}(k_2) G^{ab}_{D \overline{D};WW}
\end{equation}

For the spin zero configuration, $\xi \xi^{' \dagger} = (1 / \sqrt{2}) \mathbf{1}$, where $\mathbf{1}$ is the 2x2 identity matrix (see Ref.~\cite{Peskin:1995ev}). Using
\begin{equation}
    \leftindex^{\pm}{\epsilon}^a (k_1) = \frac{1}{\sqrt{2}} (0,1,\pm i,0), \quad \leftindex^{\pm}{\epsilon}^b (k_2) = \frac{1}{\sqrt{2}} (0,-1,\pm i,0),
\end{equation}
we obtain the amplitudes 
\begin{equation}
\begin{split}
    &\mathcal{M}_{t,+-} = \mathcal{M}_{t,-+} = 0 \\
    &\mathcal{M}_{t,++} = -\mathcal{M}_{t,--} = \sqrt{2} (i g_2)^2 G^{ab}_{D \overline{D};WW}
\end{split}
\end{equation}

For the $u$-channel, $t^a \leftrightarrow t^b$ so $G^{ab}_{D \overline{D};WW} \rightarrow G^{ba}_{D \overline{D};WW}$, $\mu \leftrightarrow \nu$ in the $\gamma$ matrices, and $\slashed{p}_1 - \slashed{k}_1 \rightarrow \slashed{p}_1 - \slashed{k}_2 = - (\slashed{p}_1 - \slashed{k}_1)$. Adding both diagrams,
\begin{equation}
    \begin{split}
        |\mathcal{M}|^2 = 4 g_2^4 (G^{ab}_{D \overline{D};WW} + G^{ba}_{D \overline{D};WW})^2.
    \end{split}
\end{equation}

From the Landau-Yang theorem, this process only occurs with spin zero~\cite{Mitridate:2017izz}. $WW$ must have $R_{FS} = 1,3,5$, but for $R = 3$ the gauge factor combination vanishes. So this process occurs in spin-0, $R_{IS}=1,5$ states.

\subsection{$D \overline{D} \rightarrow HH^*$}

The $t$-channel process of $D \overline{D} \rightarrow HH^*$ shares the same kinematics and approximations as the $D \overline{D} \rightarrow WW$ case. The amplitude is
\begin{equation}
    i \mathcal{M} = -i y_1 y_2 {\rm TR}(\sigma^3 \xi \xi^{'\dagger}).
\end{equation}
For spin zero, $\xi \xi^{'\dagger} \propto \mathbf{1}$ and $\mathcal{M}$ vanishes.

For spin one, $\xi \xi^{'\dagger} = (\vec{n}^* \cdot \vec{\sigma})/\sqrt{2}$, where $\vec{n}$ are the three bound state polarization vectors. This is only non-zero for the transverse polarization so
\begin{equation}
    i \mathcal{M} = -i 2 \sqrt{2} y_1 y_2.
\end{equation}
Because isospin is conserved, this process occurs for spin one, $R_{BS}=1,3$.

The $s$-channel process $D \overline{D} \rightarrow W \rightarrow HH^*$ is easiest to account for by adding it to the $D \overline{D} \rightarrow W \rightarrow f \overline{f}$ case considered below.

\subsection{$D \overline{D} \rightarrow f \overline{f}$}

$D \overline{D} \rightarrow f \overline{f}$ occurs through an $s$-channel process mediated by the $W$. Requiring isospin and spin conservation, this only occurs with $I_{BS} = 1, s_{BS}=1$. Because $I_{BS}$ is restricted to one value, we do not need to decompose the process using CG coefficients.

Defining 
\begin{equation}
    G_{D \overline{D};f \overline{f}}^2 = {\rm Tr}(t^a_D t^b_D) {\rm Tr}(t^a_{SM} t^b_{SM}),
\end{equation}
the amplitude is
\begin{equation}
    i \mathcal{M} = (ig_2)^2 \overline{v} \gamma^\mu u \frac{-i g_{\mu \nu}}{q^2} \overline{u} \gamma^\nu v G_{D \overline{D};f \overline{f}},
\end{equation}
where $q$ is the $W$ momentum and we have omitted the fermion momenta as well as the $SU(2)$ indices on $G_{D \overline{D};f \overline{f}}$ which will be contracted over when obtaining the cross section.

Because the process is mediated by the $W$, $f$ must be left-handed and $\overline{f}$ right-handed. Using the familiar non-relativistic approximations,
\begin{equation}
    \mathcal{M} = 2 (i g_2)^2 G_{D \overline{D};f \overline{f}}.
\end{equation}

When we compute the cross section from this amplitude, we multiply by the number of fermions, $n_f = 12$. However, we can account for the process $D \overline{D} \rightarrow W \rightarrow H H^*$ by replacing $n_f \rightarrow n_p = 25/2$, as noted in Sec.~\ref{sec:potentials and cross sections}.

\subsection{$MM \rightarrow WW$}

We can read off the $MM \rightarrow WW$ amplitude  from the $D \overline{D} \rightarrow WW$ one. We define $G^{ab}_{MM;WW}$ as in the $D \overline{D}$ case and note that we obtain an additional symmetry factor of $1/2$ in the cross section.

\subsection{$MM \rightarrow H H^*$}

The $MM \rightarrow H H^*$ can be determined from the $D \overline{D}$ case. We now have a $u$-channel diagram which has a relative minus from fermion exchange and an additional minus from $\slashed{p}_1 - \slashed{k}_1 \rightarrow \slashed{p}_1 - \slashed{k}_2$, so the diagrams add together. We also get a symmetry factor of $1/2$ from the identical particles, so overall the cross section is double the $D \overline{D}$ cross section.

\subsection{$MM \rightarrow f \overline{f}$}

The $MM \rightarrow f \overline{f}$ amplitude is found by comparison to the $D \overline{D} \rightarrow f \overline{f}$ case. We define $G_{MM; f \overline{f}}$ as in the the $D \overline{D}$ case and multiply the cross section by the symmetry factor of $1/2$.

\subsection{$DD/\overline{D} \overline{D} \rightarrow H H$}

The $DD/\overline{D} \overline{D} \rightarrow H H$ is found from the $MM \rightarrow H H^*$ example. In this case we have identical particles in the initial and final states, so the cross section gets a factor of $1/2$ compared to the $MM$ cross section

\subsection{$MD/\overline{D} \rightarrow W H/H^*$}

\begin{figure}
\begin{fmffile}{diagram11}
 \begin{fmfgraph*}(90,60)
   \fmfleft{i1,i2}
   \fmfright{o1,o2}
   \fmf{fermion, label=$p_2$,l.side=right}{i1,v1}
   \fmf{dashes,label=$k_2$,l.side=left}{o1,v1}
   \fmf{fermion,label=$p_1$,l.side=right}{v2,i2}
   \fmf{photon,label=$k_1$,l.side=left}{v2,o2}
   \fmf{plain, label=$M$,l.side=left}{v1,v2}
   \fmflabel{$D_j$}{i1}
   \fmflabel{$H$}{o1}
   \fmflabel{$M_i$}{i2}
   \fmflabel{$W^a_\mu$}{o2}
 \end{fmfgraph*}
 \,\,\,\,\,\,\,\,\,\,\,\,\,\,\,\,\,\,
 \begin{fmfgraph*}(90,60)
   \fmfleft{i1,i2}
   \fmfright{o1,o2}
   \fmf{fermion, label=$p_2$,l.side=right}{i1,v1}
   \fmf{photon,label=$k_2$,l.side=left}{o1,v1}
   \fmf{fermion,label=$p_1$,l.side=right}{v2,i2}
   \fmf{dashes,label=$k_1$,l.side=left}{v2,o2}
   \fmf{plain, label=$D$,l.side=left}{v1,v2}
   \fmflabel{$D_j$}{i1}
   \fmflabel{$W^a_\mu$}{o1}
   \fmflabel{$M_i$}{i2}
   \fmflabel{$H$}{o2}
 \end{fmfgraph*}
\end{fmffile}

\vspace{20pt}

\caption{Diagrams for $M D$ annihilation to $HW$.}
\label{fig:MD to WH diagrams}
\end{figure}
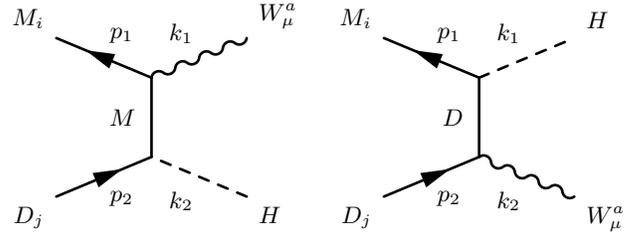

Figure~\ref{fig:MD to WH diagrams} shows the diagrams for the $MD\rightarrow W H$ case. Writing the $SU(2)_L$ indices explicitly, the diagram on the left has the structure
\begin{equation}
    i \mathcal{M} \propto \overline{v}_i (\hat{\epsilon_M})_{ij} (t^a_M)_{jk} \delta^{I^3_D(l)}_{I^3_M(k) \pm 1/2} u_l,
\end{equation}
where we have omitted the momenta arguments. Defining
\begin{equation}
    \leftindex^{\pm}_1{G}_{MD;WH} = \langle I_{IS},M_{IS}|I_M,I^3_M(i);I_D,I^3_M(j) \pm 1/2 \rangle (\hat{\epsilon}_{M} t^a_M)_{ij},
\end{equation}
and using the non-relativistic approximations
\begin{equation}
    i \mathcal{M} = -i g_2 y_1 \epsilon^*_n(k_1) {\rm Tr}(\sigma^n \xi \xi^{'\dagger}) \leftindex^{\pm}_1{G}_{MD;WH}.
\end{equation}

In spin zero, this vanishes. In spin-1, using $L,R$ to denote the $W$ handedness and $1,2$ for the initial state polarization,
\begin{equation}
    \mathcal{M}^{R,2} = \mathcal{M}^{L,1} = g_2 y_1 \sqrt{2} \leftindex^{\pm}_1{G}_{MD;WH}.
\end{equation}

The diagram on the right in Fig.~\ref{fig:MD to WH diagrams} produces the same amplitude with the substitution $\leftindex^{\pm}_1{G}_{MD;WH} \rightarrow \leftindex^{\pm}_2{G}_{MD;WH}$,
\begin{equation}
    \leftindex^{\pm}_2{G}_{MD;WH} = \langle I_{IS},M_{IS}|I_M,I^3_D(i) \pm 1/2;I_D,I^3_D(j) \rangle (t^a_D)_{ij}.
\end{equation}

Adding the two diagrams together leads to the cross section in Eq.~(\ref{eq:annihilation cross sections}). Note that when the gauge factor is squared, we sum over the $+, \,-$ as indices. 

\subsection{$MD/\overline{D} \rightarrow H/H^* \rightarrow SM$}

In unbroken $SU(2)_L$, the $H$ does not couple to $WW$ (we can also note that isospin could not be conserved in the process) so the relevant process is $MD/\overline{D} \rightarrow H/H^* \rightarrow f \overline{f} $. However, in the non-relativistic limit this is also zero.


\section{Bound-state formation amplitudes}
\label{sec:bound-state formation amplitudes}

Here we derive the amplitudes used in the calculation of bound-state formation cross sections in Sec.~\ref{sec:bound states}. 

We use $\phi_{p \ell,ij}$ for the scattering state wavefunctions and $\psi_{n' \ell' m',i'j'}$ for the bound state wavefunctions. Here $p,\ell$ are the momentum and angular momentum of the scattering state, $n \ell m$ are the quantum numbers of the bound state, and $i,j$ are $SU(2)_L$ indices. Bound state indices are primed and scattering states unprimed. The radial part of the bound state wavefunction is given in Eq.~(\ref{eq:bound state wavefunction})

The radial part of the scattering state wavefunction in the massless vector limit is approximated by
\begin{equation}
\begin{split}
    &R_\ell(r) = \frac{\sqrt{4 \pi (2 \ell + 1) S}}{\Gamma(2 \ell + 2)} e^{-ipr} (2pr)^\ell \\
    &F_1 \left( \ell + 1 + \frac{i \alpha_{\rm{eff}}}{v_{\rm{rel}}}, 2 \ell + 2, 2ipr \right) \prod_{n=1}^{\ell} \left( \ell - n + 1 - \frac{i \alpha_{\rm{eff}}}{v_{\rm{rel}}} \right),
\end{split}
\label{eq:scattering state radial wavefunction}
\end{equation}
where $F$ is the hypergeometric function and $S$ is the Sommerfeld factor. This approximation requires an additional correction for states with $\ell \neq 0$, which is the case when we consider formation of $\ell = 0$ bound states from $\ell=1$ scattering states through $W,B$ emission. The correction multiplies the resulting cross sections by
\begin{equation}
\begin{split}
    &L_\ell = \frac{w^{2 \ell}}{\prod_{k=0}^{\ell - 1} \left( (\ell - k)^2 + w^2\right)} \quad {\rm with} \quad w = \frac{m v_{\rm rel}}{\kappa m_V} ,
\end{split}
\label{eq:scattering state radial wavefunction correction}
\end{equation}
where $m_V$ is the relevant vector boson mass and $\kappa \approx 1.74$ arises from the Hulthen potential used to derive the Sommerfeld factor in the case of massive vector bosons~\cite{Mitridate:2017izz}.

We make use of the non-relativistic approximations
\begin{equation}
\begin{split}
    \overline{u}_i(p')u_j(p) &\rightarrow 2m \delta_{ij}\\
    \overline{u}_i(p') \gamma^\mu u_j(p) &\rightarrow \left( 2m \delta^{\mu 0} + (\vec{p} + \vec{p'})_{k} \delta^{\mu k} \right) \delta_{ij} \\
    \overline{u}_i(p') \gamma^{\mu} t^a_{ij} u_j(p) &\rightarrow \left( 2m \delta^{\mu 0} + (\vec{p} + \vec{p'})_{k} \delta^{\mu k} \right) t^a_{ij},
\end{split}
\end{equation}
where $k$ is a vector index and we have suppressed the spin indices because the are individually conserved. We have been slightly careless with the $SU(2)_L$ indices in the last expression, but these indices are contracted with the overlap integrals shown below. We also use the overlap integrals
\begin{equation}
\begin{split}
    \vec{J}^{ij,i'j'}_{p,nlm} &= \int d^3r \psi_{i'j'}^* \nabla \phi_{ij} \\
    \vec{T}^{ij,i'j'}_{p,nlm} &= \frac{\alpha_2 m}{2} \int d^3r \psi_{i'j'}^* \hat{r} \phi_{ij} \\
    \vec{F}^{ij,i'j'}_{p,nlm} &= \frac{m}{2} \int d^3 r \psi_{i'j'}^* \phi_{ij},
\end{split}
\end{equation}
where we have omitted indices on the wavefunctions for convenience and use the kinematics (see Fig.~\ref{fig:vector emmision diagrams})
\begin{equation}
\begin{split}
    \vec{P} = \vec{p}_1 + \vec{p}_2, \qquad \vec{K} = \vec{k}_1 + \vec{k}_2 \\
    \vec{p}_1 = \frac{\vec{P}}{2} + \vec{p}, \qquad \vec{k}_1 = \frac{\vec{K}}{2} + \vec{k} \\
    \vec{p}_2 = \frac{\vec{P}}{2} - \vec{p}, \qquad \vec{k}_2 = \frac{\vec{K}}{2} - \vec{k}.
\end{split}
\end{equation}

\subsection{$W/B$ emission}

For the $W/B$ emission diagram in the top left of Fig.~\ref{fig:vector emmision diagrams},
\begin{equation}
\begin{split}
    i \mathcal{M} = &\overline{u}_{j'}(k_2) (\hat{\epsilon}_{X_2})_{j'j} u_j(p_2) \overline{u}_{i'}(k_1) \\ &(ig_2(t^a_{X_1})_{i'k} (\hat{\epsilon}_{X_1})_{ki}  + ig_1 Y_{X_1} (\hat{\epsilon}_{X_1})_{i'i}) \gamma^{\mu} u_i(p_1) \epsilon_\mu^*(q).
\end{split}
\end{equation}

We only need the spatial part of $\mathcal{M}^\mu$. Using the non-relativistic approximations and dressing with the scattering and bound state wavefunctions
\begin{equation}
\begin{split}
    \vec{\mathcal{M}} = &2m (g_2(t^a_{X_1})_{i'k} (\hat{\epsilon}_{X_1})_{ki} + g_1 Y_{X_1} (\hat{\epsilon}_{X_1})_{i'i}) (\hat{\epsilon}_{X_2})_{j'j} \frac{1}{\sqrt{m}} \\
    & \int \frac{d^3 \vec{k}}{(2 \pi)^3} \frac{d^3 \vec{p}}{(2 \pi)^3} \psi_{i'j'}^*(\vec{k}) (\vec{p} + \vec{k}) \phi_{ij}(\vec{p}) \delta(\vec{p}_2 - \vec{k}_2) (2 \pi)^3.
\end{split}
\label{eq:W/B emission amplitude 1}
\end{equation}
Because $|\vec{q}|$ is small, $\vec{P} \approx \vec{K}$ so
\begin{equation}
\begin{split}
    \vec{\mathcal{M}} \rightarrow &4 \sqrt{m} \left( g_2(t^a_{X_1})_{i'k} (\hat{\epsilon}_{X_1})_{ki} + g_1 Y_{X_1} (\hat{\epsilon}_{X_1})_{i'i} \right) (\hat{\epsilon}_{X_2})_{j'j} \\
    &\int \frac{d^3 p}{(2 \pi)^3} \psi_{i'j'}^*(\vec{p}) \vec{p} \phi_{ij}({\vec{p}}).
    \label{eq:W/B emission amplitude 2}
\end{split}
\end{equation}
Fourier transforming gives us our result
\begin{equation}
\begin{split}
    \vec{\mathcal{M}} \rightarrow &- 4 i \sqrt{m} \left(g_2(t^a_{X_1})_{i'k} (\hat{\epsilon}_{X_1})_{ki} + g_1 Y_{X_1} (\hat{\epsilon}_{X_1})_{i'i}\right) \\
    &(\hat{\epsilon}_{X_2})_{j'j} \vec{J}^{ij,i'j'}_{p,n'l'm'}.
    \label{eq:W/B emission amplitude final}
\end{split}
\end{equation}

For emission from the $X_2$ leg of the diagram (the top right of Fig.~\ref{fig:vector emmision diagrams}, we have $\vec{p}_1 + \vec{k}_1 \rightarrow \vec{p}_2 + \vec{k}_2 = -(\vec{p} + \vec{k})$. The $\delta$-function argument gets an overall negative sign, but this does not change the result because the relative sign between $\vec{p}$ and $\vec{k}$ is unchanged. So the result is given by Eq.~(\ref{eq:W/B emission amplitude final}) with an overall minus, appropriately changing $SU(2)_L$ indices, and $Y_{X_1} \rightarrow Y_{X_2}$.

When $X_1 = X_2$, we also have $u$-channel diagrams. This introduces a relative minus sign between arguments in the $\delta$-function so $\psi^*(\vec{p}) \rightarrow \psi^*(- \vec{p})$ in Eq.~(\ref{eq:W/B emission amplitude 2}). The effect on the cross section from the symmetries of the wavefunction are discussed in Sec.~\ref{sec:bound states}.

We can also emit a $W$ boson through the three boson coupling in the bottom of Fig.~\ref{fig:vector emmision diagrams}. This amplitude, when simplified to the non-relativistic limit and dressed with the scattering and bound state wavefunctions is
\begin{equation}
    i \mathcal{\vec{M}} = i 8 \sqrt{m} g_2 (t^b_{X_1})_{i'k} (\hat{\epsilon}_{X_1})_{ki} (t^c_{X_2})_{j'l} (\hat{\epsilon}_{X_2})_{lj} f^{abc} \vec{T}^{ij,i'j'}_{p,n'l'm'}\,.
\end{equation}

\subsection{$H$ emission}

For the $H$ emission diagram on the left in Fig.~\ref{fig:MM H emmision diagrams},
\begin{equation}
\begin{split}
    &i \mathcal{M} = \\
    &(-i y_1) \overline{u}_{j'} (k_2) u_j(p_2) (\hat{\epsilon}_{M})_{j'j} \overline{u}_{i'}(k_1) u_i(p_1) (\hat{\epsilon}_{D})_{ki} \delta^{I^3_M(i')}_{I^3_D(k) \pm 1/2}.
\end{split}
\end{equation}
Dressing with the scattering and bound state wavefunctions and using non-relativistic approximations, we have
\begin{equation}
\begin{split}
    i \mathcal{M} = &-i y_1 4 m^2 (\hat{\epsilon}_{D})_{ki} \delta^{I^3_M(i')}_{I^3_D(k) \pm 1/2}(\hat{\epsilon}_{M})_{j'j} \frac{1}{\sqrt{m}} \\
    &\int \frac{d^3p}{(2 \pi)^3} \frac{d^3k}{(2 \pi)^3} \psi_{i'j'}^*(\vec{k}) \phi_{ij}(\vec{p}) \delta(\vec{p}_2 - \vec{k}_2) (2 \pi)^3.
\end{split}
\end{equation}
Using the small $|\vec{q}|$ approximation  and Fourier transforming to position space 
\begin{equation}
    i \mathcal{M} = -8 i y_1 \sqrt{m} (\hat{\epsilon}_{D})_{ki} \delta^{I^3_M(i')}_{I^3_D(k) \pm 1/2}(\hat{\epsilon}_{M})_{j'j} F^{ij,i'j'}_{p,n'l'm'}.
    \label{eq:H emission amplitude final}
\end{equation}

Swapping $M \rightleftharpoons D, H \rightarrow H^*$ changes $y_1 \rightarrow y_2$ in Eq.~(\ref{eq:H emission amplitude final}). However, this has no effect when we square the amplitude because $|y_1|=|y_2|$. Emitting from the $p_2 \rightarrow k_2$ leg only changes the overall sign of the $\delta$-function argument, so this also has no effect on the amplitude. When we have $u$-channel diagrams, the relative sign in the $\delta$-function argument changes. This changes the sign of the argument of the $BS$ wavefunction, $\psi$, and we obtain the symmetry factors discussed in Sec.~\ref{sec:bound states}.

\begin{figure}
\begin{fmffile}{diagram15}
\begin{fmfgraph*}(80,80)
   \fmfstraight
   \fmfleft{i1,i2,i3,i4,i5}
   \fmfright{o1,o2,o3,o4,o5}
   \fmf{plain}{i2,v2}
   \fmf{plain}{v2,o2}
   \fmf{plain}{i4,v4}
   \fmf{plain}{v4,o4}
   \fmf{phantom}{i3,v3}
   \fmf{dashes}{v3,o3}
   \fmf{photon,tension=0}{v3,v4}
   \fmf{dashes, tension=0}{v2,v3}
   \fmflabel{$X_{2,j}$}{i2}
   \fmflabel{$X_{2',j'}$}{o2}
   \fmflabel{$X_{1,i}$}{i4}
   \fmflabel{$X_{1,i'}$}{o4}
   \fmflabel{$H$}{o3}
\end{fmfgraph*}
\end{fmffile}
\caption{Bound-state formation with Higgs emission and scalar vector fusion.}
\label{fig:bs fusion}
\end{figure}
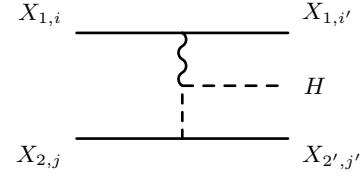

Figure~\ref{fig:bs fusion} shows a diagram which also produces bound states through $H$ emission, however it is suppressed by higher orders of the coupling constants~\cite{Oncala:2021tkz}. Therefore we ignore this bound-state formation mechanism.


\clearpage
\twocolumngrid
\bibliography{bibliography}

\begin{thebibliography}{74}%
\makeatletter
\providecommand \@ifxundefined [1]{%
 \@ifx{#1\undefined}
}%
\providecommand \@ifnum [1]{%
 \ifnum #1\expandafter \@firstoftwo
 \else \expandafter \@secondoftwo
 \fi
}%
\providecommand \@ifx [1]{%
 \ifx #1\expandafter \@firstoftwo
 \else \expandafter \@secondoftwo
 \fi
}%
\providecommand \natexlab [1]{#1}%
\providecommand \enquote  [1]{``#1''}%
\providecommand \bibnamefont  [1]{#1}%
\providecommand \bibfnamefont [1]{#1}%
\providecommand \citenamefont [1]{#1}%
\providecommand \href@noop [0]{\@secondoftwo}%
\providecommand \href [0]{\begingroup \@sanitize@url \@href}%
\providecommand \@href[1]{\@@startlink{#1}\@@href}%
\providecommand \@@href[1]{\endgroup#1\@@endlink}%
\providecommand \@sanitize@url [0]{\catcode `\\12\catcode `\$12\catcode
  `\&12\catcode `\#12\catcode `\^12\catcode `\_12\catcode `\%12\relax}%
\providecommand \@@startlink[1]{}%
\providecommand \@@endlink[0]{}%
\providecommand \url  [0]{\begingroup\@sanitize@url \@url }%
\providecommand \@url [1]{\endgroup\@href {#1}{\urlprefix }}%
\providecommand \urlprefix  [0]{URL }%
\providecommand \Eprint [0]{\href }%
\providecommand \doibase [0]{http://dx.doi.org/}%
\providecommand \selectlanguage [0]{\@gobble}%
\providecommand \bibinfo  [0]{\@secondoftwo}%
\providecommand \bibfield  [0]{\@secondoftwo}%
\providecommand \translation [1]{[#1]}%
\providecommand \BibitemOpen [0]{}%
\providecommand \bibitemStop [0]{}%
\providecommand \bibitemNoStop [0]{.\EOS\space}%
\providecommand \EOS [0]{\spacefactor3000\relax}%
\providecommand \BibitemShut  [1]{\csname bibitem#1\endcsname}%
\let\auto@bib@innerbib\@empty
\bibitem [{\citenamefont {Bertone}\ and\ \citenamefont
  {Hooper}(2018)}]{Bertone_2018}%
  \BibitemOpen
  \bibfield  {author} {\bibinfo {author} {\bibfnamefont {G.}~\bibnamefont
  {Bertone}}\ and\ \bibinfo {author} {\bibfnamefont {D.}~\bibnamefont
  {Hooper}},\ }\href {\doibase 10.1103/revmodphys.90.045002} {\bibfield
  {journal} {\bibinfo  {journal} {Reviews of Modern Physics}\ }\textbf
  {\bibinfo {volume} {90}} (\bibinfo {year} {2018}),\
  10.1103/revmodphys.90.045002}\BibitemShut {NoStop}%
\bibitem [{\citenamefont {Carr}\ \emph {et~al.}(2017)\citenamefont {Carr},
  \citenamefont {Raidal}, \citenamefont {Tenkanen}, \citenamefont {Vaskonen},\
  and\ \citenamefont {Veerm{\"a}e}}]{Carr:2017jsz}%
  \BibitemOpen
  \bibfield  {author} {\bibinfo {author} {\bibfnamefont {B.}~\bibnamefont
  {Carr}}, \bibinfo {author} {\bibfnamefont {M.}~\bibnamefont {Raidal}},
  \bibinfo {author} {\bibfnamefont {T.}~\bibnamefont {Tenkanen}}, \bibinfo
  {author} {\bibfnamefont {V.}~\bibnamefont {Vaskonen}}, \ and\ \bibinfo
  {author} {\bibfnamefont {H.}~\bibnamefont {Veerm{\"a}e}},\ }\href {\doibase
  10.1103/PhysRevD.96.023514} {\bibfield  {journal} {\bibinfo  {journal} {Phys.
  Rev. D}\ }\textbf {\bibinfo {volume} {96}},\ \bibinfo {pages} {023514}
  (\bibinfo {year} {2017})},\ \Eprint {http://arxiv.org/abs/1705.05567}
  {arXiv:1705.05567 [astro-ph.CO]} \BibitemShut {NoStop}%
\bibitem [{\citenamefont {Ge}\ \emph {et~al.}(2019)\citenamefont {Ge},
  \citenamefont {Lawson},\ and\ \citenamefont {Zhitnitsky}}]{Ge:2019voa}%
  \BibitemOpen
  \bibfield  {author} {\bibinfo {author} {\bibfnamefont {S.}~\bibnamefont
  {Ge}}, \bibinfo {author} {\bibfnamefont {K.}~\bibnamefont {Lawson}}, \ and\
  \bibinfo {author} {\bibfnamefont {A.}~\bibnamefont {Zhitnitsky}},\ }\href
  {\doibase 10.1103/PhysRevD.99.116017} {\bibfield  {journal} {\bibinfo
  {journal} {Phys. Rev. D}\ }\textbf {\bibinfo {volume} {99}},\ \bibinfo
  {pages} {116017} (\bibinfo {year} {2019})},\ \Eprint
  {http://arxiv.org/abs/1903.05090} {arXiv:1903.05090 [hep-ph]} \BibitemShut
  {NoStop}%
\bibitem [{\citenamefont {Cirelli}\ \emph {et~al.}(2019)\citenamefont
  {Cirelli}, \citenamefont {Gouttenoire}, \citenamefont {Petraki},\ and\
  \citenamefont {Sala}}]{Cirelli:2018iax}%
  \BibitemOpen
  \bibfield  {author} {\bibinfo {author} {\bibfnamefont {M.}~\bibnamefont
  {Cirelli}}, \bibinfo {author} {\bibfnamefont {Y.}~\bibnamefont
  {Gouttenoire}}, \bibinfo {author} {\bibfnamefont {K.}~\bibnamefont
  {Petraki}}, \ and\ \bibinfo {author} {\bibfnamefont {F.}~\bibnamefont
  {Sala}},\ }\href {\doibase 10.1088/1475-7516/2019/02/014} {\bibfield
  {journal} {\bibinfo  {journal} {JCAP}\ }\textbf {\bibinfo {volume} {02}},\
  \bibinfo {pages} {014} (\bibinfo {year} {2019})},\ \Eprint
  {http://arxiv.org/abs/1811.03608} {arXiv:1811.03608 [hep-ph]} \BibitemShut
  {NoStop}%
\bibitem [{\citenamefont {Berges}\ \emph {et~al.}(2019)\citenamefont {Berges},
  \citenamefont {Chatrchyan},\ and\ \citenamefont {Jaeckel}}]{Berges:2019dgr}%
  \BibitemOpen
  \bibfield  {author} {\bibinfo {author} {\bibfnamefont {J.}~\bibnamefont
  {Berges}}, \bibinfo {author} {\bibfnamefont {A.}~\bibnamefont {Chatrchyan}},
  \ and\ \bibinfo {author} {\bibfnamefont {J.}~\bibnamefont {Jaeckel}},\ }\href
  {\doibase 10.1088/1475-7516/2019/08/020} {\bibfield  {journal} {\bibinfo
  {journal} {JCAP}\ }\textbf {\bibinfo {volume} {08}},\ \bibinfo {pages} {020}
  (\bibinfo {year} {2019})},\ \Eprint {http://arxiv.org/abs/1903.03116}
  {arXiv:1903.03116 [hep-ph]} \BibitemShut {NoStop}%
\bibitem [{\citenamefont {Blennow}\ \emph {et~al.}(2019)\citenamefont
  {Blennow}, \citenamefont {Fernandez-Martinez}, \citenamefont
  {Olivares-Del~Campo}, \citenamefont {Pascoli}, \citenamefont
  {Rosauro-Alcaraz},\ and\ \citenamefont {Titov}}]{Blennow:2019fhy}%
  \BibitemOpen
  \bibfield  {author} {\bibinfo {author} {\bibfnamefont {M.}~\bibnamefont
  {Blennow}}, \bibinfo {author} {\bibfnamefont {E.}~\bibnamefont
  {Fernandez-Martinez}}, \bibinfo {author} {\bibfnamefont {A.}~\bibnamefont
  {Olivares-Del~Campo}}, \bibinfo {author} {\bibfnamefont {S.}~\bibnamefont
  {Pascoli}}, \bibinfo {author} {\bibfnamefont {S.}~\bibnamefont
  {Rosauro-Alcaraz}}, \ and\ \bibinfo {author} {\bibfnamefont {A.~V.}\
  \bibnamefont {Titov}},\ }\href {\doibase 10.1140/epjc/s10052-019-7060-5}
  {\bibfield  {journal} {\bibinfo  {journal} {Eur. Phys. J. C}\ }\textbf
  {\bibinfo {volume} {79}},\ \bibinfo {pages} {555} (\bibinfo {year} {2019})},\
  \Eprint {http://arxiv.org/abs/1903.00006} {arXiv:1903.00006 [hep-ph]}
  \BibitemShut {NoStop}%
\bibitem [{\citenamefont {Arcadi}\ \emph {et~al.}(2020)\citenamefont {Arcadi},
  \citenamefont {Djouadi},\ and\ \citenamefont {Raidal}}]{Arcadi:2019lka}%
  \BibitemOpen
  \bibfield  {author} {\bibinfo {author} {\bibfnamefont {G.}~\bibnamefont
  {Arcadi}}, \bibinfo {author} {\bibfnamefont {A.}~\bibnamefont {Djouadi}}, \
  and\ \bibinfo {author} {\bibfnamefont {M.}~\bibnamefont {Raidal}},\ }\href
  {\doibase 10.1016/j.physrep.2019.11.003} {\bibfield  {journal} {\bibinfo
  {journal} {Phys. Rept.}\ }\textbf {\bibinfo {volume} {842}},\ \bibinfo
  {pages} {1} (\bibinfo {year} {2020})},\ \Eprint
  {http://arxiv.org/abs/1903.03616} {arXiv:1903.03616 [hep-ph]} \BibitemShut
  {NoStop}%
\bibitem [{\citenamefont {Allen}\ \emph {et~al.}(2011)\citenamefont {Allen},
  \citenamefont {Evrard},\ and\ \citenamefont {Mantz}}]{Allen:2011zs}%
  \BibitemOpen
  \bibfield  {author} {\bibinfo {author} {\bibfnamefont {S.~W.}\ \bibnamefont
  {Allen}}, \bibinfo {author} {\bibfnamefont {A.~E.}\ \bibnamefont {Evrard}}, \
  and\ \bibinfo {author} {\bibfnamefont {A.~B.}\ \bibnamefont {Mantz}},\ }\href
  {\doibase 10.1146/annurev-astro-081710-102514} {\bibfield  {journal}
  {\bibinfo  {journal} {Ann. Rev. Astron. Astrophys.}\ }\textbf {\bibinfo
  {volume} {49}},\ \bibinfo {pages} {409} (\bibinfo {year} {2011})},\ \Eprint
  {http://arxiv.org/abs/1103.4829} {arXiv:1103.4829 [astro-ph.CO]} \BibitemShut
  {NoStop}%
\bibitem [{\citenamefont {Salucci}(2019)}]{Salucci:2018hqu}%
  \BibitemOpen
  \bibfield  {author} {\bibinfo {author} {\bibfnamefont {P.}~\bibnamefont
  {Salucci}},\ }\href {\doibase 10.1007/s00159-018-0113-1} {\bibfield
  {journal} {\bibinfo  {journal} {Astron. Astrophys. Rev.}\ }\textbf {\bibinfo
  {volume} {27}},\ \bibinfo {pages} {2} (\bibinfo {year} {2019})},\ \Eprint
  {http://arxiv.org/abs/1811.08843} {arXiv:1811.08843 [astro-ph.GA]}
  \BibitemShut {NoStop}%
\bibitem [{\citenamefont {Aghanim}\ \emph {et~al.}(2020)\citenamefont {Aghanim}
  \emph {et~al.}}]{Planck:2018vyg}%
  \BibitemOpen
  \bibfield  {author} {\bibinfo {author} {\bibfnamefont {N.}~\bibnamefont
  {Aghanim}} \emph {et~al.} (\bibinfo {collaboration} {Planck}),\ }\href
  {\doibase 10.1051/0004-6361/201833910} {\bibfield  {journal} {\bibinfo
  {journal} {Astron. Astrophys.}\ }\textbf {\bibinfo {volume} {641}},\ \bibinfo
  {pages} {A6} (\bibinfo {year} {2020})},\ \bibinfo {note} {[Erratum:
  Astron.Astrophys. 652, C4 (2021)]},\ \Eprint
  {http://arxiv.org/abs/1807.06209} {arXiv:1807.06209 [astro-ph.CO]}
  \BibitemShut {NoStop}%
\bibitem [{\citenamefont {Simon}(2019)}]{Simon:2019nxf}%
  \BibitemOpen
  \bibfield  {author} {\bibinfo {author} {\bibfnamefont {J.~D.}\ \bibnamefont
  {Simon}},\ }\href {\doibase 10.1146/annurev-astro-091918-104453} {\bibfield
  {journal} {\bibinfo  {journal} {Ann. Rev. Astron. Astrophys.}\ }\textbf
  {\bibinfo {volume} {57}},\ \bibinfo {pages} {375} (\bibinfo {year} {2019})},\
  \Eprint {http://arxiv.org/abs/1901.05465} {arXiv:1901.05465 [astro-ph.GA]}
  \BibitemShut {NoStop}%
\bibitem [{\citenamefont {Cirelli}\ \emph {et~al.}(2024)\citenamefont
  {Cirelli}, \citenamefont {Strumia},\ and\ \citenamefont
  {Zupan}}]{Cirelli:2024ssz}%
  \BibitemOpen
  \bibfield  {author} {\bibinfo {author} {\bibfnamefont {M.}~\bibnamefont
  {Cirelli}}, \bibinfo {author} {\bibfnamefont {A.}~\bibnamefont {Strumia}}, \
  and\ \bibinfo {author} {\bibfnamefont {J.}~\bibnamefont {Zupan}},\
  }\href@noop {} {\  (\bibinfo {year} {2024})},\ \Eprint
  {http://arxiv.org/abs/2406.01705} {arXiv:2406.01705 [hep-ph]} \BibitemShut
  {NoStop}%
\bibitem [{\citenamefont {Cirelli}\ \emph {et~al.}(2006)\citenamefont
  {Cirelli}, \citenamefont {Fornengo},\ and\ \citenamefont
  {Strumia}}]{Cirelli:2005uq}%
  \BibitemOpen
  \bibfield  {author} {\bibinfo {author} {\bibfnamefont {M.}~\bibnamefont
  {Cirelli}}, \bibinfo {author} {\bibfnamefont {N.}~\bibnamefont {Fornengo}}, \
  and\ \bibinfo {author} {\bibfnamefont {A.}~\bibnamefont {Strumia}},\ }\href
  {\doibase 10.1016/j.nuclphysb.2006.07.012} {\bibfield  {journal} {\bibinfo
  {journal} {Nucl. Phys. B}\ }\textbf {\bibinfo {volume} {753}},\ \bibinfo
  {pages} {178} (\bibinfo {year} {2006})},\ \Eprint
  {http://arxiv.org/abs/hep-ph/0512090} {arXiv:hep-ph/0512090} \BibitemShut
  {NoStop}%
\bibitem [{\citenamefont {Cirelli}\ \emph {et~al.}(2007)\citenamefont
  {Cirelli}, \citenamefont {Strumia},\ and\ \citenamefont
  {Tamburini}}]{Cirelli:2007xd}%
  \BibitemOpen
  \bibfield  {author} {\bibinfo {author} {\bibfnamefont {M.}~\bibnamefont
  {Cirelli}}, \bibinfo {author} {\bibfnamefont {A.}~\bibnamefont {Strumia}}, \
  and\ \bibinfo {author} {\bibfnamefont {M.}~\bibnamefont {Tamburini}},\ }\href
  {\doibase 10.1016/j.nuclphysb.2007.07.023} {\bibfield  {journal} {\bibinfo
  {journal} {Nucl. Phys. B}\ }\textbf {\bibinfo {volume} {787}},\ \bibinfo
  {pages} {152} (\bibinfo {year} {2007})},\ \Eprint
  {http://arxiv.org/abs/0706.4071} {arXiv:0706.4071 [hep-ph]} \BibitemShut
  {NoStop}%
\bibitem [{\citenamefont {Steigman}\ and\ \citenamefont
  {Turner}(1985)}]{Steigman:1984ac}%
  \BibitemOpen
  \bibfield  {author} {\bibinfo {author} {\bibfnamefont {G.}~\bibnamefont
  {Steigman}}\ and\ \bibinfo {author} {\bibfnamefont {M.~S.}\ \bibnamefont
  {Turner}},\ }\href {\doibase 10.1016/0550-3213(85)90537-1} {\bibfield
  {journal} {\bibinfo  {journal} {Nucl. Phys. B}\ }\textbf {\bibinfo {volume}
  {253}},\ \bibinfo {pages} {375} (\bibinfo {year} {1985})}\BibitemShut
  {NoStop}%
\bibitem [{\citenamefont {Bertone}\ \emph {et~al.}(2005)\citenamefont
  {Bertone}, \citenamefont {Hooper},\ and\ \citenamefont
  {Silk}}]{Bertone:2004pz}%
  \BibitemOpen
  \bibfield  {author} {\bibinfo {author} {\bibfnamefont {G.}~\bibnamefont
  {Bertone}}, \bibinfo {author} {\bibfnamefont {D.}~\bibnamefont {Hooper}}, \
  and\ \bibinfo {author} {\bibfnamefont {J.}~\bibnamefont {Silk}},\ }\href
  {\doibase 10.1016/j.physrep.2004.08.031} {\bibfield  {journal} {\bibinfo
  {journal} {Phys. Rept.}\ }\textbf {\bibinfo {volume} {405}},\ \bibinfo
  {pages} {279} (\bibinfo {year} {2005})},\ \Eprint
  {http://arxiv.org/abs/hep-ph/0404175} {arXiv:hep-ph/0404175} \BibitemShut
  {NoStop}%
\bibitem [{\citenamefont {Steigman}\ \emph {et~al.}(2012)\citenamefont
  {Steigman}, \citenamefont {Dasgupta},\ and\ \citenamefont
  {Beacom}}]{Steigman:2012nb}%
  \BibitemOpen
  \bibfield  {author} {\bibinfo {author} {\bibfnamefont {G.}~\bibnamefont
  {Steigman}}, \bibinfo {author} {\bibfnamefont {B.}~\bibnamefont {Dasgupta}},
  \ and\ \bibinfo {author} {\bibfnamefont {J.~F.}\ \bibnamefont {Beacom}},\
  }\href {\doibase 10.1103/PhysRevD.86.023506} {\bibfield  {journal} {\bibinfo
  {journal} {Phys. Rev. D}\ }\textbf {\bibinfo {volume} {86}},\ \bibinfo
  {pages} {023506} (\bibinfo {year} {2012})},\ \Eprint
  {http://arxiv.org/abs/1204.3622} {arXiv:1204.3622 [hep-ph]} \BibitemShut
  {NoStop}%
\bibitem [{\citenamefont {Arcadi}\ \emph {et~al.}(2018)\citenamefont {Arcadi},
  \citenamefont {Dutra}, \citenamefont {Ghosh}, \citenamefont {Lindner},
  \citenamefont {Mambrini}, \citenamefont {Pierre}, \citenamefont {Profumo},\
  and\ \citenamefont {Queiroz}}]{Arcadi:2017kky}%
  \BibitemOpen
  \bibfield  {author} {\bibinfo {author} {\bibfnamefont {G.}~\bibnamefont
  {Arcadi}}, \bibinfo {author} {\bibfnamefont {M.}~\bibnamefont {Dutra}},
  \bibinfo {author} {\bibfnamefont {P.}~\bibnamefont {Ghosh}}, \bibinfo
  {author} {\bibfnamefont {M.}~\bibnamefont {Lindner}}, \bibinfo {author}
  {\bibfnamefont {Y.}~\bibnamefont {Mambrini}}, \bibinfo {author}
  {\bibfnamefont {M.}~\bibnamefont {Pierre}}, \bibinfo {author} {\bibfnamefont
  {S.}~\bibnamefont {Profumo}}, \ and\ \bibinfo {author} {\bibfnamefont
  {F.~S.}\ \bibnamefont {Queiroz}},\ }\href {\doibase
  10.1140/epjc/s10052-018-5662-y} {\bibfield  {journal} {\bibinfo  {journal}
  {Eur. Phys. J. C}\ }\textbf {\bibinfo {volume} {78}},\ \bibinfo {pages} {203}
  (\bibinfo {year} {2018})},\ \Eprint {http://arxiv.org/abs/1703.07364}
  {arXiv:1703.07364 [hep-ph]} \BibitemShut {NoStop}%
\bibitem [{\citenamefont {Roszkowski}\ \emph {et~al.}(2018)\citenamefont
  {Roszkowski}, \citenamefont {Sessolo},\ and\ \citenamefont
  {Trojanowski}}]{Roszkowski:2017nbc}%
  \BibitemOpen
  \bibfield  {author} {\bibinfo {author} {\bibfnamefont {L.}~\bibnamefont
  {Roszkowski}}, \bibinfo {author} {\bibfnamefont {E.~M.}\ \bibnamefont
  {Sessolo}}, \ and\ \bibinfo {author} {\bibfnamefont {S.}~\bibnamefont
  {Trojanowski}},\ }\href {\doibase 10.1088/1361-6633/aab913} {\bibfield
  {journal} {\bibinfo  {journal} {Rept. Prog. Phys.}\ }\textbf {\bibinfo
  {volume} {81}},\ \bibinfo {pages} {066201} (\bibinfo {year} {2018})},\
  \Eprint {http://arxiv.org/abs/1707.06277} {arXiv:1707.06277 [hep-ph]}
  \BibitemShut {NoStop}%
\bibitem [{\citenamefont {Smirnov}(2023)}]{Smirnov:2022tcg}%
  \BibitemOpen
  \bibfield  {author} {\bibinfo {author} {\bibfnamefont {J.}~\bibnamefont
  {Smirnov}},\ }\href {\doibase 10.21468/SciPostPhysProc.12.003} {\bibfield
  {journal} {\bibinfo  {journal} {SciPost Phys. Proc.}\ }\textbf {\bibinfo
  {volume} {12}},\ \bibinfo {pages} {003} (\bibinfo {year} {2023})},\ \Eprint
  {http://arxiv.org/abs/2212.14361} {arXiv:2212.14361 [hep-ph]} \BibitemShut
  {NoStop}%
\bibitem [{\citenamefont {Hisano}\ \emph {et~al.}(2004)\citenamefont {Hisano},
  \citenamefont {Matsumoto},\ and\ \citenamefont {Nojiri}}]{Hisano:2003ec}%
  \BibitemOpen
  \bibfield  {author} {\bibinfo {author} {\bibfnamefont {J.}~\bibnamefont
  {Hisano}}, \bibinfo {author} {\bibfnamefont {S.}~\bibnamefont {Matsumoto}}, \
  and\ \bibinfo {author} {\bibfnamefont {M.~M.}\ \bibnamefont {Nojiri}},\
  }\href {\doibase 10.1103/PhysRevLett.92.031303} {\bibfield  {journal}
  {\bibinfo  {journal} {Phys. Rev. Lett.}\ }\textbf {\bibinfo {volume} {92}},\
  \bibinfo {pages} {031303} (\bibinfo {year} {2004})},\ \Eprint
  {http://arxiv.org/abs/hep-ph/0307216} {arXiv:hep-ph/0307216} \BibitemShut
  {NoStop}%
\bibitem [{\citenamefont {Hisano}\ \emph {et~al.}(2005)\citenamefont {Hisano},
  \citenamefont {Matsumoto}, \citenamefont {Nojiri},\ and\ \citenamefont
  {Saito}}]{Hisano:2004ds}%
  \BibitemOpen
  \bibfield  {author} {\bibinfo {author} {\bibfnamefont {J.}~\bibnamefont
  {Hisano}}, \bibinfo {author} {\bibfnamefont {S.}~\bibnamefont {Matsumoto}},
  \bibinfo {author} {\bibfnamefont {M.~M.}\ \bibnamefont {Nojiri}}, \ and\
  \bibinfo {author} {\bibfnamefont {O.}~\bibnamefont {Saito}},\ }\href
  {\doibase 10.1103/PhysRevD.71.063528} {\bibfield  {journal} {\bibinfo
  {journal} {Phys. Rev. D}\ }\textbf {\bibinfo {volume} {71}},\ \bibinfo
  {pages} {063528} (\bibinfo {year} {2005})},\ \Eprint
  {http://arxiv.org/abs/hep-ph/0412403} {arXiv:hep-ph/0412403} \BibitemShut
  {NoStop}%
\bibitem [{\citenamefont {Hisano}\ \emph {et~al.}(2007)\citenamefont {Hisano},
  \citenamefont {Matsumoto}, \citenamefont {Nagai}, \citenamefont {Saito},\
  and\ \citenamefont {Senami}}]{Hisano:2006nn}%
  \BibitemOpen
  \bibfield  {author} {\bibinfo {author} {\bibfnamefont {J.}~\bibnamefont
  {Hisano}}, \bibinfo {author} {\bibfnamefont {S.}~\bibnamefont {Matsumoto}},
  \bibinfo {author} {\bibfnamefont {M.}~\bibnamefont {Nagai}}, \bibinfo
  {author} {\bibfnamefont {O.}~\bibnamefont {Saito}}, \ and\ \bibinfo {author}
  {\bibfnamefont {M.}~\bibnamefont {Senami}},\ }\href {\doibase
  10.1016/j.physletb.2007.01.012} {\bibfield  {journal} {\bibinfo  {journal}
  {Phys. Lett. B}\ }\textbf {\bibinfo {volume} {646}},\ \bibinfo {pages} {34}
  (\bibinfo {year} {2007})},\ \Eprint {http://arxiv.org/abs/hep-ph/0610249}
  {arXiv:hep-ph/0610249} \BibitemShut {NoStop}%
\bibitem [{\citenamefont {Arkani-Hamed}\ \emph {et~al.}(2009)\citenamefont
  {Arkani-Hamed}, \citenamefont {Finkbeiner}, \citenamefont {Slatyer},\ and\
  \citenamefont {Weiner}}]{Arkani-Hamed:2008hhe}%
  \BibitemOpen
  \bibfield  {author} {\bibinfo {author} {\bibfnamefont {N.}~\bibnamefont
  {Arkani-Hamed}}, \bibinfo {author} {\bibfnamefont {D.~P.}\ \bibnamefont
  {Finkbeiner}}, \bibinfo {author} {\bibfnamefont {T.~R.}\ \bibnamefont
  {Slatyer}}, \ and\ \bibinfo {author} {\bibfnamefont {N.}~\bibnamefont
  {Weiner}},\ }\href {\doibase 10.1103/PhysRevD.79.015014} {\bibfield
  {journal} {\bibinfo  {journal} {Phys. Rev. D}\ }\textbf {\bibinfo {volume}
  {79}},\ \bibinfo {pages} {015014} (\bibinfo {year} {2009})},\ \Eprint
  {http://arxiv.org/abs/0810.0713} {arXiv:0810.0713 [hep-ph]} \BibitemShut
  {NoStop}%
\bibitem [{\citenamefont {Cassel}(2010)}]{Cassel:2009wt}%
  \BibitemOpen
  \bibfield  {author} {\bibinfo {author} {\bibfnamefont {S.}~\bibnamefont
  {Cassel}},\ }\href {\doibase 10.1088/0954-3899/37/10/105009} {\bibfield
  {journal} {\bibinfo  {journal} {J. Phys. G}\ }\textbf {\bibinfo {volume}
  {37}},\ \bibinfo {pages} {105009} (\bibinfo {year} {2010})},\ \Eprint
  {http://arxiv.org/abs/0903.5307} {arXiv:0903.5307 [hep-ph]} \BibitemShut
  {NoStop}%
\bibitem [{\citenamefont {March-Russell}\ and\ \citenamefont
  {West}(2009)}]{March-Russell:2008klu}%
  \BibitemOpen
  \bibfield  {author} {\bibinfo {author} {\bibfnamefont {J.~D.}\ \bibnamefont
  {March-Russell}}\ and\ \bibinfo {author} {\bibfnamefont {S.~M.}\ \bibnamefont
  {West}},\ }\href {\doibase 10.1016/j.physletb.2009.04.010} {\bibfield
  {journal} {\bibinfo  {journal} {Phys. Lett. B}\ }\textbf {\bibinfo {volume}
  {676}},\ \bibinfo {pages} {133} (\bibinfo {year} {2009})},\ \Eprint
  {http://arxiv.org/abs/0812.0559} {arXiv:0812.0559 [astro-ph]} \BibitemShut
  {NoStop}%
\bibitem [{\citenamefont {von Harling}\ and\ \citenamefont
  {Petraki}(2014)}]{vonHarling:2014kha}%
  \BibitemOpen
  \bibfield  {author} {\bibinfo {author} {\bibfnamefont {B.}~\bibnamefont {von
  Harling}}\ and\ \bibinfo {author} {\bibfnamefont {K.}~\bibnamefont
  {Petraki}},\ }\href {\doibase 10.1088/1475-7516/2014/12/033} {\bibfield
  {journal} {\bibinfo  {journal} {JCAP}\ }\textbf {\bibinfo {volume} {12}},\
  \bibinfo {pages} {033} (\bibinfo {year} {2014})},\ \Eprint
  {http://arxiv.org/abs/1407.7874} {arXiv:1407.7874 [hep-ph]} \BibitemShut
  {NoStop}%
\bibitem [{\citenamefont {An}\ \emph {et~al.}(2016)\citenamefont {An},
  \citenamefont {Wise},\ and\ \citenamefont {Zhang}}]{An:2016gad}%
  \BibitemOpen
  \bibfield  {author} {\bibinfo {author} {\bibfnamefont {H.}~\bibnamefont
  {An}}, \bibinfo {author} {\bibfnamefont {M.~B.}\ \bibnamefont {Wise}}, \ and\
  \bibinfo {author} {\bibfnamefont {Y.}~\bibnamefont {Zhang}},\ }\href
  {\doibase 10.1103/PhysRevD.93.115020} {\bibfield  {journal} {\bibinfo
  {journal} {Phys. Rev. D}\ }\textbf {\bibinfo {volume} {93}},\ \bibinfo
  {pages} {115020} (\bibinfo {year} {2016})},\ \Eprint
  {http://arxiv.org/abs/1604.01776} {arXiv:1604.01776 [hep-ph]} \BibitemShut
  {NoStop}%
\bibitem [{\citenamefont {Mitridate}\ \emph
  {et~al.}(2017{\natexlab{a}})\citenamefont {Mitridate}, \citenamefont {Redi},
  \citenamefont {Smirnov},\ and\ \citenamefont {Strumia}}]{Mitridate:2017izz}%
  \BibitemOpen
  \bibfield  {author} {\bibinfo {author} {\bibfnamefont {A.}~\bibnamefont
  {Mitridate}}, \bibinfo {author} {\bibfnamefont {M.}~\bibnamefont {Redi}},
  \bibinfo {author} {\bibfnamefont {J.}~\bibnamefont {Smirnov}}, \ and\
  \bibinfo {author} {\bibfnamefont {A.}~\bibnamefont {Strumia}},\ }\href
  {\doibase 10.1088/1475-7516/2017/05/006} {\bibfield  {journal} {\bibinfo
  {journal} {JCAP}\ }\textbf {\bibinfo {volume} {05}},\ \bibinfo {pages} {006}
  (\bibinfo {year} {2017}{\natexlab{a}})},\ \Eprint
  {http://arxiv.org/abs/1702.01141} {arXiv:1702.01141 [hep-ph]} \BibitemShut
  {NoStop}%
\bibitem [{\citenamefont {Bottaro}\ \emph {et~al.}(2022)\citenamefont
  {Bottaro}, \citenamefont {Buttazzo}, \citenamefont {Costa}, \citenamefont
  {Franceschini}, \citenamefont {Panci}, \citenamefont {Redigolo},\ and\
  \citenamefont {Vittorio}}]{Bottaro:2021snn}%
  \BibitemOpen
  \bibfield  {author} {\bibinfo {author} {\bibfnamefont {S.}~\bibnamefont
  {Bottaro}}, \bibinfo {author} {\bibfnamefont {D.}~\bibnamefont {Buttazzo}},
  \bibinfo {author} {\bibfnamefont {M.}~\bibnamefont {Costa}}, \bibinfo
  {author} {\bibfnamefont {R.}~\bibnamefont {Franceschini}}, \bibinfo {author}
  {\bibfnamefont {P.}~\bibnamefont {Panci}}, \bibinfo {author} {\bibfnamefont
  {D.}~\bibnamefont {Redigolo}}, \ and\ \bibinfo {author} {\bibfnamefont
  {L.}~\bibnamefont {Vittorio}},\ }\href {\doibase
  10.1140/epjc/s10052-021-09917-9} {\bibfield  {journal} {\bibinfo  {journal}
  {Eur. Phys. J. C}\ }\textbf {\bibinfo {volume} {82}},\ \bibinfo {pages} {31}
  (\bibinfo {year} {2022})},\ \Eprint {http://arxiv.org/abs/2107.09688}
  {arXiv:2107.09688 [hep-ph]} \BibitemShut {NoStop}%
\bibitem [{\citenamefont {Bloch}\ \emph {et~al.}(2025)\citenamefont {Bloch},
  \citenamefont {Bottaro}, \citenamefont {Redigolo},\ and\ \citenamefont
  {Vittorio}}]{Bloch:2024suj}%
  \BibitemOpen
  \bibfield  {author} {\bibinfo {author} {\bibfnamefont {I.~M.}\ \bibnamefont
  {Bloch}}, \bibinfo {author} {\bibfnamefont {S.}~\bibnamefont {Bottaro}},
  \bibinfo {author} {\bibfnamefont {D.}~\bibnamefont {Redigolo}}, \ and\
  \bibinfo {author} {\bibfnamefont {L.}~\bibnamefont {Vittorio}},\ }\href
  {\doibase 10.1007/JHEP08(2025)216} {\bibfield  {journal} {\bibinfo  {journal}
  {JHEP}\ }\textbf {\bibinfo {volume} {08}},\ \bibinfo {pages} {216} (\bibinfo
  {year} {2025})},\ \Eprint {http://arxiv.org/abs/2410.02723} {arXiv:2410.02723
  [hep-ph]} \BibitemShut {NoStop}%
\bibitem [{\citenamefont {Baudis}(2024)}]{Baudis:2024jnk}%
  \BibitemOpen
  \bibfield  {author} {\bibinfo {author} {\bibfnamefont {L.}~\bibnamefont
  {Baudis}},\ }\href {\doibase 10.1016/j.nuclphysb.2024.116473} {\bibfield
  {journal} {\bibinfo  {journal} {Nucl. Phys. B}\ }\textbf {\bibinfo {volume}
  {1003}},\ \bibinfo {pages} {116473} (\bibinfo {year} {2024})},\ \Eprint
  {http://arxiv.org/abs/2404.19524} {arXiv:2404.19524 [astro-ph.IM]}
  \BibitemShut {NoStop}%
\bibitem [{\citenamefont {Abdukerim}\ \emph {et~al.}(2025)\citenamefont
  {Abdukerim} \emph {et~al.}}]{PANDA-X:2024dlo}%
  \BibitemOpen
  \bibfield  {author} {\bibinfo {author} {\bibfnamefont {A.}~\bibnamefont
  {Abdukerim}} \emph {et~al.} (\bibinfo {collaboration} {PANDA-X, PandaX}),\
  }\href {\doibase 10.1007/s11433-024-2539-y} {\bibfield  {journal} {\bibinfo
  {journal} {Sci. China Phys. Mech. Astron.}\ }\textbf {\bibinfo {volume}
  {68}},\ \bibinfo {pages} {221011} (\bibinfo {year} {2025})},\ \Eprint
  {http://arxiv.org/abs/2402.03596} {arXiv:2402.03596 [hep-ex]} \BibitemShut
  {NoStop}%
\bibitem [{\citenamefont {Goodman}\ and\ \citenamefont
  {Witten}(1985)}]{Goodman:1984dc}%
  \BibitemOpen
  \bibfield  {author} {\bibinfo {author} {\bibfnamefont {M.~W.}\ \bibnamefont
  {Goodman}}\ and\ \bibinfo {author} {\bibfnamefont {E.}~\bibnamefont
  {Witten}},\ }\href {\doibase 10.1103/PhysRevD.31.3059} {\bibfield  {journal}
  {\bibinfo  {journal} {Phys. Rev. D}\ }\textbf {\bibinfo {volume} {31}},\
  \bibinfo {pages} {3059} (\bibinfo {year} {1985})}\BibitemShut {NoStop}%
\bibitem [{\citenamefont {Akerib}\ \emph {et~al.}(2006)\citenamefont {Akerib}
  \emph {et~al.}}]{CDMS:2005rss}%
  \BibitemOpen
  \bibfield  {author} {\bibinfo {author} {\bibfnamefont {D.~S.}\ \bibnamefont
  {Akerib}} \emph {et~al.} (\bibinfo {collaboration} {CDMS}),\ }\href {\doibase
  10.1103/PhysRevLett.96.011302} {\bibfield  {journal} {\bibinfo  {journal}
  {Phys. Rev. Lett.}\ }\textbf {\bibinfo {volume} {96}},\ \bibinfo {pages}
  {011302} (\bibinfo {year} {2006})},\ \Eprint
  {http://arxiv.org/abs/astro-ph/0509259} {arXiv:astro-ph/0509259} \BibitemShut
  {NoStop}%
\bibitem [{\citenamefont {Mahbubani}\ and\ \citenamefont
  {Senatore}(2006)}]{Mahbubani:2005pt}%
  \BibitemOpen
  \bibfield  {author} {\bibinfo {author} {\bibfnamefont {R.}~\bibnamefont
  {Mahbubani}}\ and\ \bibinfo {author} {\bibfnamefont {L.}~\bibnamefont
  {Senatore}},\ }\href {\doibase 10.1103/PhysRevD.73.043510} {\bibfield
  {journal} {\bibinfo  {journal} {Phys. Rev. D}\ }\textbf {\bibinfo {volume}
  {73}},\ \bibinfo {pages} {043510} (\bibinfo {year} {2006})},\ \Eprint
  {http://arxiv.org/abs/hep-ph/0510064} {arXiv:hep-ph/0510064} \BibitemShut
  {NoStop}%
\bibitem [{\citenamefont {D'Eramo}(2007)}]{DEramo:2007anh}%
  \BibitemOpen
  \bibfield  {author} {\bibinfo {author} {\bibfnamefont {F.}~\bibnamefont
  {D'Eramo}},\ }\href {\doibase 10.1103/PhysRevD.76.083522} {\bibfield
  {journal} {\bibinfo  {journal} {Phys. Rev. D}\ }\textbf {\bibinfo {volume}
  {76}},\ \bibinfo {pages} {083522} (\bibinfo {year} {2007})},\ \Eprint
  {http://arxiv.org/abs/0705.4493} {arXiv:0705.4493 [hep-ph]} \BibitemShut
  {NoStop}%
\bibitem [{\citenamefont {Enberg}\ \emph {et~al.}(2007)\citenamefont {Enberg},
  \citenamefont {Fox}, \citenamefont {Hall}, \citenamefont {Papaioannou},\ and\
  \citenamefont {Papucci}}]{Enberg:2007rp}%
  \BibitemOpen
  \bibfield  {author} {\bibinfo {author} {\bibfnamefont {R.}~\bibnamefont
  {Enberg}}, \bibinfo {author} {\bibfnamefont {P.~J.}\ \bibnamefont {Fox}},
  \bibinfo {author} {\bibfnamefont {L.~J.}\ \bibnamefont {Hall}}, \bibinfo
  {author} {\bibfnamefont {A.~Y.}\ \bibnamefont {Papaioannou}}, \ and\ \bibinfo
  {author} {\bibfnamefont {M.}~\bibnamefont {Papucci}},\ }\href {\doibase
  10.1088/1126-6708/2007/11/014} {\bibfield  {journal} {\bibinfo  {journal}
  {JHEP}\ }\textbf {\bibinfo {volume} {11}},\ \bibinfo {pages} {014} (\bibinfo
  {year} {2007})},\ \Eprint {http://arxiv.org/abs/0706.0918} {arXiv:0706.0918
  [hep-ph]} \BibitemShut {NoStop}%
\bibitem [{\citenamefont {Cohen}\ \emph {et~al.}(2012)\citenamefont {Cohen},
  \citenamefont {Kearney}, \citenamefont {Pierce},\ and\ \citenamefont
  {Tucker-Smith}}]{Cohen:2011ec}%
  \BibitemOpen
  \bibfield  {author} {\bibinfo {author} {\bibfnamefont {T.}~\bibnamefont
  {Cohen}}, \bibinfo {author} {\bibfnamefont {J.}~\bibnamefont {Kearney}},
  \bibinfo {author} {\bibfnamefont {A.}~\bibnamefont {Pierce}}, \ and\ \bibinfo
  {author} {\bibfnamefont {D.}~\bibnamefont {Tucker-Smith}},\ }\href {\doibase
  10.1103/PhysRevD.85.075003} {\bibfield  {journal} {\bibinfo  {journal} {Phys.
  Rev. D}\ }\textbf {\bibinfo {volume} {85}},\ \bibinfo {pages} {075003}
  (\bibinfo {year} {2012})},\ \Eprint {http://arxiv.org/abs/1109.2604}
  {arXiv:1109.2604 [hep-ph]} \BibitemShut {NoStop}%
\bibitem [{\citenamefont {Cheung}\ and\ \citenamefont
  {Sanford}(2014)}]{Cheung:2013dua}%
  \BibitemOpen
  \bibfield  {author} {\bibinfo {author} {\bibfnamefont {C.}~\bibnamefont
  {Cheung}}\ and\ \bibinfo {author} {\bibfnamefont {D.}~\bibnamefont
  {Sanford}},\ }\href {\doibase 10.1088/1475-7516/2014/02/011} {\bibfield
  {journal} {\bibinfo  {journal} {JCAP}\ }\textbf {\bibinfo {volume} {02}},\
  \bibinfo {pages} {011} (\bibinfo {year} {2014})},\ \Eprint
  {http://arxiv.org/abs/1311.5896} {arXiv:1311.5896 [hep-ph]} \BibitemShut
  {NoStop}%
\bibitem [{\citenamefont {Calibbi}\ \emph {et~al.}(2015)\citenamefont
  {Calibbi}, \citenamefont {Mariotti},\ and\ \citenamefont
  {Tziveloglou}}]{Calibbi:2015nha}%
  \BibitemOpen
  \bibfield  {author} {\bibinfo {author} {\bibfnamefont {L.}~\bibnamefont
  {Calibbi}}, \bibinfo {author} {\bibfnamefont {A.}~\bibnamefont {Mariotti}}, \
  and\ \bibinfo {author} {\bibfnamefont {P.}~\bibnamefont {Tziveloglou}},\
  }\href {\doibase 10.1007/JHEP10(2015)116} {\bibfield  {journal} {\bibinfo
  {journal} {JHEP}\ }\textbf {\bibinfo {volume} {10}},\ \bibinfo {pages} {116}
  (\bibinfo {year} {2015})},\ \Eprint {http://arxiv.org/abs/1505.03867}
  {arXiv:1505.03867 [hep-ph]} \BibitemShut {NoStop}%
\bibitem [{\citenamefont {Freitas}\ \emph {et~al.}(2015)\citenamefont
  {Freitas}, \citenamefont {Westhoff},\ and\ \citenamefont
  {Zupan}}]{Freitas:2015hsa}%
  \BibitemOpen
  \bibfield  {author} {\bibinfo {author} {\bibfnamefont {A.}~\bibnamefont
  {Freitas}}, \bibinfo {author} {\bibfnamefont {S.}~\bibnamefont {Westhoff}}, \
  and\ \bibinfo {author} {\bibfnamefont {J.}~\bibnamefont {Zupan}},\ }\href
  {\doibase 10.1007/JHEP09(2015)015} {\bibfield  {journal} {\bibinfo  {journal}
  {JHEP}\ }\textbf {\bibinfo {volume} {09}},\ \bibinfo {pages} {015} (\bibinfo
  {year} {2015})},\ \Eprint {http://arxiv.org/abs/1506.04149} {arXiv:1506.04149
  [hep-ph]} \BibitemShut {NoStop}%
\bibitem [{\citenamefont {Banerjee}\ \emph {et~al.}(2016)\citenamefont
  {Banerjee}, \citenamefont {Matsumoto}, \citenamefont {Mukaida},\ and\
  \citenamefont {Tsai}}]{Banerjee:2016hsk}%
  \BibitemOpen
  \bibfield  {author} {\bibinfo {author} {\bibfnamefont {S.}~\bibnamefont
  {Banerjee}}, \bibinfo {author} {\bibfnamefont {S.}~\bibnamefont {Matsumoto}},
  \bibinfo {author} {\bibfnamefont {K.}~\bibnamefont {Mukaida}}, \ and\
  \bibinfo {author} {\bibfnamefont {Y.-L.~S.}\ \bibnamefont {Tsai}},\ }\href
  {\doibase 10.1007/JHEP11(2016)070} {\bibfield  {journal} {\bibinfo  {journal}
  {JHEP}\ }\textbf {\bibinfo {volume} {11}},\ \bibinfo {pages} {070} (\bibinfo
  {year} {2016})},\ \Eprint {http://arxiv.org/abs/1603.07387} {arXiv:1603.07387
  [hep-ph]} \BibitemShut {NoStop}%
\bibitem [{\citenamefont {Dedes}\ and\ \citenamefont
  {Karamitros}(2014)}]{Dedes:2014hga}%
  \BibitemOpen
  \bibfield  {author} {\bibinfo {author} {\bibfnamefont {A.}~\bibnamefont
  {Dedes}}\ and\ \bibinfo {author} {\bibfnamefont {D.}~\bibnamefont
  {Karamitros}},\ }\href {\doibase 10.1103/PhysRevD.89.115002} {\bibfield
  {journal} {\bibinfo  {journal} {Phys. Rev. D}\ }\textbf {\bibinfo {volume}
  {89}},\ \bibinfo {pages} {115002} (\bibinfo {year} {2014})},\ \Eprint
  {http://arxiv.org/abs/1403.7744} {arXiv:1403.7744 [hep-ph]} \BibitemShut
  {NoStop}%
\bibitem [{\citenamefont {Beneke}\ \emph {et~al.}(2017)\citenamefont {Beneke},
  \citenamefont {Bharucha}, \citenamefont {Hryczuk}, \citenamefont
  {Recksiegel},\ and\ \citenamefont {Ruiz-Femenia}}]{Beneke:2016jpw}%
  \BibitemOpen
  \bibfield  {author} {\bibinfo {author} {\bibfnamefont {M.}~\bibnamefont
  {Beneke}}, \bibinfo {author} {\bibfnamefont {A.}~\bibnamefont {Bharucha}},
  \bibinfo {author} {\bibfnamefont {A.}~\bibnamefont {Hryczuk}}, \bibinfo
  {author} {\bibfnamefont {S.}~\bibnamefont {Recksiegel}}, \ and\ \bibinfo
  {author} {\bibfnamefont {P.}~\bibnamefont {Ruiz-Femenia}},\ }\href {\doibase
  10.1007/JHEP01(2017)002} {\bibfield  {journal} {\bibinfo  {journal} {JHEP}\
  }\textbf {\bibinfo {volume} {01}},\ \bibinfo {pages} {002} (\bibinfo {year}
  {2017})},\ \Eprint {http://arxiv.org/abs/1611.00804} {arXiv:1611.00804
  [hep-ph]} \BibitemShut {NoStop}%
\bibitem [{\citenamefont {Tait}\ and\ \citenamefont {Yu}(2016)}]{Tait:2016qbg}%
  \BibitemOpen
  \bibfield  {author} {\bibinfo {author} {\bibfnamefont {T.~M.~P.}\
  \bibnamefont {Tait}}\ and\ \bibinfo {author} {\bibfnamefont {Z.-H.}\
  \bibnamefont {Yu}},\ }\href {\doibase 10.1007/JHEP03(2016)204} {\bibfield
  {journal} {\bibinfo  {journal} {JHEP}\ }\textbf {\bibinfo {volume} {03}},\
  \bibinfo {pages} {204} (\bibinfo {year} {2016})},\ \Eprint
  {http://arxiv.org/abs/1601.01354} {arXiv:1601.01354 [hep-ph]} \BibitemShut
  {NoStop}%
\bibitem [{\citenamefont {Lopez~Honorez}\ \emph {et~al.}(2018)\citenamefont
  {Lopez~Honorez}, \citenamefont {Tytgat}, \citenamefont {Tziveloglou},\ and\
  \citenamefont {Zaldivar}}]{LopezHonorez:2017zrd}%
  \BibitemOpen
  \bibfield  {author} {\bibinfo {author} {\bibfnamefont {L.}~\bibnamefont
  {Lopez~Honorez}}, \bibinfo {author} {\bibfnamefont {M.~H.~G.}\ \bibnamefont
  {Tytgat}}, \bibinfo {author} {\bibfnamefont {P.}~\bibnamefont {Tziveloglou}},
  \ and\ \bibinfo {author} {\bibfnamefont {B.}~\bibnamefont {Zaldivar}},\
  }\href {\doibase 10.1007/JHEP04(2018)011} {\bibfield  {journal} {\bibinfo
  {journal} {JHEP}\ }\textbf {\bibinfo {volume} {04}},\ \bibinfo {pages} {011}
  (\bibinfo {year} {2018})},\ \Eprint {http://arxiv.org/abs/1711.08619}
  {arXiv:1711.08619 [hep-ph]} \BibitemShut {NoStop}%
\bibitem [{\citenamefont {Oncala}\ and\ \citenamefont
  {Petraki}(2021{\natexlab{a}})}]{Oncala:2021tkz}%
  \BibitemOpen
  \bibfield  {author} {\bibinfo {author} {\bibfnamefont {R.}~\bibnamefont
  {Oncala}}\ and\ \bibinfo {author} {\bibfnamefont {K.}~\bibnamefont
  {Petraki}},\ }\href {\doibase 10.1007/JHEP06(2021)124} {\bibfield  {journal}
  {\bibinfo  {journal} {JHEP}\ }\textbf {\bibinfo {volume} {06}},\ \bibinfo
  {pages} {124} (\bibinfo {year} {2021}{\natexlab{a}})},\ \Eprint
  {http://arxiv.org/abs/2101.08666} {arXiv:2101.08666 [hep-ph]} \BibitemShut
  {NoStop}%
\bibitem [{\citenamefont {Oncala}\ and\ \citenamefont
  {Petraki}(2021{\natexlab{b}})}]{Oncala:2021swy}%
  \BibitemOpen
  \bibfield  {author} {\bibinfo {author} {\bibfnamefont {R.}~\bibnamefont
  {Oncala}}\ and\ \bibinfo {author} {\bibfnamefont {K.}~\bibnamefont
  {Petraki}},\ }\href {\doibase 10.1007/JHEP08(2021)069} {\bibfield  {journal}
  {\bibinfo  {journal} {JHEP}\ }\textbf {\bibinfo {volume} {08}},\ \bibinfo
  {pages} {069} (\bibinfo {year} {2021}{\natexlab{b}})},\ \Eprint
  {http://arxiv.org/abs/2101.08667} {arXiv:2101.08667 [hep-ph]} \BibitemShut
  {NoStop}%
\bibitem [{\citenamefont {Asadi}\ \emph {et~al.}(2017)\citenamefont {Asadi},
  \citenamefont {Baumgart}, \citenamefont {Fitzpatrick}, \citenamefont
  {Krupczak},\ and\ \citenamefont {Slatyer}}]{Asadi:2016ybp}%
  \BibitemOpen
  \bibfield  {author} {\bibinfo {author} {\bibfnamefont {P.}~\bibnamefont
  {Asadi}}, \bibinfo {author} {\bibfnamefont {M.}~\bibnamefont {Baumgart}},
  \bibinfo {author} {\bibfnamefont {P.~J.}\ \bibnamefont {Fitzpatrick}},
  \bibinfo {author} {\bibfnamefont {E.}~\bibnamefont {Krupczak}}, \ and\
  \bibinfo {author} {\bibfnamefont {T.~R.}\ \bibnamefont {Slatyer}},\ }\href
  {\doibase 10.1088/1475-7516/2017/02/005} {\bibfield  {journal} {\bibinfo
  {journal} {JCAP}\ }\textbf {\bibinfo {volume} {02}},\ \bibinfo {pages} {005}
  (\bibinfo {year} {2017})},\ \Eprint {http://arxiv.org/abs/1610.07617}
  {arXiv:1610.07617 [hep-ph]} \BibitemShut {NoStop}%
\bibitem [{\citenamefont {Smirnov}\ and\ \citenamefont
  {Beacom}(2019)}]{Smirnov:2019ngs}%
  \BibitemOpen
  \bibfield  {author} {\bibinfo {author} {\bibfnamefont {J.}~\bibnamefont
  {Smirnov}}\ and\ \bibinfo {author} {\bibfnamefont {J.~F.}\ \bibnamefont
  {Beacom}},\ }\href {\doibase 10.1103/PhysRevD.100.043029} {\bibfield
  {journal} {\bibinfo  {journal} {Phys. Rev. D}\ }\textbf {\bibinfo {volume}
  {100}},\ \bibinfo {pages} {043029} (\bibinfo {year} {2019})},\ \Eprint
  {http://arxiv.org/abs/1904.11503} {arXiv:1904.11503 [hep-ph]} \BibitemShut
  {NoStop}%
\bibitem [{\citenamefont {Hambye}\ \emph {et~al.}(2009)\citenamefont {Hambye},
  \citenamefont {Ling}, \citenamefont {Lopez~Honorez},\ and\ \citenamefont
  {Rocher}}]{Hambye:2009pw}%
  \BibitemOpen
  \bibfield  {author} {\bibinfo {author} {\bibfnamefont {T.}~\bibnamefont
  {Hambye}}, \bibinfo {author} {\bibfnamefont {F.~S.}\ \bibnamefont {Ling}},
  \bibinfo {author} {\bibfnamefont {L.}~\bibnamefont {Lopez~Honorez}}, \ and\
  \bibinfo {author} {\bibfnamefont {J.}~\bibnamefont {Rocher}},\ }\href
  {\doibase 10.1007/JHEP05(2010)066} {\bibfield  {journal} {\bibinfo  {journal}
  {JHEP}\ }\textbf {\bibinfo {volume} {07}},\ \bibinfo {pages} {090} (\bibinfo
  {year} {2009})},\ \bibinfo {note} {[Erratum: JHEP 05, 066 (2010)]},\ \Eprint
  {http://arxiv.org/abs/0903.4010} {arXiv:0903.4010 [hep-ph]} \BibitemShut
  {NoStop}%
\bibitem [{\citenamefont {Billard}\ \emph {et~al.}(2014)\citenamefont
  {Billard}, \citenamefont {Strigari},\ and\ \citenamefont
  {Figueroa-Feliciano}}]{Billard:2013qya}%
  \BibitemOpen
  \bibfield  {author} {\bibinfo {author} {\bibfnamefont {J.}~\bibnamefont
  {Billard}}, \bibinfo {author} {\bibfnamefont {L.}~\bibnamefont {Strigari}}, \
  and\ \bibinfo {author} {\bibfnamefont {E.}~\bibnamefont
  {Figueroa-Feliciano}},\ }\href {\doibase 10.1103/PhysRevD.89.023524}
  {\bibfield  {journal} {\bibinfo  {journal} {Phys. Rev. D}\ }\textbf {\bibinfo
  {volume} {89}},\ \bibinfo {pages} {023524} (\bibinfo {year} {2014})},\
  \Eprint {http://arxiv.org/abs/1307.5458} {arXiv:1307.5458 [hep-ph]}
  \BibitemShut {NoStop}%
\bibitem [{\citenamefont {Fan}\ and\ \citenamefont
  {Reece}(2013)}]{Fan:2013faa}%
  \BibitemOpen
  \bibfield  {author} {\bibinfo {author} {\bibfnamefont {J.}~\bibnamefont
  {Fan}}\ and\ \bibinfo {author} {\bibfnamefont {M.}~\bibnamefont {Reece}},\
  }\href {\doibase 10.1007/JHEP10(2013)124} {\bibfield  {journal} {\bibinfo
  {journal} {JHEP}\ }\textbf {\bibinfo {volume} {10}},\ \bibinfo {pages} {124}
  (\bibinfo {year} {2013})},\ \Eprint {http://arxiv.org/abs/1307.4400}
  {arXiv:1307.4400 [hep-ph]} \BibitemShut {NoStop}%
\bibitem [{\citenamefont {Cohen}\ \emph {et~al.}(2013)\citenamefont {Cohen},
  \citenamefont {Lisanti}, \citenamefont {Pierce},\ and\ \citenamefont
  {Slatyer}}]{Cohen:2013ama}%
  \BibitemOpen
  \bibfield  {author} {\bibinfo {author} {\bibfnamefont {T.}~\bibnamefont
  {Cohen}}, \bibinfo {author} {\bibfnamefont {M.}~\bibnamefont {Lisanti}},
  \bibinfo {author} {\bibfnamefont {A.}~\bibnamefont {Pierce}}, \ and\ \bibinfo
  {author} {\bibfnamefont {T.~R.}\ \bibnamefont {Slatyer}},\ }\href {\doibase
  10.1088/1475-7516/2013/10/061} {\bibfield  {journal} {\bibinfo  {journal}
  {JCAP}\ }\textbf {\bibinfo {volume} {10}},\ \bibinfo {pages} {061} (\bibinfo
  {year} {2013})},\ \Eprint {http://arxiv.org/abs/1307.4082} {arXiv:1307.4082
  [hep-ph]} \BibitemShut {NoStop}%
\bibitem [{\citenamefont {Rodd}\ \emph {et~al.}(2024)\citenamefont {Rodd},
  \citenamefont {Safdi},\ and\ \citenamefont {Xu}}]{Rodd:2024qsi}%
  \BibitemOpen
  \bibfield  {author} {\bibinfo {author} {\bibfnamefont {N.~L.}\ \bibnamefont
  {Rodd}}, \bibinfo {author} {\bibfnamefont {B.~R.}\ \bibnamefont {Safdi}}, \
  and\ \bibinfo {author} {\bibfnamefont {W.~L.}\ \bibnamefont {Xu}},\ }\href
  {\doibase 10.1103/PhysRevD.110.043003} {\bibfield  {journal} {\bibinfo
  {journal} {Phys. Rev. D}\ }\textbf {\bibinfo {volume} {110}},\ \bibinfo
  {pages} {043003} (\bibinfo {year} {2024})},\ \Eprint
  {http://arxiv.org/abs/2405.13104} {arXiv:2405.13104 [hep-ph]} \BibitemShut
  {NoStop}%
\bibitem [{\citenamefont {Safdi}\ and\ \citenamefont
  {Xu}(2025)}]{Safdi:2025sfs}%
  \BibitemOpen
  \bibfield  {author} {\bibinfo {author} {\bibfnamefont {B.~R.}\ \bibnamefont
  {Safdi}}\ and\ \bibinfo {author} {\bibfnamefont {W.~L.}\ \bibnamefont {Xu}},\
  }\href@noop {} {\  (\bibinfo {year} {2025})},\ \Eprint
  {http://arxiv.org/abs/2507.15934} {arXiv:2507.15934 [hep-ph]} \BibitemShut
  {NoStop}%
\bibitem [{\citenamefont {Aghaie}\ \emph {et~al.}(2025)\citenamefont {Aghaie},
  \citenamefont {Dondarini}, \citenamefont {Marino},\ and\ \citenamefont
  {Panci}}]{Aghaie:2025iyn}%
  \BibitemOpen
  \bibfield  {author} {\bibinfo {author} {\bibfnamefont {M.}~\bibnamefont
  {Aghaie}}, \bibinfo {author} {\bibfnamefont {A.}~\bibnamefont {Dondarini}},
  \bibinfo {author} {\bibfnamefont {G.}~\bibnamefont {Marino}}, \ and\ \bibinfo
  {author} {\bibfnamefont {P.}~\bibnamefont {Panci}},\ }\href@noop {} {\
  (\bibinfo {year} {2025})},\ \Eprint {http://arxiv.org/abs/2507.17607}
  {arXiv:2507.17607 [hep-ph]} \BibitemShut {NoStop}%
\bibitem [{\citenamefont {Baumgart}\ \emph {et~al.}(2026)\citenamefont
  {Baumgart}, \citenamefont {Bottaro}, \citenamefont {Redigolo}, \citenamefont
  {Rodd},\ and\ \citenamefont {Slatyer}}]{Baumgart:2025dov}%
  \BibitemOpen
  \bibfield  {author} {\bibinfo {author} {\bibfnamefont {M.}~\bibnamefont
  {Baumgart}}, \bibinfo {author} {\bibfnamefont {S.}~\bibnamefont {Bottaro}},
  \bibinfo {author} {\bibfnamefont {D.}~\bibnamefont {Redigolo}}, \bibinfo
  {author} {\bibfnamefont {N.~L.}\ \bibnamefont {Rodd}}, \ and\ \bibinfo
  {author} {\bibfnamefont {T.~R.}\ \bibnamefont {Slatyer}},\ }\href {\doibase
  10.1007/JHEP02(2026)213} {\bibfield  {journal} {\bibinfo  {journal} {JHEP}\
  }\textbf {\bibinfo {volume} {02}},\ \bibinfo {pages} {213} (\bibinfo {year}
  {2026})},\ \Eprint {http://arxiv.org/abs/2507.15937} {arXiv:2507.15937
  [hep-ph]} \BibitemShut {NoStop}%
\bibitem [{\citenamefont {Hisano}\ \emph {et~al.}(2015)\citenamefont {Hisano},
  \citenamefont {Ishiwata},\ and\ \citenamefont {Nagata}}]{Hisano:2015rsa}%
  \BibitemOpen
  \bibfield  {author} {\bibinfo {author} {\bibfnamefont {J.}~\bibnamefont
  {Hisano}}, \bibinfo {author} {\bibfnamefont {K.}~\bibnamefont {Ishiwata}}, \
  and\ \bibinfo {author} {\bibfnamefont {N.}~\bibnamefont {Nagata}},\ }\href
  {\doibase 10.1007/JHEP06(2015)097} {\bibfield  {journal} {\bibinfo  {journal}
  {JHEP}\ }\textbf {\bibinfo {volume} {06}},\ \bibinfo {pages} {097} (\bibinfo
  {year} {2015})},\ \Eprint {http://arxiv.org/abs/1504.00915} {arXiv:1504.00915
  [hep-ph]} \BibitemShut {NoStop}%
\bibitem [{\citenamefont {Mitridate}\ \emph
  {et~al.}(2017{\natexlab{b}})\citenamefont {Mitridate}, \citenamefont {Redi},
  \citenamefont {Smirnov},\ and\ \citenamefont {Strumia}}]{Mitridate:2017oky}%
  \BibitemOpen
  \bibfield  {author} {\bibinfo {author} {\bibfnamefont {A.}~\bibnamefont
  {Mitridate}}, \bibinfo {author} {\bibfnamefont {M.}~\bibnamefont {Redi}},
  \bibinfo {author} {\bibfnamefont {J.}~\bibnamefont {Smirnov}}, \ and\
  \bibinfo {author} {\bibfnamefont {A.}~\bibnamefont {Strumia}},\ }\href
  {\doibase 10.1007/JHEP10(2017)210} {\bibfield  {journal} {\bibinfo  {journal}
  {JHEP}\ }\textbf {\bibinfo {volume} {10}},\ \bibinfo {pages} {210} (\bibinfo
  {year} {2017}{\natexlab{b}})},\ \Eprint {http://arxiv.org/abs/1707.05380}
  {arXiv:1707.05380 [hep-ph]} \BibitemShut {NoStop}%
\bibitem [{\citenamefont {Dondi}\ \emph {et~al.}(2020)\citenamefont {Dondi},
  \citenamefont {Sannino},\ and\ \citenamefont {Smirnov}}]{Dondi:2019olm}%
  \BibitemOpen
  \bibfield  {author} {\bibinfo {author} {\bibfnamefont {N.~A.}\ \bibnamefont
  {Dondi}}, \bibinfo {author} {\bibfnamefont {F.}~\bibnamefont {Sannino}}, \
  and\ \bibinfo {author} {\bibfnamefont {J.}~\bibnamefont {Smirnov}},\ }\href
  {\doibase 10.1103/PhysRevD.101.103010} {\bibfield  {journal} {\bibinfo
  {journal} {Phys. Rev. D}\ }\textbf {\bibinfo {volume} {101}},\ \bibinfo
  {pages} {103010} (\bibinfo {year} {2020})},\ \Eprint
  {http://arxiv.org/abs/1905.08810} {arXiv:1905.08810 [hep-ph]} \BibitemShut
  {NoStop}%
\bibitem [{\citenamefont {Garny}\ and\ \citenamefont
  {Heisig}(2022)}]{Garny:2021qsr}%
  \BibitemOpen
  \bibfield  {author} {\bibinfo {author} {\bibfnamefont {M.}~\bibnamefont
  {Garny}}\ and\ \bibinfo {author} {\bibfnamefont {J.}~\bibnamefont {Heisig}},\
  }\href {\doibase 10.1103/PhysRevD.105.055004} {\bibfield  {journal} {\bibinfo
   {journal} {Phys. Rev. D}\ }\textbf {\bibinfo {volume} {105}},\ \bibinfo
  {pages} {055004} (\bibinfo {year} {2022})},\ \Eprint
  {http://arxiv.org/abs/2112.01499} {arXiv:2112.01499 [hep-ph]} \BibitemShut
  {NoStop}%
\bibitem [{\citenamefont {Navas}\ \emph {et~al.}(2024)\citenamefont {Navas}
  \emph {et~al.}}]{ParticleDataGroup:2024cfk}%
  \BibitemOpen
  \bibfield  {author} {\bibinfo {author} {\bibfnamefont {S.}~\bibnamefont
  {Navas}} \emph {et~al.} (\bibinfo {collaboration} {Particle Data Group}),\
  }\href {\doibase 10.1103/PhysRevD.110.030001} {\bibfield  {journal} {\bibinfo
   {journal} {Phys. Rev. D}\ }\textbf {\bibinfo {volume} {110}},\ \bibinfo
  {pages} {030001} (\bibinfo {year} {2024})}\BibitemShut {NoStop}%
\bibitem [{\citenamefont {Sakurai}(1967)}]{sakurai1967advanced}%
  \BibitemOpen
  \bibfield  {author} {\bibinfo {author} {\bibfnamefont {J.}~\bibnamefont
  {Sakurai}},\ }\href@noop {} {\emph {\bibinfo {title} {Advanced Quantum
  Mechanics}}}\ (\bibinfo  {publisher} {Addison-Wesley Publishing Company},\
  \bibinfo {year} {1967})\BibitemShut {NoStop}%
\bibitem [{\citenamefont {Berestetskii}\ \emph {et~al.}(1982)\citenamefont
  {Berestetskii}, \citenamefont {Lifshitz},\ and\ \citenamefont
  {Pitaevskii}}]{Berestetskii:1982qgu}%
  \BibitemOpen
  \bibfield  {author} {\bibinfo {author} {\bibfnamefont {V.~B.}\ \bibnamefont
  {Berestetskii}}, \bibinfo {author} {\bibfnamefont {E.~M.}\ \bibnamefont
  {Lifshitz}}, \ and\ \bibinfo {author} {\bibfnamefont {L.~P.}\ \bibnamefont
  {Pitaevskii}},\ }\href@noop {} {\emph {\bibinfo {title} {{Quantum
  Electrodynamics}}}},\ \bibinfo {series} {Course of Theoretical Physics},
  Vol.~\bibinfo {volume} {4}\ (\bibinfo  {publisher} {Pergamon Press},\
  \bibinfo {address} {Oxford},\ \bibinfo {year} {1982})\BibitemShut {NoStop}%
\bibitem [{\citenamefont {Peskin}\ and\ \citenamefont
  {Schroeder}(1995)}]{Peskin:1995ev}%
  \BibitemOpen
  \bibfield  {author} {\bibinfo {author} {\bibfnamefont {M.~E.}\ \bibnamefont
  {Peskin}}\ and\ \bibinfo {author} {\bibfnamefont {D.~V.}\ \bibnamefont
  {Schroeder}},\ }\href {\doibase 10.1201/9780429503559} {\emph {\bibinfo
  {title} {{An Introduction to quantum field theory}}}}\ (\bibinfo  {publisher}
  {Addison-Wesley},\ \bibinfo {address} {Reading, USA},\ \bibinfo {year}
  {1995})\BibitemShut {NoStop}%
\bibitem [{\citenamefont {Schiff}(1968)}]{schiff1968quantum}%
  \BibitemOpen
  \bibfield  {author} {\bibinfo {author} {\bibfnamefont {L.}~\bibnamefont
  {Schiff}},\ }\href@noop {} {\emph {\bibinfo {title} {Quantum Mechanics: 3rd
  Edition}}}\ (\bibinfo  {publisher} {McGraw Hill},\ \bibinfo {year}
  {1968})\BibitemShut {NoStop}%
\bibitem [{\citenamefont {Cohen-Tannoudji}\ \emph {et~al.}(2019)\citenamefont
  {Cohen-Tannoudji}, \citenamefont {Diu},\ and\ \citenamefont
  {Lalo{\"e}}}]{cohen2019quantum}%
  \BibitemOpen
  \bibfield  {author} {\bibinfo {author} {\bibfnamefont {C.}~\bibnamefont
  {Cohen-Tannoudji}}, \bibinfo {author} {\bibfnamefont {B.}~\bibnamefont
  {Diu}}, \ and\ \bibinfo {author} {\bibfnamefont {F.}~\bibnamefont
  {Lalo{\"e}}},\ }\href@noop {} {\emph {\bibinfo {title} {Quantum Mechanics,
  Volume 2: Angular Momentum, Spin, and Approximation Methods}}}\ (\bibinfo
  {publisher} {Wiley},\ \bibinfo {year} {2019})\BibitemShut {NoStop}%
\bibitem [{\citenamefont {Bethe}\ and\ \citenamefont
  {Salpeter}(1957)}]{bethe1957}%
  \BibitemOpen
  \bibfield  {author} {\bibinfo {author} {\bibfnamefont {H.~A.}\ \bibnamefont
  {Bethe}}\ and\ \bibinfo {author} {\bibfnamefont {E.~E.}\ \bibnamefont
  {Salpeter}},\ }\href@noop {} {\emph {\bibinfo {title} {Quantum Mechanics of
  One and Two-Electron Atoms}}}\ (\bibinfo  {publisher} {Springer Berlin,
  Heidelberg},\ \bibinfo {year} {1957})\BibitemShut {NoStop}%
\bibitem [{\citenamefont {Griest}\ and\ \citenamefont
  {Seckel}(1991)}]{Griest:1990kh}%
  \BibitemOpen
  \bibfield  {author} {\bibinfo {author} {\bibfnamefont {K.}~\bibnamefont
  {Griest}}\ and\ \bibinfo {author} {\bibfnamefont {D.}~\bibnamefont
  {Seckel}},\ }\href {\doibase 10.1103/PhysRevD.43.3191} {\bibfield  {journal}
  {\bibinfo  {journal} {Phys. Rev. D}\ }\textbf {\bibinfo {volume} {43}},\
  \bibinfo {pages} {3191} (\bibinfo {year} {1991})}\BibitemShut {NoStop}%
\bibitem [{\citenamefont {Aalbers}\ \emph {et~al.}(2025)\citenamefont {Aalbers}
  \emph {et~al.}}]{LZ:2024zvo}%
  \BibitemOpen
  \bibfield  {author} {\bibinfo {author} {\bibfnamefont {J.}~\bibnamefont
  {Aalbers}} \emph {et~al.} (\bibinfo {collaboration} {LZ}),\ }\href {\doibase
  10.1103/4dyc-z8zf} {\bibfield  {journal} {\bibinfo  {journal} {Phys. Rev.
  Lett.}\ }\textbf {\bibinfo {volume} {135}},\ \bibinfo {pages} {011802}
  (\bibinfo {year} {2025})},\ \Eprint {http://arxiv.org/abs/2410.17036}
  {arXiv:2410.17036 [hep-ex]} \BibitemShut {NoStop}%
\bibitem [{\citenamefont {Bo}\ \emph {et~al.}(2025)\citenamefont {Bo} \emph
  {et~al.}}]{PandaX:2024qfu}%
  \BibitemOpen
  \bibfield  {author} {\bibinfo {author} {\bibfnamefont {Z.}~\bibnamefont {Bo}}
  \emph {et~al.} (\bibinfo {collaboration} {PandaX}),\ }\href {\doibase
  10.1103/PhysRevLett.134.011805} {\bibfield  {journal} {\bibinfo  {journal}
  {Phys. Rev. Lett.}\ }\textbf {\bibinfo {volume} {134}},\ \bibinfo {pages}
  {011805} (\bibinfo {year} {2025})},\ \Eprint
  {http://arxiv.org/abs/2408.00664} {arXiv:2408.00664 [hep-ex]} \BibitemShut
  {NoStop}%
\bibitem [{\citenamefont {Billard}\ \emph {et~al.}(2012)\citenamefont
  {Billard}, \citenamefont {Mayet},\ and\ \citenamefont
  {Santos}}]{Billard:2011zj}%
  \BibitemOpen
  \bibfield  {author} {\bibinfo {author} {\bibfnamefont {J.}~\bibnamefont
  {Billard}}, \bibinfo {author} {\bibfnamefont {F.}~\bibnamefont {Mayet}}, \
  and\ \bibinfo {author} {\bibfnamefont {D.}~\bibnamefont {Santos}},\ }\href
  {\doibase 10.1103/PhysRevD.85.035006} {\bibfield  {journal} {\bibinfo
  {journal} {Phys. Rev. D}\ }\textbf {\bibinfo {volume} {85}},\ \bibinfo
  {pages} {035006} (\bibinfo {year} {2012})},\ \Eprint
  {http://arxiv.org/abs/1110.6079} {arXiv:1110.6079 [astro-ph.CO]} \BibitemShut
  {NoStop}%
\end{thebibliography}%

\end{document}